\documentclass[aps,prd,twocolumn,superscriptaddress,floatfix]{revtex4-2}

\usepackage{amsmath}
\usepackage{amssymb}
\usepackage{bm}
\usepackage{microtype}
\usepackage{hyperref}
\usepackage{siunitx}
\usepackage{xcolor}

\definecolor{red}{rgb}{0,0,0}

\usepackage{textcmds}

\usepackage{overpic}
\makeatletter
\define@key{Gin}{permille}[]{%
  \setkeys{Gin}{rel=\@m}%
}
\makeatother

\usepackage{etoolbox}
\apptocmd{\sloppy}{\hbadness 10000\relax}{}{}

\renewcommand{\eqref}[1]{Eq.~(\ref{#1})}
\newcommand*\Laplace{\mathop{}\!\mathbin\bigtriangleup}
\renewcommand{\phi}{\varphi}
\newcommand{\norm}[1]{\left\lVert#1\right\rVert}

\DeclareSIUnit\parsec{pc}

\begin{document}
% number in the upper righthand corner of the title page in preprint mode.
% Multiple \preprint commands are allowed.
% Use the 'preprintnumbers' class option to override journal defaults
% to display numbers if necessary
% \preprint{}

\title{%
    One-Dimensional Fuzzy Dark Matter Models:\texorpdfstring{\\}{ }
    Structure Growth and Asymptotic Dynamics
}

% repeat the \author .. \affiliation  etc. as needed
% \email, \thanks, \homepage, \altaffiliation all apply to the current
% author. Explanatory text should go in the []'s, actual e-mail
% address or url should go in the {}'s for \email and \homepage.
% Please use the appropriate macro foreach each type of information

% \affiliation command applies to all authors since the last
% \affiliation command. The \affiliation command should follow the
% other information
% \affiliation can be followed by \email, \homepage, \thanks as well.
\author{Tim Zimmermann}
\email[]{zimmermann@thphys.uni-heidelberg.de}
\affiliation{%
    Institut f\"ur Theoretische Physik, 
    Philosophenweg 16, 
    69120 Heidelberg, 
    Germany
}

\author{Nico Schwersenz}
\email[]{schwersenz@thphys.uni-heidelberg.de}
\affiliation{%
    Institut f\"ur Theoretische Physik, 
    Philosophenweg 16, 
    69120 Heidelberg, 
    Germany
}

\author{Massimo Pietroni}
\email[]{massimo.pietroni@unipr.it}
\affiliation{%
    Dipartimento di Scienze Matematiche, Fisiche e Informatiche, 
    Universit\'a di Parma,
    Campus Universitario, Parco Area delle Scienze n. 7/a, 
    43124 Parma, 
    Italy
}
\affiliation{%
    INFN, Sezione di Milano Bicocca, Gruppo Collegato di Parma, 
    43124 Parma, 
    Italy
}

\author{Sandro Wimberger}
\email[]{sandromarcel.wimberger@unipr.it}
\affiliation{%
    Dipartimento di Scienze Matematiche, Fisiche e Informatiche, 
    Universit\'a di Parma,
    Campus Universitario, Parco Area delle Scienze n. 7/a, 
    43124 Parma, 
    Italy
}
\affiliation{%
    INFN, Sezione di Milano Bicocca, Gruppo Collegato di Parma, 
    43124 Parma, 
    Italy
}

\date{\today}

\begin{abstract}
This paper investigates the feasibility of simulating Fuzzy Dark Matter (FDM)
with a reduced number of spatial dimensions. Our aim is to set up a realistic,
yet numerically inexpensive, toy model in $(1+1)$-dimensional space time, that ---
under well controlled system conditions --- is capable of realizing important aspects
of the full-fledged $(3+1)$-FDM phenomenology by means of one-dimensional analogues.
Based on the coupled, nonlinear and nonlocal $(3+1)$-\textcolor{red}{Schr\"odinger}-
Poisson equation under periodic boundary conditions,
we derive two distinct one-dimensional models that differ in their
transversal matter distribution and consequently in their nonlocal interaction
along the single dimension of interest.
We show that these discrepancies change the relaxation process of initial states as 
well as the asymptotic, i.e., thermalized and virialized, equilibrium state. 
Our investigation includes the dynamical evolution of artificial initial 
conditions for non-expanding space, as well as cosmological initial conditions in 
expanding space. 
The findings of this work are relevant for the interpretation of numerical simulation 
data modelling nonrelativistic fuzzy cold dark matter in reduced dimensions, 
in the quest for testing such models and for possible laboratory implementations of 
them.
\end{abstract}

\maketitle

\section{\label{sec:intro}Introduction}

Nonlinear Schr\"odinger equations are ubiquitous in physics. Let us just think
of interacting many-body problems in nonrelativistic condensed-matter theory that 
are reduced to a mean-field approximation that usually ends up in a nonlinear 
effective Schr\"odinger equation, see e.g. 
\cite{Fetter2003, Mahan2000, Dalfovo1999, Stringari2016}. A specific form of a such 
a nonlinear equation is the Schr\"odinger-Poisson, also known as Schr\"odinger-Newton 
equation, describing a scalar massive quantum particle in its own gravitations field. 
It finds many applications, e.g., in nonlinear optics 
\cite{Picozzi2011, Segev2015, Roger2016, Navarrete2017}, decoherence theory 
\cite{Diosi1984}, and, of course, in cosmology \cite{Ruffini1969}, whenever a 
nonrelativistic description of the particle suffices. 
Ref. \cite{Paredes2020} presents a recent review on the subject.

Here, we are interested in the Schr\"odinger-Poisson (SP) equation as an
alternative dark matter model to the established cold dark matter (CDM) paradigm. 
Hu {\em et al.} \cite{Hu2000} dubbed the matter so described fuzzy cold dark matter, 
or simply fuzzy dark matter (FDM), a notion that we will follow throughout the paper.

The most compelling feature of FDM obeying SP in four dimensional $(3+1)$
space-time is its distinct behavior on large and small spatial scales:
Assuming a dark matter particle mass of $m \approx \SI{e-22}{\electronvolt}$ ---
canonical for FDM --- cosmic structure growth under FDM is in accordance with CDM 
on super-galactic scales, e.g., identical matter power evolution or halo densities 
\cite{Schive2014}, sub-galactic scales are influenced by quantum mechanical effects.
In particular, one expects the existence of a solitonic state, with a flat, 
high-density core region, that is roughly speaking obtained from a balance of quantum 
pressure and gravity, \cite{Hu2000}. Structure smaller than the
solitonic core is suppressed, or smoothed out, by the uncertainty principle.
Thus, FDM may provide a natural solution to the small scale crisis of CDM 
\cite{Bullock2017}, and in particular its cusp-core problem,
\cite{Moore1994, deBlok2010, Mocz2017}, without the need of adding
sophisticated baryonic feedback processes.
From a fundamental physics point of view, SP simulations could predict the mass 
scales of the bosonic particles possibly constituting dark matter, of course in 
comparison with and constrained by experimental observation data 
\cite{Porayko2018, Amorisco2018, Lidz2018, Niemeyer2020}.
We note in passing that SP simulations may also be interpreted as an alternative 
sampling approach of the CDM phase space evolution that may be controlled by the
phase space resolution $\hbar/m$, see e.g. \cite{Widrow1993, Uhlemann2014,
Kopp2017,Mocz2018,Eberhardt2020}.

Numerical considerations of the $(3+1)$-SP equation, see e.g. 
\cite{Guzman2004, Guzman2006, Schive2014, Schwabe2016, Mocz2017}, 
identify the soliton state as a dynamical attractor in the time 
evolution of FDM. More precisely, overdense regions collapse under their self-gravity 
and during this process radiate away excess matter. 
The result of this process, sometimes dubbed gravitational cooling,
\cite{Seidel1994}, is a relaxed quantum matter distribution in which a solitonic core 
is immersed in a \textcolor{red}{\q{sea of fluctuations}}. 
The latter follows a power-law decay, with a radial profile, known as NFW profile, 
predicted in corresponding many-body simulations of CDM \cite{Navarro1996} to scale 
as $\rho \propto r^{-3}$ at large radii.

This paper addresses the question whether the time-asymptotic behavior just
described, or at least a similar scenario thereof, were recovered if only one spatial 
degree of freedom is available. 
In other words, can we derive a one-dimensional, yet sufficiently realistic, toy 
model that realizes one-dimensional analogues of the $(3+1)$-FDM structure formation, 
and in particular the mentioned solitonic core and its role as dynamical dynamical 
attractors? 
This includes, both the evolution to and the possible 
reaching of the asymptotic state, hence the relaxation process itself as well as its 
final product. Are the relaxation mechanisms in one-dimensional FDM the same as for 
3D FDM, i.e., is there a one-dimensional analogue of gravitational cooling? 
What are suitable astrophysical, quantum mechanical or statistical measures to
judge if the asymptotic state has been reached?
These questions have several motivations. 
Firstly, it is a priori not clear whether a model with 
reduced dimension will lead to the same evolution as the full-fledged three 
dimensional one, simply because assumptions on the matter configuration in the 
transversal dimensions must be made and, in higher dimensions, interdimensional 
coupling and redistribution of mass is, in principle, possible due to the
non-separable nonlinearity \cite{Modugno2004}. 
Secondly, it is needless to say that models with reduced dimensions lend themselves to 
much more detailed investigations even on relatively long spatial and temporal scales. 
Reliable high-precision numerical simulations are simply more efficient in one than 
in higher dimensions. 
Consequently, larger patches of the system's parameter space can
be covered with manageable resources. The reduced numerical complexity also allows
to generate large ensembles of simulations once a, in some sense optimal,
parameter set was identified, thereby reducing statistical uncertainties
without the need to invoke assumptions like ergodicity. Obviously, both aspects, 
i.e., parameter space coverage and statistical reliability, are vital for comparison 
with observational data.

The way of the reduction of dimensionality will certainly be important. 
Anticipating our results, we will see that the relaxation process in the evolution as 
well as in the finally reached states do depend strongly on how matter is
organized in the transversal dimensions.
The usual approach of assuming a uniform matter distribution in the transverse 
dimension, and hence in practice just forgetting about the other degrees of freedom, 
does not yield a correspondence with the $(3+1)$-FDM predictions. 
This motivated our second approach in which we confine the transverse degrees of 
freedom. 
Then, most of these predictions are actually seen in the reduced one-dimensional 
model as well. Our findings are based on a relative simple but highly flexible, 
efficient and precise numerical method to integrate both dimensionality-reduced SP 
equations. 
This allows us access to long temporal evolutions in a constant as well as in an 
expanding universe.

The paper is organized as follows: the next Sec. reviews the theory of FDM based on 
the SP equation. We discuss therein the importance of the reduction to $(1+1)$ 
(space+time) dimensions, and we compare, in particular, the two ways of doing this: 
first, we assume a uniform extension in the transverse directions, then secondly a 
cigar-shaped transversal confinement. Sec. \ref{sec:numerics} gives details on our 
numerical methods and its properties and quality. Sec. \ref{sec:results} reports our 
central results and compares the outcome of different numerical simulations of the 
two versions of reduction to $(1+1)$. 
The last Sec. \ref{sec:concl} concludes the paper with a short outlook on open 
questions and future work.

\section{\label{sec:theory}Theoretical background}

Point of departure for our discussion is the dynamics of Fuzzy Dark Matter (FDM) in
$(3+1)$ dimensions. To this end, we introduce its governing equation, comment
on its origin and mention different interpretations of the FDM model. 
After a quick review of the hydrodynamic formalism in the linear evolution regime, 
the remainder of this section is devoted to the problem of dimensional
reduction. More precisely, two competing one dimensional FDM representations are
derived and their physical discrepancies and similarities are highlighted. This
includes an in-depth discussion on the preparation and properties of the $(1+1)$-FDM 
ground state. We close this theory section by commenting on possible relaxation 
mechanisms and introduce suitable metrics for $(1+1)$-FDM to quantify if an 
equilibrated system state is reached.

\subsection{\label{sub:fdm3D}%
    Fuzzy Dark Matter in \texorpdfstring{$(3+1)$}{3D} Dimensions
}
\subsubsection{\label{subsub:governing_eq}Governing Equations}
For the sake of simplicity, envision a universe with negligible contribution of
baryonic matter and radiation towards the cosmic energy budget.
Furthermore, let dark energy be given by means of a time independent energy
density $\rho_\Lambda c^2$. 
Under these assumptions only dark matter dynamics plays a nontrivial role.

Starting from a massive, minimally coupled, scalar field, one derives, e.g.
\cite{Chavanis2012}, the $(3+1)$-Schr\"odinger-Poisson (SP) equation as governing 
equation of $(3+1)$-FDM applicable in the nonrelativistic limit $c\to\infty$:
\begin{align}
    \label{eq:sp3d}
    \begin{split}
        i\hbar\partial_t\Psi 
        &=\left[- \frac{\hbar^2}{2ma^2}\Laplace + m\Phi \right]\Psi \;, \\
        \Laplace \Phi &= \frac{4\pi G}{a}\left(|\Psi|^2 - \rho_m \right)  \;,
    \end{split}
    &&\bm x \in \Omega \;,
\end{align}
with $\Omega = \Omega_1 \times \Omega_2 \times \Omega_3 \subset \mathbb{R}^3$
and $\Omega_i = [0, L_i)$. 
Here, $\Psi$ denotes the nonrelativistic FDM field coupled to its own 
gravitational potential $\Phi$. Fields are evaluated at comoving position $\bm x
= \left(x_1, x_2, x_3\right)^\intercal$ and cosmic time $t$. Densities are
measured with respect to a comoving volume, so that
$\rho_m$ coincides with the present day total matter density. 
$m$ denotes the FDM particle mass. 
In accordance with our initial assumptions on the composition of the
cosmic energy budget, the scale factor $a(t)$ obeys the flat space,
radiation free Friedmann equation:
\begin{equation*}
    \left(\frac{\dot a}{a}\right)^2 
    = \frac{8\pi G}{3}\left(\rho_m a^{-3} + \rho_\Lambda\right) \;.
\end{equation*}
$G$ is Newton's constant.

Analyzing the asymptotic behavior of the energy momentum tensor associated with
$\Psi$ suggests the identification:
\begin{equation}
    \label{eq:psi_rho}
    |\Psi(\bm x,t)|^2 \stackrel{!}{=} \rho_m(\bm x,t) 
    \equiv \rho_m + \delta \rho(\bm x,t) \;,
\end{equation}
i.e. the scalar field encapsulated both the time-independent background density
$\rho_m$ and deviations from it, $\delta \rho(\bm x,t)$.
Since density deviations vanish upon 
averaging over $\Omega$, \eqref{eq:psi_rho} translates directly into a
normalization condition on $\Psi$:
\begin{equation}
    \label{eq:normalization}
    \rho_m = 
    \frac{1}{L_1L_2L_3}\int_\Omega \text{d}^3x |\Psi(\bm x,t)|^2 = \mathrm{const}\;.
\end{equation}
We emphasize \eqref{eq:normalization} is a physically relevant constraint as
\eqref{eq:sp3d} is \emph{nonlinear}. Therefore changing the
normalization of $\Psi$ will lead to a different time evolution.

Eq. (\ref{eq:sp3d}) still requires suitable boundary conditions
ooup
The natural choice for modelling an infinite system 
is to impose the following periodic boundary conditions:
\begin{align}
    \begin{split}
        \partial^m_{x_i}\Psi(0,x_j) &= \partial^m_{x_i}\Psi(L_i,x_j) \;, \\
        \partial^m_{x_i}\Phi(0,x_j) &= \partial^m_{x_i}\Phi(L_i,x_j) \;,
    \end{split}
    &
    \label{eq:pbc}
    \begin{split}
        m &\in \{0,1\} \;, \\
        i &\in \{1,2,3\} \;, \\
        j &\in \{1,2,3\} \setminus \{i\} \;.
    \end{split}
\end{align}

\subsubsection{\label{subsub:interpretation}%
    Interpretations of \texorpdfstring{$(3+1)$}{3D}-Schr\"odinger-Poisson
}

It is instructive to shed some light onto the physical character of the 
non-relativistic scalar $\Psi$. Although \eqref{eq:sp3d}
has the mathematical structure of Schr\"odinger's equation, there is a priori
nothing quantum mechanical about the problem.
Hence, the least arcane way to interpret \eqref{eq:sp3d} is in a literal sense, 
i.e. as the Euler-Lagrange equation of a classical Lagrangian and
$\hbar$ as a constant with dimensions of an action but with a numerical value
not constrained to Planck's constant.

That said, there is significant value in finding (formal) correspondences
between $(3+1)$-SP and other, potentially non-cosmological, theories as it 
enlarges the number of available tools with which FDM can be analyzed.

\paragraph{\label{par:bec} Cosmic Bose-Einstein Condensate}
In fact, \eqref{eq:sp3d} can also be identified with the evolution equation of a 
self-gravitating Bose-Einstein condensate with negligible local self-interaction
and $\Psi$ as the condensate \emph{wave function}.
The authors of \cite{Woo2009} substantiates this claim by comparing the critical 
temperature of an ultralight boson gas undergoing pair production 
with the cosmic microwave background temperature. 

A quantum mechanical derivation for \eqref{eq:sp3d} could then depart 
from a second quantized many-body Hamiltonian which is subsequently reduced to an 
effective Hamiltonian for the order parameter, i.e. the condensate wave function 
$\Psi$, \cite{Stringari2016}. 
The result would be again \eqref{eq:sp3d} with $\hbar$ as Planck's constant.

We return to this quantum mechanical interpretation in Sec.~
\ref{sec:numerics} when the time integration is formulated as approximation to time 
evolution operator or in Sec.~\ref{subsub:virialization} 
when the quantum virial theorem is used to analyze the long term FDM dynamics.

\paragraph{\label{par:scm} Smoothed CDM Dynamics}
On the other hand, if we accept \eqref{eq:sp3d} as an abstract evolutionary problem
and forget momentarily about its interpretation as alternative dark
matter model, the dynamics of $\Psi$ can be associated with a \emph{smoothed}
version of the Vlasov-Poisson equation (VP) --- the
phase space description of CDM, \cite{Widrow1993,Uhlemann2014,Kopp2017}. 

More precisely, if $f_V(\bm x, \bm p)$ denotes the solution to VP and 
$f_W(\bm x, \bm p)$ the Wigner phase space distribution, constructed from $\Psi$ via:
\begin{equation*}
    f_W(\bm x, \bm p) = \int \text{d}^3x' 
    \Psi\left(\bm x-\frac{\bm x'}{2}\right)
    \Psi^*\left(\bm x+\frac{\bm x'}{2}\right)
    e^{\frac{i}{\hbar}\bm p \cdot \bm x'} \;,
\end{equation*}
then the evolution of the smoothed, or convolved, distributions,
\begin{equation}
        \label{eq:smooth_f}
        \bar f_{V/W} (\bm x, \bm p)
        = \exp
        \left(
            -\frac{\bm x ^2}{2\sigma_x^2}-\frac{2 \sigma_x^2}{\hbar^2} \bm p^2
        \right) * f_{V/W}\;,
\end{equation}
obeys, \cite{Uhlemann2014}:
\begin{equation*}
    \partial_t(\bar f_W - \bar f_V) 
    =
    \frac{\hbar^2}{24}\partial_{x_i}\partial_{x_j}
    \nabla_x \bar V
    \partial_{p_i}\partial_{p_j}
    \nabla_p \bar f_W + \mathcal{O}(\hbar^4, \hbar^2\sigma_x^2) \;.
\end{equation*}
Here, $\hbar$ is \emph{not} Planck's constant and $\hbar/m$ acts as a
model parameter that sets the maximum phase space resolution. Then, $\sigma_x$ is a
artificial smoothing scale in comoving position space.

In that sense, \eqref{eq:sp3d} can be understood as an alternative sampling of the CDM
distribution compared to the $N$-body approach: Instead of following the evolution
of $N$ test particles sampling $f_V$, we coarse grain the phase space
distribution directly and use $\Psi$ as a dynamical proxy for its evolution.
See Sec.~\ref{subsub:thermalization} for an application of this interpretation.

\subsubsection{\label{subsub:linear_regime}Dynamics in the Linear Regime}
The behavior of FDM in the linear growth regime is well-established in the
literature, see e.g. \cite{Woo2009,Chavanis2012,Li2019}. 
We are therefore brief and only focus on aspects relevant for our interpretation of 
the numerical simulations later on in Sec.~\ref{sub:cosmo}.

In short, Madelung's ansatz, \cite{Madelung1927}, of decomposing the wave function 
into $\Psi(\bm x,t) = \sqrt{\rho(\bm x,t)}\exp\left(i\frac{S(\bm
x,t)}{\hbar}\right)$, turns \eqref{eq:sp3d} into the Euler-Poisson (EP)
equation including an additional (quantum) pressure term: 
\begin{subequations}
\begin{align}
    \label{eq:continuity}
        \partial_t\rho + \frac{1}{a} \nabla \cdot (\rho \bm v) &= 0 \;, \\
    \label{eq:poisson}
        \Laplace \Phi &= \frac{4\pi G}{a}\left(\rho - \rho_m \right) \;, \\
    \label{eq:euler}
    \partial_t \bm v + \frac{1}{a}(\bm v \cdot \nabla )\bm v + H\bm v 
    &= - \frac{1}{a}\nabla \Phi 
    + \frac{\hbar^2}{2m^2a^3}\nabla\left(
    \frac{\Laplace \sqrt \rho}{\sqrt\rho}\right)\;,
\end{align}
\end{subequations}
\textcolor{red}{%
    Note that in order to pass from SP to EP we assumed $\psi\neq0$ and 
    identified the phase function $S(\bm x,t)$ as potential of the 
    peculiar velocity $\bm v$:
    \begin{equation}
        \label{eq:v}
        \bm v \equiv \frac{1}{ma}\nabla S = 
        \frac{\hbar}{m a|\Psi|^2} \text{Im}\left(\Psi^*\nabla\Psi\right) \;.
    \end{equation}
    Adding this condition to EP makes all solutions to
    \eqref{eq:continuity}-(\ref{eq:euler}) irrotational, 
    i.e., $\nabla \times \bm v = 0$, by construction. 
    It is worth pointing out that SP and EP, if considered in isolation, 
    support solutions with non-vanishing circulation. These so called vortices play an
    instrumental role for the energy transport in classical, turbulent flows
    \cite{Kolmogorov91} and Bose-Einstein condensates 
    \cite{Kobayashi2005,Baggaley2012}. 
    The distinctive property of vortices governed by SP is their quantized
    circulation. This physical constraint is missing in the
    classical, hydrodynamical equations and needs to be added by hand to EP in
    order to establish a formal SP--EP equivalence, see Ref. \cite{Wallstrom1994}.
}

Linearizing \eqref{eq:continuity}-(\ref{eq:euler}) up to first order in $\bm v$ and
$\rho(\bm x,t) = \rho_m\left(1 + \delta(\bm x,t)\right)$ 
yields a damped oscillator equation for the density contrast $\delta$ in the 
reciprocal domain:
\begin{gather}
    \label{eq:osci_jeans}
    0=\partial^2_t{\hat\delta} + 2H(t)\partial_t{\hat\delta}
    - \frac{4\pi G\rho_m}{a^3} 
    \left[
        1 - \left(\frac{k}{k_J(a)}\right)^{4} 
    \right] \hat\delta
    \\
    \label{eq:jeans_FCDM}
    \text{with} \quad
    k_J(a)= \left(\frac{16\pi G \rho_m m^2 a}{\hbar^2}\right)^{\frac{1}{4}} \;,
\end{gather}
and $\hat\delta = \hat \delta(k,t)$.
As for CDM, no mode coupling occurs and all perturbations evolve
independently under FDM evolution. However, in contrast to CDM, large and small
scale modes behave differently. Most notably, \eqref{eq:jeans_FCDM} defines
a time-dependent critical length scale, $\lambda_J(a) = 2\pi/k_J(a)$, --- the
Jeans scale--- below which the quantum pressure
counteracts gravity so that density perturbations do not collapse under their
self-gravity.

The linear Jeans scale can also be understood as a consequence of Heisenberg's
uncertainty principle, \cite{Hu2000}, $m\sigma_r\sigma_v \simeq \hbar$. In
hydrodynamic terms $\sigma_v$ may be interpreted as a velocity dispersion and a
simple way to estimate it in the linear regime is to follow a particle trapped inside 
a gravitational well of a matter distribution with density $\rho_m$:
\begin{equation*}
    \sigma_v \approx \frac{r}{t_\text{dyn}} \approx ax \sqrt{G\rho_ma^{-3}} \;,
\end{equation*}
with $t_\text{dyn}$ as dynamical time scale estimate. Thus:
\begin{equation*}
    \sigma_x \simeq \frac{\hbar}{mx\sqrt{G\rho_ma}} \;.
\end{equation*}
Setting $x=\sigma_x$ yields \eqref{eq:jeans_FCDM} up to a numerical constant of
$\mathcal{O}(1)$.

The interpretation then is that the source of the quantum pressure is Heisenberg’s 
uncertainty principle which induces an increasing velocity dispersion in the 
FDM condensate once particles are confined to a space region that is comparable
to $\sigma_x$.

Although, the Jeans length of \eqref{eq:jeans_FCDM} is a purely linear concept,
the uncertainty principle is not. One should therefore expect a distinct
signature in the matter power spectrum, even deeply in the nonlinear regime. We
return to this in Sec.~\ref{subsub:small_scale}.

Once a perturbation mode $\hat\delta(k,t)$ leaves the linear regime,
\eqref{eq:osci_jeans} is not applicable anymore. One then expects 
mode coupling to take place and thus a redistribution of power across all
scales that participate in the nonlinear evolution. The implications of this
effect will be analyzed in Sec.~\ref{subsub:mid_scale}.

\subsection{\label{sub:reduction}%
    Fuzzy Dark Matter in \texorpdfstring{$(1+1)$}{1D} Dimensions
}

Let us now turn the attention to one dimensional approximations to $(3+1)$-SP
and how such models may be deduced from \eqref{eq:sp3d}.
A convenient starting point for the dimension reduction procedure is to subsume
\eqref{eq:normalization}-(\ref{eq:pbc}) into a single dimensionless, nonlinear
Schr\"odinger equation (NLSE). This simplifies the discussion and is 
achieved by (i) absorbing the solution of Poisson's equation by means of a
convolution integral and (ii) adopting dimensionless quantities. One arrives at:
\begin{equation}
    \label{eq:nlse3d}
    i\hbar\partial_t\Psi 
    =\left[
        -\frac{1}{2}\Laplace + a(t) \left(G^{\text{ppp}}_{\Laplace_3}*|\Psi|^2\right)
    \right]\Psi \quad \bm x \in \Omega \;,
\end{equation}
where we defined:
\begin{gather*}
    \bm x' \equiv \left(\frac{m}{\hbar}\right)^{\frac{1}{2}} 
    \left[\frac{3}{2}H_0^2\Omega_{m}\right]^{\frac{1}{4}} \bm x \;, \quad
    \text{d}t' \equiv 
    \frac{1}{a^2}\left[\frac{3}{2}H_0^2\Omega_{m}\right]^{\frac{1}{2}}\text{d}t \;,\\
    \Psi'(\bm x', t') \equiv \frac{\Psi(\bm x', t')}{\sqrt{\rho_m}} \;, \quad
    V(\bm x', t') \equiv
    a\frac{m}{\hbar}\left[\frac{3}{2}H_0^2\Omega_{m}\right]^{-\frac{1}{2}}\Phi
\end{gather*}
and dropped all primes subsequently. The nonlinear potential is then given by:
\begin{equation*}
    G^{\text{ppp}}_{\Laplace_3}*|\Psi|^2\ 
    = \int_\Omega \text{d}^3x' G^{\text{ppp}}_{\Laplace_3}(\bm x-\bm x')|\Psi(\bm
    x')|^2 \;,
\end{equation*}
and $G^{\text{ppp}}_{\Laplace_3}$ denotes the Green's function of the $d=3$ Poisson
equation augmented with periodic boundary conditions in all three dimensions.
Eq. (\ref{eq:nlse3d}) takes the form of a non-autonomous, i.e. explicitly
time-dependent, NLSE with a long range (i.e. non-local) interaction kernel. 

We stress $G^{\text{ppp}}_{\Laplace_3}(\bm x, \bm x')$ is \emph{not} 
the canonical $1/r$-potential as it lacks the required periodicity 
and only applies under free space boundary conditions,
$\lim_{|\bm x|\to\infty} |\bm x|V(\bm x) = -\frac{1}{4\pi}$, see \cite{Kellogg1967}.
Instead, we have:
\begin{equation}
    \label{eq:G_3d_periodic} 
    G^{\text{ppp}}_{\Laplace_3}(\bm x, \bm x') 
    =\frac{1}{L_1L_2L_3} \sum_{\norm{\bm n}>0}\frac{-1}{\bm k^2}
    e^{i \bm k \cdot (\bm x - \bm x')}  \;,
\end{equation}
with ${\bm n \in \mathbb{Z}^3}$, ${\bm k \in \mathbb{R}^3}$ and 
${k_i= \frac{2\pi}{L_i} n_i}$. 

The Newtonian $1/r$-potential may be recovered as free-space limit in two stages. 
By first taking $L_{1,2}\to\infty$ \cite{Marshall2000} recasts
\eqref{eq:G_3d_periodic} into:
\begin{widetext}
\begin{equation}
    \label{eq:G_laplace_3d_mixed}
    G^{\text{ppp}}_{\Laplace_3}(\bm x,\bm x') \xrightarrow{L_{1,2} \to \infty}
    G^{\text{ffp}}_{\Laplace_3}(\bm x,\bm x') 
    = 
    \frac{1}{2\pi L_3}\log|\bm x_\perp-\bm x'_\perp|
    -\frac{1}{\pi L_3}\sum_{m=1}^{\infty}
    \cos\left(k_m(x_3-x_3')\right)
    K_0 \biggl(k_m|\bm x_\perp-\bm x_\perp'| \biggr) \;,
\end{equation}
\end{widetext}
with $K_0(x)$ denoting the $0^\text{th}$-modified Bessel function of the second
kind and $\bm x_\perp=(x_1,x_2)^\intercal$. 
We return to this mixed boundary condition Green's function in Sec.~
\ref{par:confinment}.
Finally, take $L_3 \to \infty$ so that the Riemann sum in
\eqref{eq:G_laplace_3d_mixed} approaches an analytically solvable integral,
\cite{Gradshteyn2014}:
\begin{equation*}
        G^{\text{ffp}}_{\Laplace_3}(\bm x,\bm x') 
        \xrightarrow{L_{3} \to \infty}
        G^{\text{fff}}_{\Laplace_3}(\bm x,\bm x') 
        = -\frac{1}{4\pi|\bm x - \bm x'|}  \;,
\end{equation*}
yielding the expected result.

The naive way of carrying out the dimension reduction of \eqref{eq:nlse3d} is to 
simply drop all partial derivatives in $x_1,x_2$-direction.
This appears to be the common approach in low-dimensional studies
on FDM \cite{Widrow1993, Woo2009,Kopp2017,Garny2018,Zimmermann2019,Garny2020} and
turns out to be true assuming we demand a \emph{uniform} matter distribution along the
neglected dimensions. The approach is equally applicable for $d = 1,2$ and leads
to the $(d+1)$-SP equation.
  
Maintaining Poisson’s equation as field equation has implications on 
how gravity acts in lower dimensions since the periodic Green’s function in
\eqref{eq:G_3d_periodic} depends on the dimensionality of the Laplace operator. 
We stress even if the aforementioned violation of the periodic boundary
conditions would not exist for the Newtonian potential, it would still be 
impossible to simply enforce a $1/r$-interaction kernel in one 
dimension. Its singularity at the origin remains too strong 
and consequently yields an ill-defined convolution kernel. 
Hence, we ask whether a reduction exists that approximately 
preserves the three dimensional interaction with only one spatial degree of freedom. 
This is realized by \emph{strongly confining} matter orthogonal to the dimension
in which the evolution is observed. 

\subsubsection{\label{subsub:general}General Reduction Procedure}
We adapt the discussion under free-space conditions outlined in \cite{Bao2013}
to the periodic situation at hand. To this end, the NLSE in
\eqref{eq:nlse3d} is augmented by an artificial, external potential 
$\mathcal{V_\text{ext}}(\bm x_\perp; \epsilon) = 
\frac{1}{\epsilon^2}V_\text{ext}\left(\frac{\bm x_\perp}{\epsilon}\right)$ 
in the orthogonal $\bm x_\perp$-plane that is controlled by a confinement parameter 
$\epsilon$. The extended Hamiltonian $H = H_{x_3} + H^\epsilon_\perp$ is then 
decomposed into:
\begin{align*}
    H_{x_3} &= -\frac{1}{2}\partial^2_{x_3} 
    + a(t)\left(G_{\Laplace_3}^{\text{\_\_p}} * |\Psi|^2\right) \;, \\
    H^\epsilon_\perp &= \frac{1}{\epsilon^2}
    \left[
        -\frac{1}{2}
        \Laplace_{\widetilde\perp} + V_\text{ext}(\bm x_{\widetilde\perp})
    \right]
    =\frac{1}{\epsilon^2}H_{\widetilde\perp} \;,
\end{align*}
with $\bm x_{\widetilde\perp} = \bm x_\perp/\epsilon$.
Notice how we keep the boundary conditions of the interaction kernel
$G_{\Laplace_3}^{\text{\_\_p}}$ deliberately unspecified in the $\bm x_\perp$-plane.
These will be set in accordance with the external potential, 
i.e. if $V_\text{ext}(\bm x_\perp)$ is periodic on $\Omega_1 \times \Omega_2$,
the long range interaction is $G_{\Laplace_3}^{\text{ppp}}$, as
in \eqref{eq:G_3d_periodic}. If, however, the external potential assumes
free-space conditions, the reduction departs from the mixed condition Green's function
$G_{\Laplace_3}^{\text{ffp}}$ in \eqref{eq:G_laplace_3d_mixed}.

Next, find the eigensystem $\{\chi_k^\epsilon, \lambda_k^\epsilon\}$ to the linear
eigenvalue problem 
$H^\epsilon_{\perp}\chi^\epsilon_k(\bm x_{\perp}) 
= \lambda^\epsilon_k\chi^\epsilon_k(\bm x_{\perp})$ 
and enforce a factorization of the full fledged
wave function $\Psi(\bm x, t)$ according to $\Psi(\bm x,
t)=\psi(x_3,t)\chi_k^\epsilon(\bm x_\perp)e^{-i\lambda_k^\epsilon t}$.
Insert this ansatz into the NLSE with the extended Hamiltonian, multiply by 
$\left(\chi_k^\epsilon\right)^*$ and integrate over the $\bm x_\perp$-plane. 
The result is a one dimensional NLSE for $\psi(x_3,t)$: 
\begin{equation}
    \label{eq:nlse1d}
    i\hbar\partial_t\psi
    =\left[
        -\frac{1}{2}\partial^2_{x_3} + a(t) \left(U^{\text{p}}*|\psi|^2\right)
    \right]\psi \quad x_3 \in \Omega_3\;,
\end{equation}
alongside a one dimensional, long range interaction, $U^\text{p}$:
\begin{equation}
    \begin{split}
    \label{eq:U1d_generic}
    U^{\text{p}}(x_3,x'_3) 
    = 
    \frac{1}{\norm{\chi^\epsilon_k}_2^2}
    \int\displaylimits_{\Omega_1\times\Omega_2}
    \hspace{-1em}
    \text{d}^2x'_\perp
    \hspace{-1em}
    \int\displaylimits_{\Omega_1\times\Omega_2}
    \hspace{-1em}
    \text{d}^2x_\perp
    |\chi^\epsilon_k(\bm x_\perp)|^2|\chi^\epsilon_k(\bm x'_\perp)|^2
    \\\qquad\times
    G^{\text{\_\_p}}_{\Laplace_3}(\bm x_\perp, x_3, \bm x'_\perp, x'_3)
    \end{split}
\end{equation}

Since the spectrum of \eqref{eq:G_3d_periodic} is spherical
symmetric around $\bm k = 0$ it is clear that, irrespective of the confining
potential in the transversal plane, the one dimensional interaction kernel
remains even and depends only on the relative distance: ${U^{\text{p}}(x_3, x_3') = U^{\text{p}}(|x_3-x_3'|)}$.

\subsubsection{\label{subsub:comp}Uniform vs. Confined Transversal Density}
Let us restrict the discussion to two distinct external potentials that either induce a
complete delocalization of matter in the orthogonal plane or constrain all
matter around the $x_3$-direction. In what follows, both of these models will be
coined $(1+1)$-FDM collectively. If a distinction has to be made, a more precise
denotation will be used.

\paragraph{\label{par:uniform}Uniform Matter} 
Set $V_\text{ext}=0$. The corresponding eigenstates
are plane waves and matter is therefore assumed to be organized in homogeneous
matter sheets parallel to the $\bm x_\perp$-plane.
Moreover, the external potential is trivially $L_3$-periodic so that
$G^{\text{\_\_p}}_{\Laplace_3} = G^{\text{ppp}}_{\Laplace_3}$. Eq.
(\ref{eq:U1d_generic}) then evaluates to, \cite{Gradshteyn2014}:
\begin{equation}
    \label{eq:G_laplace_1d_periodic}
    G^{\text{p}}_{\Laplace_1}(x_3,x_3') 
    = \frac{1}{2}|x_3-x'_3| 
    - \frac{1}{2}\left[\frac{(x_3-x_3')^2}{L_3}+\frac{L_3}{6}\right] \;,
\end{equation}
i.e. the Green's function of the d=1 Poisson equation under periodic boundary
conditions. 

Eq. (\ref{eq:nlse1d}) together with the interaction of 
\eqref{eq:G_laplace_1d_periodic} is denoted $(1+1)$-SP.

\paragraph{\label{par:confinment}Confined Matter} 
We enforce an integrable harmonic confining potential
$V_\text{ext}(\bm x_\perp) = \frac{1}{2}\bm x_\perp^2$ and consider only the ground
state $\chi_0(\bm x_\perp)\propto\exp( -\bm x_\perp^2/2 )$ for the dynamics in
the orthogonal plane. Our choice is motivated by cigar-shaped confinements 
often used for trapping Bose-Einstein condensates, see e.g. \cite{Dalfovo1999,Bloch2008,Olshanii1998,Salasnich2002,Wimberger2005,Wimberger2005b}.
Neither the external potential nor its ground state
are periodic. Hence, we set $G^{\text{\_\_p}}_{\Laplace_3} =
G^{\text{ffp}}_{\Laplace_3}$ in eq. (\ref{eq:U1d_generic}) and find:
\begin{equation}
    \label{eq:G_LAM_1d_periodic}
    U^{\text{p}}_{\text{conf}}(x_3, x_3') = 
    \frac{-1}{4\pi L_3}\sum_{m}
    U\left(1,1,\frac{1}{2}k_m^2\epsilon^2\right)
    e^{ik_m(x_3-x_3')}\;,
\end{equation}
with $m\in\mathbb{Z}\setminus 0$ and $U(a,b,x)$ denoting the confluent hypergeometric
function of the second kind, \cite{Owen1965}. 

\begin{figure}
    \includegraphics{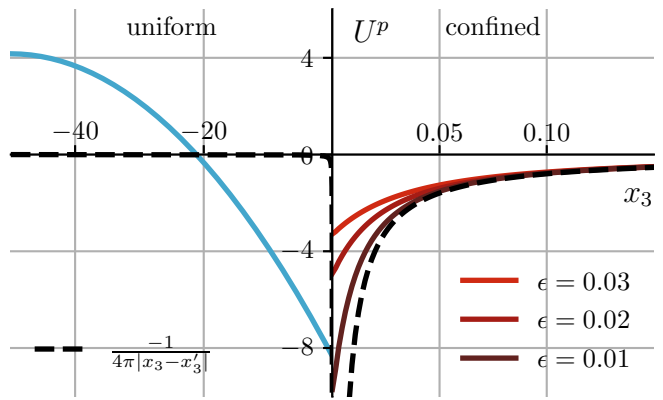}
\caption{\label{fig:kernel_comparison}
    (Color online)
    Comparison of the one dimensional, long range interactions of Sec.~
    \ref{par:uniform} and Sec.
    \ref{par:confinment} induced by a source particle located at $x_3 = 0$ in a
    periodic box of $L_3=100$.
    The negative x-space illustrates the situation for a uniform
    matter distribution in the transversal plane with $G^\text{p}_{\Laplace_1}$,
    \eqref{eq:G_laplace_1d_periodic}, as gravitational potential (blue line). 
    The positive x-space depicts the gravitational potential 
    $U^{\text{p}}_{\text{conf}}$, \eqref{eq:G_LAM_1d_periodic},
    obtained under harmonic confinement for various confinement 
    strengths $\epsilon$ (red lines). While both $G^\text{p}_{\Laplace_1}$ and 
    $U^{\text{p}}_{\text{conf}}$ are finite at the origin only
    $U^{\text{p}}_{\text{conf}}$ approaches the Newtonian gravitational
    potential (black, dashed line) in the far field. Note the different limits of
    the negative and positive $x_3$-axis: The negative half space covers
    the interval $[-L_3/2, 0]$, the positive half space is truncated at
    $x_3=0.15$ to make the asymptotic behavior of $U^{\text{p}}_{\text{conf}}$
    better observable.
}
\end{figure}

Consider Fig.~\ref{fig:kernel_comparison} for a graphical comparison of the
long range interaction induced by a test particle at $x_3'=0$ with and without
confinement. All potentials are shown for a periodic box of $L_3=100$.
Apart from the required finiteness of both kernels at the origin, the behavior of 
$U^{\text{p}}_{\text{conf}}$ and $G^{\text{p}}_{\Laplace_1}$ is quite disparate
in the far-field region.  
A key difference lies in the effective interaction range $R$ which may be defined as:
\begin{equation}
    \label{eq:range}
    \partial_r U^p(R) \equiv \delta \text{max}|\partial_r U^p|
    \quad \text{with} \quad
    0 < \delta \ll 1 \;.
\end{equation}
While the gravitational potential under confinement is rather
localized with $R \sim \epsilon$ and quickly approaches the desired Newtonian
potential (black, dashed line) at large distances $|x_3-x_3'|$, the effective interaction range for
$G^{\text{p}}_{\Laplace_1}$ evaluates to 
$R = \frac{1}{2}\left(1-\delta\right)L_3$ and is therefore comparable to the box
size. As discussed in Sec.~\ref{subsub:organization}, this has implications on 
which asymptotic states are accessible in the $(1+1)$-SP case.

\subsubsection{\label{subsub:symmetries}Symmetries and Conserved Quantities}

Naturally, one is interested in conserved quantities of \eqref{eq:nlse1d}
for even kernels but irrespective of its exact form.
By analyzing the action:
\begin{equation}
    \label{eq:NLSE_action}
        \mathcal{S} 
        = \int\text{d}t\int\displaylimits_{\Omega_3}\text{d}x_3
        (i \psi^* \partial_t \psi -
        \mathcal{H}[\psi, \psi^*, \partial_{x_3} \psi, \partial_{x_3} \psi^*, t])\;,
\end{equation}
with the Hamiltonian density:
\begin{equation*}
    \mathcal{H} = \frac{1}{2}\left(|\partial_{x_3} \psi|^2 + a(t)(U^\text{p} *
    |\psi|^2)|\psi|^2 \right)
\end{equation*}
generating \eqref{eq:nlse1d} upon variation, 
it is straightforward to show that such a system obeys mass and momentum conservation.
Moreover, \eqref{eq:NLSE_action} is invariant under galilean boosts of the form:
\begin{equation}
    \label{eq:boost}
    \psi(x,t) \to e^{i(vx_3 - \frac{1}{2}v^2t)}\psi(x_3-vt,t) \;.
\end{equation}

If we consider a \emph{static} space-time, i.e. $a(t) = \mathrm{const.}$,
\eqref{eq:NLSE_action} is also time translation invariant and the total energy,
\begin{equation}
    \label{eq:energy_conservation}
    E[\psi] = \int\displaylimits_{\Omega_3} \text{d}x_3
        \mathcal{H}[\psi, \psi^*, \partial_{x_3} \psi, \partial_{x_3} \psi^*, t]
        = \langle T \rangle + \frac{a}{2} \langle V \rangle
        \;,
\end{equation}
is conserved as well. 
This is of course \emph{not} true once space-time is allowed to expand. 

A symmetry unique to $(1+1)$-SP is the following scaling transformation:  
If $\psi(x_3,t)$ solves  \eqref{eq:nlse1d} with $U^\text{p} =
G^\text{p}_{\Laplace_1}$ then so does:
\begin{equation}
    \label{eq:scaling_symmetry}
    \widetilde\psi(x_3,t) 
    = \lambda^2\psi(\lambda x_3, \lambda^2 t) \;, \quad \lambda
    \in \mathbb{R}^{+} \;.
\end{equation}
Equivalent scaling symmetries exist for $d=2,3$ spatial dimensions \cite{Guzman2006}.

\subsubsection{\label{subsub:solitons}%
    Properties of the \texorpdfstring{$(1+1)$}{1D}-FDM Ground State
}

We already alluded in our introductory remarks to the importance of stationary
states of $(3+1)$-SP, especially the role of its stable ground state that acts as 
dynamical attractor and realizes a flat-density core inside relaxed dark matter 
structures. Let us extend this discussion to the previously derived
one dimensional FDM models by analyzing properties of the 
ground state to \eqref{eq:nlse1d} that influence the asymptotic $(1+1)$-FDM dynamics. 

For practical purposes, ground states of mass 
$N[\psi] = \int \text{d}x_3 |\psi|^2 = M$ may be prepared by
choosing an interaction kernel and minimizing the grand canonical energy
$E_\text{grand}[\psi] = E[\psi] + \mu(\tau)N[\psi]$ by means of a gradient descent 
with $\mu(\tau)$ as chemical potential at descent parameter $\tau$. The reader
is referred to \cite{Bao2004, Bao2006} for numerical details and
\cite{Choquard2008} for a rigorous analysis on the existence and uniqueness of a
minimizer to \eqref{eq:energy_conservation} for $(1+1)$-SP under free-space
conditions. We note in passing that the chosen numerical implementation of the gradient
descent makes the energy minimization approach equivalent to the well-known
imaginary time propagation method \cite{Pang2006,Wimberger2005b}.

\begin{figure}
    \includegraphics{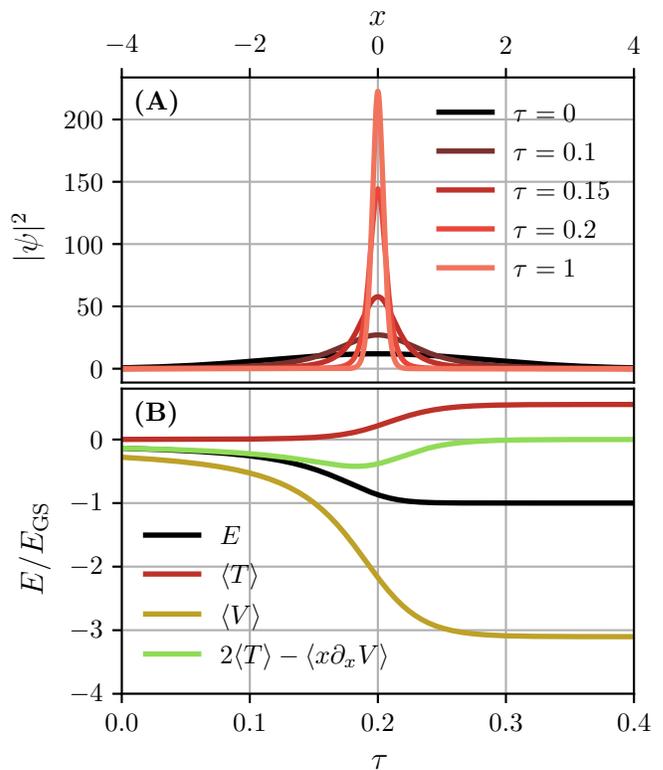}
    \caption{\label{fig:gradient_descent}%
        (Color online)
        Exemplary gradient descent for ${U^\text{p} = U^\text{p}_\text{conf}}$, 
        i.e. under confinement, with ground state mass ${M=50}$, ${a=1}$ and 
        ${\epsilon=10^{-2}}$. 
        Panel (A): 
        Densities at various stages of the minimization procedure. 
        Panel (B): 
        Evolution of kinetic (red), potential (yellow) and total energy (black) 
        normalized to the ground state energy $E_\text{GS}$. 
        Also shown in green is the free-space, quantum virial theorem, discussed in 
        Sec.~\ref{subsub:virialization}. 
    }
\end{figure}

An exemplary gradient descent under transversal, harmonic confinement is
depicted in Fig.~\ref{fig:gradient_descent}. As time progresses the initial
gaussian distribution focuses more and more in position space until the descent converges
to a spatially localized structure at $\tau=1$.

Irrespective whether $U^\text{p} = U^\text{p}_\text{conf}$ or 
$U^\text{p} = G^\text{p}_{\Laplace_1}$, the $(1+1)$-FDM ground state shows the
following properties:

\paragraph{\label{par:stationary_solution}Solitary Wave}
It is easy to see that the equation governing the gradient descent, namely
${\partial_\tau\psi=-\frac{1}{2}\frac{\delta}{\delta\psi}E_\text{grand}}$,
reduces to a stationary form of \eqref{eq:nlse1d},
\begin{equation}
    \label{eq:stationary_nlse}
    \mu(\tau^*) \psi_\text{GS} 
    = \left[
        -\frac{1}{2}\partial^2_{x_3}
        + a (U^\text{p} * |\psi_\text{GS}|^2)
    \right]\psi_\text{GS} \;,
\end{equation}
once the energetic minimum at $\tau = \tau^*$ is reached. 

Therefore, the ground state may be written in its canonical, linear quantum
mechanics form, $\psi_\text{GS}(x_3, t) = \psi_\text{GS}(x_3)e^{-i\mu(\tau^*)t}$, 
and we conclude $\psi_\text{GS}$ is
a \emph{solitary wave}, i.e. a localized solution to a nonlinear equation with time 
independent envelope $|\psi|^2$. Obviously, its persistent form is also preserved
for uniformly travelling configurations obtained via \eqref{eq:boost}.

Two remarks are in order:
Firstly, the above discussion derives $\psi_\text{GS}$
by minimizing the total energy functional. 
An alternative approach, e.g. \cite{Guzman2004}, is to interpret the nonlinear
eigenvalue problem of \eqref{eq:stationary_nlse} as boundary value problem
and solve it via a shooting method.
Secondly, recall the scaling symmetry of \eqref{eq:scaling_symmetry} unique to
$(d+1)$-SP. Applying it to a single ground state 
gives access to an entire family of ground states parameterized by their total
mass $M$. Thus, for $(1+1)$-SP, it is sufficient to conduct the energy
minimization only once for an arbitrary reference mass $M_\text{ref}$. All other
energy minimizers then follow from rescaling with \eqref{eq:scaling_symmetry}. 
Lacking an equivalent symmetry, this is of course not true for the confinement model.

\paragraph{\label{par:collision}Inelastic Collisions}
So far, we demonstrated the solitary character of the $(1+1)$-FDM ground state.
Naturally, we are interested whether $\psi_\text{GS}$ can also be identified as
a solitonic solution.

Strictly speaking, the concept of a soliton calls for a rigorous mathematical
definition. To keep technical details to a minimum, we instead follow
\cite{Drazin1989} and characterize a soliton as a solitary wave that is invariant
under interactions with other solitons. Put differently, despite the nonlinear 
evolution, solitons obey a superposition principle and neither mass nor energy
should be exchanged during the interaction.

\begin{figure*}
    \includegraphics{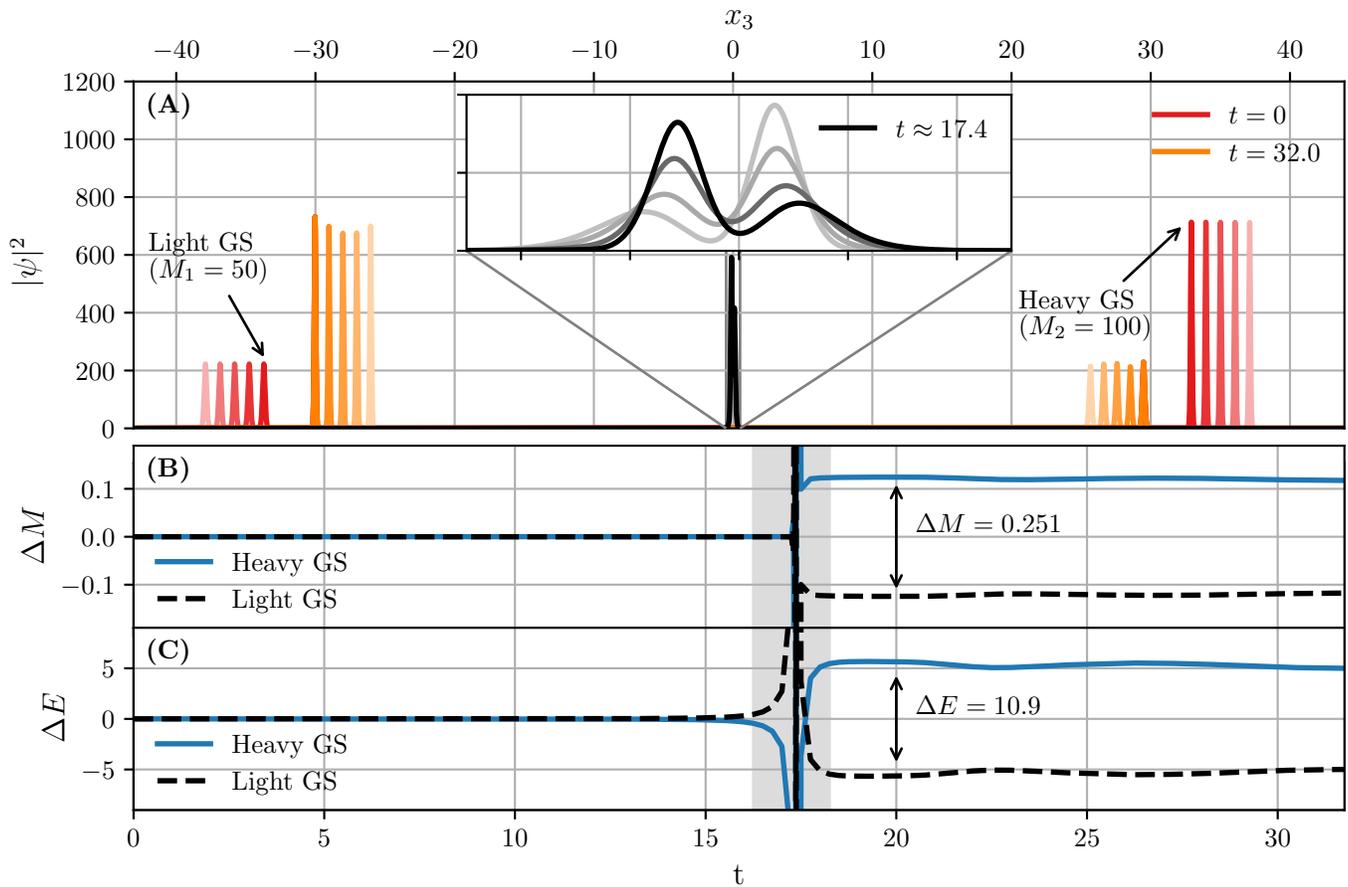}
    \caption{\label{fig:collision}%
    (Color online)
    Inelastic interaction of an asymmetric high-mass, low-mass ground state
    configuration under strong, transversal confinement ($\epsilon=10^{-2}$) and $a=1$.
    The evolution starts from \eqref{eq:collision_ic}.
    Panel (A): 
    Density evolution. Initially, both densities
    travel as solitary waves (red), pass through each other (black, inset) and
    continue to propagate in a quasi-solitary movement after the interaction (orange). 
    By this we mean a state for
    which neither linear dispersive nor nonlinear focusing effects induce a
    permanent deformation of the density. Instead, one observes an oscillation
    around a solitary wave. Similar oscillatory behavior was found for
    $(3+1)$-SP, \cite{Guzman2004}, once the ground state density is perturbed.
    Panel (B): 
    Time evolution of the mass deviation of both ground 
    states relative to their initial masses cf. \eqref{eq:mass_evolution}.
    Post-interaction, it is the high-mass ground state gaining additional matter
    at the expense of the lighter ground state.
    Panel (C): 
    Time evolution of the total energy deviation of both ground 
    states relative to their respective initial total energies.
    As for the mass, the interaction induces an energy transfer from the low to high
    mass ground state. Notice, how the quasi-solitary behavior of the post-interaction
    densities is also seen in an oscillation of the energy deviation.
    Since the association of mass and energy contained in the positive and negative
    box half to a particular solitary wave is ambiguous during the interaction,
    we deem data in the gray shaded interval of panel (B) and (C) as
    not reliable.
}
\end{figure*}

Figure \ref{fig:collision} investigates the behavior of an asymmetric
configuration of two confined ground states with initial masses $M_1=50$ and 
$M_2=100$ and $\epsilon=10^{-2}$, boosted onto a collision course with:
\begin{equation}
    \label{eq:collision_ic}
    \psi_0(x_3) 
    = \psi_{\mathrm{GS}, 1} \left(x_3 + x_0\right)e^{ivx_3} 
    + \psi_{\mathrm{GS}, 2} \left(x_3 - x_0\right)e^{-ivx_3}\;.
\end{equation}

Figure \ref{fig:collision}(A) depicts the matter density in a pre-collision time 
window, an instance during the ground state interaction ($t\approx17.4$) as well as a
post-collision time window around $t=32$. It is evident that the superposition
principle is not satisfied. 
After the interaction took place both ground states propagate in a
quasi-solitary fashion in which linear dispersive and
nonlinear focusing effects are not exactly balanced anymore. Instead one finds
a periodic expansion and re-contraction of the matter distribution once the 
dispersion or non-local, nonlinearity dominates. 
Similar oscillatory behavior was found for the
$(3+1)$-SP ground state, \cite{Guzman2004}, once the density is perturbed.
Nevertheless, on average the post-interaction configuration is still comprised
of two solitary, or stationary, states. 

To investigate whether these post-interaction, solitary waves are different from
their initial composition, we analyze the deviation in mass  
(Fig.~\ref{fig:collision}(B)) and total energy (Fig.~\ref{fig:collision}(C)) from 
their initial values. 
For instance, the mass difference for the $M_1$-ground is inferred as:
\begin{equation}
    \label{eq:mass_evolution}
    \Delta M(t)= \hspace{-1em}
    \int
    \displaylimits_{
        -\frac{L_3}{2}\Theta(t_\text{coll}-t)
    }^{
        \frac{L_3}{2}\Theta(t-t_\text{coll})
    } \hspace{-2em}
    \text{d}x_3|\psi(x_3, t)|^2 - M_1 \;,
\end{equation}
with $t_\text{coll}$ as collision time naively inferred from the uniform velocity
$v$ at $t=0$.

One finds, a symmetric mass and energy gap after the interaction: Both
mass and energy were transferred from the low to high mass solitary wave.
Clearly, such a matter and energy transfer should not exist if the confined 
ground state were a true soliton. 
We note although the reported energy and mass differences are small they are 
robust under variation of $\Delta t$, see Sec.~\ref{sub:evolution} for more
information. Qualitatively similar results were found for $(1+1)$-SP. 
Therefore, $(1+1)$-FDM ground states --- at least the ones
considered here --- are \emph{not} solitons in the strict sense of the word, but
interact inelastically by exchanging mass and energy during encounters, 
typically reshuffling them from the low-mass to the high-mass solitary wave. 

Of particular interest is the case of multiple successive interactions which,
thanks to the periodicity of the box, is easily observed by increasing the
integration time. The reader is referred to Sec.~\ref{subsub:relaxation_artificial}
for more details.

\paragraph{\label{par:mass_size_relation}Mass-Size Relation}
As we will see, understanding the discrepancies between the attained
asymptotic states of $(1+1)$-SP and the confinement model, Sec.
\ref{subsub:organization}, hinges on the ratio
$R(L_3)/\sigma(M)$, i.e. the interaction range $R$ given a periodic box of size
$L_3$ compared to the spatial extent $\sigma$ of a mass $M$ ground state. 

Deriving $\sigma(M)$ is particularly simple in case of the unconfined FDM model 
as Madelung's ansatz relates $(1+1)$-SP to a one dimensional version of the
hydrodynamic description of \eqref{eq:v}-(\ref{eq:euler}). In the ground state's rest
frame these reduce to the condition of hydrostatic equilibrium. 
Dimensional analysis then yields $\sigma \propto M^{-\frac{1}{3}}$.

\begin{figure}
    \includegraphics{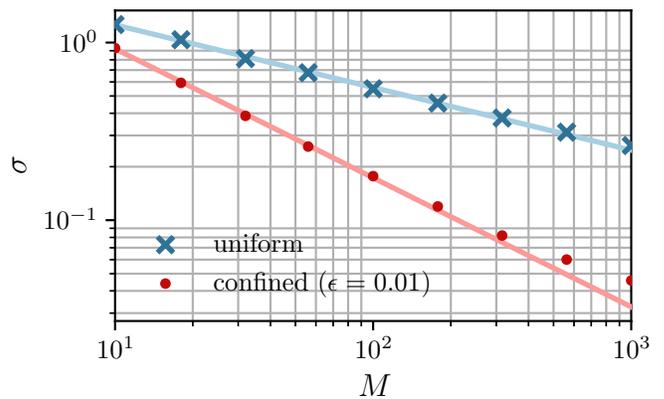}
    \caption{\label{fig:spatial_extend}%    
        (Color online)
        Spatial extent of the one dimensional FDM ground states as a function of mass
        $M$ at $a=1$. Each data point corresponds to a solitary wave prepared by the
        gradient descent shown in Fig.~\ref{fig:gradient_descent}. The spatial
        extent of the matter distribution is extracted according to 
        \eqref{eq:soliton_size}.
        For $U^\text{p} = G^\text{p}_{\Laplace_1}$ (blue) we find 
        $\sigma(M) = 2.8 \cdot M^{-0.34}$ and consequently good agreement with 
        dimensional analysis. Under strong harmonic confinement, 
        i.e. $U^\text{p} = U^\text{p}_\text{conf}$ and $\epsilon = 10^{-2}$ (red) 
        one deduces $\sigma(M) = 4.9 \cdot M^{-0.72}$ over two orders of magnitude in 
        $M$.
}
\end{figure}

The situation is more involved under harmonic confinement due to the missing PDE
for the gravitational potential. Thus, the spatial extent is deduced 
numerically by defining:
\begin{equation}
    \label{eq:soliton_size}
    0.99 M \equiv \int_{-\sigma}^{\sigma}\text{d}x_3 |\psi_\text{GS}|^2 \;,
\end{equation}
and extracting $\sigma$ for various, spatially centered ground states of mass $M$. 
Figure \ref{fig:spatial_extend} depicts the result for a static space-time with $a=1$ 
for both $(1+1)$-FDM models. While $(1+1)$-SP shows satisfactory agreement with the 
dimensional analysis, $\sigma(M) = 2.8 \cdot M^{-0.34}$, a strongly confined matter 
density at $\epsilon=10^{-2}$ results in a narrower ground state distribution at 
equal mass $M$. In this case $\sigma(M) = 4.9\cdot M^{-0.72}$.

\subsection{\label{sub:relaxation}Relaxation Mechanisms and Equilibrium Properties}

It is a priori not clear what dynamical mechanisms drive $(1+1)$-FDM into its 
asymptotic equilibrium configuration let alone whether both reduction models
obey the same relaxation processes --- recall the discrepancies in the
interactions of $(1+1)$-SP and the confinement scenario.

Given the approximative CDM interpretation of FDM in Sec.~\ref{par:scm},
classical, non-collisional relaxation mechanism may be a viable option,
in particular a combination of phase mixing and violent relaxation, see
\cite{Lynden-Bell1967, Binney2004}. 
These processes induce a filamentation of the phase space dynamics alongside a
redistribution of energy inside the self-gravitating structure due to its
fluctuating gravitational potential.

On the other hand, $(3+1)$-FDM-typical mechanisms like gravitational cooling, 
\cite{Seidel1994}, may be recovered even in one dimension, 
allowing collapsing matter structures to relax into an equilibrated state by 
radiating away excess energy in form of small scale matter waves.

Which relaxation channels are realized is discussed in Sec.
\ref{subsub:relaxation_artificial}. Here, we ask what properties the
equilibrated system configuration should have and how they may be measured such
that the progress on the overall system evolution can be quantified.

\subsubsection{\label{subsub:virialization}Virial Equilibrium}

Application of Ehrenfest's Theorem for the virial operator $\hat G = \hat p\hat x$
gives rise to a quantum analogue of the scalar virial theorem. For bounded
dynamics, i.e. $\langle\hat G\rangle(t) < \infty$, and \emph{periodic} boundary 
conditions it reads:
\begin{multline}
    \label{eq:virial_general}
    0=2(\langle T\rangle)_\infty-\left(a\langle x_3\partial_{x_3}
    V\rangle\right)_\infty \\
    + 
    \frac{L}{2}
    \left(
        (\psi^*(0, t)\partial^2_{x_3}\psi(0,t) -
        |\partial_{x_3}\psi(0,t)|^2
    \right)_\infty
\end{multline}
with $(A)_\infty = \lim_{t\to\infty} \int_0^t\text{d}t'A(t')$. 
Relaxation into virial equilibrium, i.e. the regime where
\eqref{eq:virial_general} is (approximately) satisfied, is then to be understood 
as a consequence of the evolution under Schr\"odinger’s equation. 
That said, any finite quantum system would virialize in the limit $t\to\infty$.

A couple of remarks are in order. Firstly, note \eqref{eq:virial_general}
only holds in the limit $t\to\infty$. A notable exception are stationary
states, like the $(1+1)$-FDM ground states of Sec.~\ref{subsub:solitons}, 
which obey \eqref{eq:virial_general} without time averaging cf.
\cite{Weislinger2009}. If in addition fluctuations are present, we may assess
virialization of the total system by assuming \eqref{eq:virial_general} were
approximately achieved after a finite thermalization time.

Secondly, we draw special attention to the boundary term,
\begin{equation}
    \label{eq:boundary_term}
    B(t) = \psi^*(0, t)\partial^2_{x_3}\psi(0,t) - |\partial_{x_3}\psi(0,t)|^2 \;,
\end{equation}
in
\eqref{eq:virial_general}. It emerges from the necessity to extend the domain of
the Hamiltonian onto states like $x\psi$ which are not periodic but appear once
Ehrenfest's theorem is applied to $\hat G$, see \cite{Esteve2012}. Obviously,
the boundary term is negligible, if $\psi$ decays rapidly towards the box boundaries, 
like in Fig.~\ref{fig:gradient_descent}, or when artificial absorbing
potentials are used to limit the physically relevant domain size, see e.g. 
\cite{Schwabe2016, Mocz2017}.

\subsubsection{\label{subsub:thermalization}Maximum Entropy}

From a statistical physics viewpoint, one generally expects the system to
maximize its entropy.
In fact, the idea of entropy maximization is close to the original approach of
\cite{Lynden-Bell1967}, showing that mixing processes under Vlasov-Poisson 
imply a quasi-stationary phase-space distribution that maximizes the system's 
entropy on a macroscopic, i.e. coarse-grained level.

To adopt this idea for FDM, we use Husimi's distribution, i.e. $\bar f_W$ in 
\eqref{eq:smooth_f} and define a FDM entropy functional resembling Boltzmann's
entropy, \cite{Wehrl1979}:
\begin{equation}
    \label{eq:wehrl}
    S[\bar f_W] = - \frac{1}{2\pi}\int \text{d}x_3\text{d}k_3 \bar f_W\log \bar f_W \;.
\end{equation}
Thus, an equilibrated, thermalized system state is reached, once 
$\Delta S = S(t) - S(0)$ saturates.
We note this approach was also proposed by \cite{Kopp2017}.

\section{\label{sec:numerics}Numerical Method}

We briefly explain a simple, yet accurate spatial discretization of
\eqref{eq:nlse1d} and sketch an approximation to its time evolution operator.
The main properties of the presented method are summarized.
For more information, especially on the method's behavior under expanding
space-time conditions, the reader is referred to Appendix
\ref{sec:convergence}.  Additional information on general NLSE numerics can be found in
\cite{Antoine2013}. A recent survey of existing numerical techniques on our subject is given by \cite{Zhang2019}.

\subsection{\label{sub:spatial} Spatial Discretization}

Since \eqref{eq:nlse1d} is defined on a periodic domain and involves only
second derivatives in space, expansion of $\psi$ in a truncated
momentum eigenstate basis is a natural way to discretize \eqref{eq:nlse1d} in momentum
space and diagonalize the kinetic part of the Hamiltonian.
Discreteness in real space is then achieved by evaluating the momentum state
expansion of $\psi$ on sites $\{x_j\}_{j=0\dots N-1}$ with uniform spacing 
$\Delta x = L/N$. 
This translates \eqref{eq:nlse1d} into the finite dimensional, ordinary differential 
equation:
\begin{equation}
    \label{eq:fdm1d_ode}
    i\partial_t \bm \psi(t) 
    = \Bigg[
        \underbrace{
            \mathcal{F}^{\dagger}
            \frac{\bm{k_3} \bm{k_3}^\intercal}{2}
            \mathcal{F}
        }_{\hat H_K}
        + 
        \underbrace{
            \vphantom{\frac{\bm{k_3} \cdot \bm{k_3}^\intercal}{2}} 
            a(t) \bm V(|\bm \psi(t)|^2)]
        }_{\hat H_V(t, |\bm\psi|^2)}
    \Bigg]\bm\psi(t) \;,
\end{equation}
with $\bm \psi_j = \psi(x_j,t)$, $(\bm{k_3})_n=\frac{2\pi}{L_3}n$ and
$\mathcal{F}$ denoting the change of basis matrix from the real space to the momentum
basis. In practice the action of $\mathcal{F}$ and its inverse
$\mathcal{F}^{\dagger}$ on $\bm \psi$ are implemented as discrete fast Fourier transform.
%and accelerated by means of a fast Fourier transform.

The nonlinear, non-local potential is absorbed into the diagonal matrix 
$\bm V(|\bm\psi|^2)$ and follows directly from the convolution theorem,
\begin{equation*}
    \text{Diag}\left[\bm V(|\bm\psi(t)|^2)\right] = 
    \mathcal{F}^{\dagger}\widehat{\bm U^{\pi}}\mathcal{F}|\bm \psi(t)|^2\;,
\end{equation*}
with the, in momentum space diagonal, kernel coefficient matrix 
$\widehat{\bm U^{\pi}}$:
\begin{equation*}
    \text{Diag}\left[\widehat{\bm U^{\pi}}\right]
    =
    \begin{cases}
        0 & \hspace{-0.4em}n=0 \\
        -\frac{1}{(k_3)_n^2} 
          &\hspace{-0.4em}n\neq0 \hspace{0.7em} (1+1)\text{ SP}\\
        -\frac{1}{4\pi}U(1,1,\frac{1}{2}(k_3)_n^2\epsilon^2)
          &\hspace{-0.4em}n\neq0 \hspace{0.7em} \text{confined}\;.
    \end{cases}
\end{equation*}
\subsection{\label{sub:evolution} Time Evolution Operator}
Starting from \eqref{eq:fdm1d_ode} it remains to find an approximation to the
time evolution operator $\bm \psi(t + t_0) = \hat U_{K+V}(t_0, t_0 + \Delta
t)\bm\psi(t_0)$.
For this, the idea of operator splitting is employed --- a common choice for
integrating NLSEs. Thus, we first find
(approximate) evolution operators for the kinetic, $\hat H_K$, and potential
Hamiltonian, $\hat H_V$, individually and combine them into an
approximation for $\hat U_{K+V}$ afterwards.

The solution to the kinetic problem is trivial and reads:
\begin{equation*}
    \hat U_K(\Delta t) \equiv \hat U_K(t_0, t_0 + \Delta t) 
    = 
    \mathcal{F}^{\dagger}
    \exp\left[-i\frac{\bm{k_3} \bm{k_3}^\intercal}{2}\Delta t\right] 
    \mathcal{F}\;.
\end{equation*}

For the potential sub-problem we recall $\hat H_V$ is diagonal in real space, 
implying ${[\hat H_V(t), \hat H_V(t')] = 0}$. 
Thus, its time evolution operator can be written as
${\hat U_V(\Delta t)=\exp\big[-i\int_{t_0}^{t_0 + \Delta t}\text{d}t \hat
H_V(t,|\psi(t)|^2)\big]}$ 
and it remains to approximate the time integral over $\hat H_V$.

Fortunately, it is easily verified that evolution under the nonlinear
Hamiltonian $\hat H_V$ satisfies $\frac{\text{d}}{\text{d}t}|\psi|^2 = 0$.
It is therefore sufficient to substitute 
${\hat H_V(t, |\psi(t)|^2) \to \hat H_V(t,|\psi(t_0)|^2)}$
which (i) reduces the task of approximating the time integral over $\hat H_V$
to approximating the integral over the scale factor $a(t)$ and (ii) allows for
an explicit treatment of the nonlinearity, i.e. without the need to solve a
nonlinear system of equations. 

Application of the midpoint rule yields the
following unitary approximation to $\hat U_V(\Delta t)$:
\begin{equation*}
    \begin{split}
    \hat U_V(\Delta t) 
    &= 
    \mathcal{\hat U}_V(\Delta t, t_0) +\mathcal{O}(\Delta t^3) \\
    &= \exp\hspace{-0.2em}\left[
        -i a\hspace{-0.2em} \left(
            t_0\hspace{-0.2em}+\hspace{-0.2em}\frac{\Delta t}{2}
        \right)
        \bm V(|\bm \psi(t_0)|^2)\Delta t
    \right]+\mathcal{O}(\Delta t^3).
    \end{split}
\end{equation*}

Finally, after composing both evolution operators in a second order Strang
scheme, we arrive at the approximation to $\hat U_{K+V}$:
\begin{equation*}
    \hat U_{K+V} 
    = \hat U_K\left(\frac{\Delta t}{2}\right) 
    \circ \mathcal{\hat U}_V(\Delta t, t_0) \circ 
    \hat U_K\left(\frac{\Delta t}{2}\right) + \mathcal{O}(\Delta t^3).
\end{equation*}

\subsection{\label{sub:summary_of_properties}Summary of Properties}

The presented method is a simple extension to the kick-drift-kick scheme of
\cite{Mocz2017}. In fact, for static space-times both methods are equivalent.
Aside from its implementational simplicity and resemblance of the symplectic
Leap-frog method, it is unitary by design, explicit, second-order accurate
in time, spectrally accurate in space (assuming $\psi$ is smooth) and
provides a convenient unified approach for both dimension-reduction models. 
The computational complexity per integration step is $\mathcal{O}(N\log N)$ due to the fast Fourier
transformations, requiring $\mathcal{O}(N)$ memory.

Furthermore, for time-independent, nonlinear coupling constants, the approximate
evolution operator is time-symmetric, shows unconditionally stable
numerical behavior and conserves energy, \eqref{eq:energy_conservation},
approximately with a bounded error, \cite{Blanes2016}.

For non-static background cosmologies, time-symmetry and energy conservation are
broken by the continuous problem. Concerning stability, our tests indicate an 
exponentially growing error at high redshifts. We expect this result to
be to intrinsic to constant time step integration methods under space-time
expansion. 
Further informations on the convergence properties of our numerical method 
are given in Appendix \ref{sec:convergence}.

\section{\label{sec:results}Results of numerical simulations}

\subsection{\label{sub:cosmo}Structure Growth under \texorpdfstring{$(1+1)$}{1D}-SP}
We investigate the mean cosmic structure growth in an ensemble of FDM-only 
universes obeying $(1+1)$-SP. The purpose of this study is to explain
characteristic properties of the nonlinear FDM matter power spectrum
$P(k) = \langle|\hat\delta_k|^2\rangle$.

\subsubsection{\label{subsub:setup_cosmo}Simulation Setup}
To this end, we follow the evolution of $\mathcal{N}=100$ realisations of a
gaussian random field $\delta(x, a_0)$ in a flat, radiation free 
FLRW background cosmology with 
$\Omega_m = \Omega_\text{DM} + \Omega_\text{baryon} = 0.3$, 
$H_0 = \SI{68}{\km\per\second\per\mega\parsec}$ and power spectrum:
\begin{equation}
    \label{eq:power_init}
    P(k, a_0) = D_0^2(a_0) T^2_\text{dim}(k) T^2_{\text{FDM}}(k)P_\text{CDM}(k)\;,
\end{equation}
with $k=|k_3|$.
Here, $D_0^2(a_0)$ denotes the linear growth factor, see \cite{Dodelson2003},
normalized to unity at $z=0$, $P_\text{CDM}(k)$ the linear CDM power spectrum 
produced by \texttt{CAMB}, see \cite{Lewis1999}, at redshift $z=0$, 
$T^{2}_\text{FDM}$(k) the CDM to FDM transfer function of \cite{Hu2000} and 
$T^2_\text{dim}(k)=\frac{k^2}{2\pi}$ a transfer function reducing the spectrum's 
dimensionality to one spatial degree of freedom.
The latter follows from demanding a dimension independent real space variance.

The initial phase function, $S(x,a_0)$, is obtained from \eqref{eq:continuity} 
by solving:
\begin{equation*}
    \partial^2_x S(x,a_0) =
    -ma^2_0H(a_0)\delta(x, a_0) \;.
\end{equation*}
Once $\{\delta(x, a_0), S(x,a_0)\}$ are known, the initial wave function follows from 
Madelung's ansatz, see \ref{subsub:linear_regime}. 

Each realisation starts from $z_0 = 100$ and is integrated until $z=0$. 
For the FDM mass two fiducial values, namely $m_1 = \SI{e-23}{\electronvolt}$
and $m_2 = \SI{e-22}{\electronvolt}$ are analyzed.
To guarantee sufficient resolution of $P(k)$ at $z=0$, the number of uniform spatial 
grid points is set to $N=2^{22}$, implying for $L=\SI{100}{\mega\parsec}$ a step size 
of $\Delta x = \SI{23.8}{\parsec}$.

\subsubsection{\label{subsub:overall_P_evo}%
    Overall Evolution of the Matter Power Spectrum
}

\begin{figure}
    \includegraphics{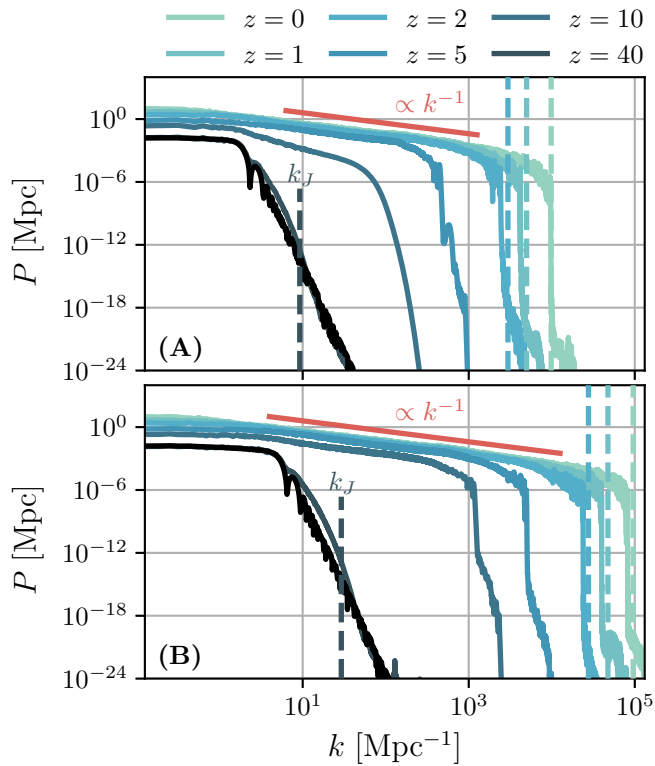}
    \caption{\label{fig:sp_matterpower}%
        (Color online)
        Evolution of the matter power spectrum 
        ${P(k) = \langle|\hat\delta_k|^2\rangle}$ inferred from the simulation 
        ensemble specified in Sec.~\ref{subsub:setup_cosmo} with 
        ${m_1 = \SI{e-23}{\electronvolt}}$ in (A) and 
        ${m_2 = \SI{e-22}{\electronvolt}}$ in (B).
        The black solid lines show the initial power spectrum, \eqref{eq:power_init}. 
        After small scales pass the time dependent Jeans scale $k_J(a)$, their
        associated perturbation modes $\hat\delta_k$ quickly grow in magnitude
        and couple to other nonlinearly evolving modes. At late times, this
        leads to a distinct shape of $P(k)$ with two characteristic regimes:
        At high $k$ the comoving uncertainty principle induces a sharp power 
        suppression past $k_\text{s}(a)$, see \eqref{eq:kcut} and dashed, 
        vertical lines. By contrast, modes with intermediate values of $k$, 
        induce a scale free power spectrum with 
        ${P(k) \propto 1/k}$ cf. Sec.~\ref{subsub:mid_scale}.
    }
\end{figure}

Figure \ref{fig:sp_matterpower} illustrates the evolution of the matter power
spectrum for ${m_1 = \SI{e-23}{\electronvolt}}$, Fig.~\ref{fig:sp_matterpower}(A), and 
${m_2 = \SI{e-22}{\electronvolt}}$, Fig.~\ref{fig:sp_matterpower}(B), at various 
redshifts $z$. In both panels, 
the solid black line represents the linearly rescaled and 
dimensionally reduced reference spectrum of \eqref{eq:power_init} from which the 
initial conditions of each realisation are drawn. 
It is characterized by a flat, large scale regime quickly transitioning into a steep 
power law suppression around $k_J(a)$.
Its exact functional behavior is encapsulated in the FDM transfer function 
$T_\text{FDM}(k)$. 

For both mass parameters the evolution of $P(k)$ may be summarized as
follows: Early on, all modes behave linearly, i.e. evolve according to
\eqref{eq:osci_jeans}. Recall that complex modes $\hat\delta_k$ with 
$k>k_J(a)$ are stabilized by quantum pressure and are therefore confined to
a damped, oscillatory motion with no increase magnitude $|\hat\delta_k|$. 
Consequently, as long as all modes evolve linearly, one expects the power spectrum 
to stay close to its initial shape for $k\gtrapprox k_J(a)$. 
The situation at $z=40$ recovers this behavior for both
mass parameters but is best seen in Fig.~\ref{fig:sp_matterpower}(B) in which $P(k)$ 
slowly detaches itself from the power law suppression regime for $k<k_J$. 
Recall linear FDM modes evolve independent but differently. 
It is therefore no surprise that the initial shape of $P(k)$ is lost.

As time progresses, $k_J$ proceeds to propagate outward until all modes of
interest are destabilized and collapse under their own gravity. It is then that
the linearized description of \eqref{eq:osci_jeans} breaks down and nonlinear
mode coupling is expected to set in --- the independence of each $\hat\delta_k$
is lost. 
Driven by the focusing, nonlinear interaction, the result is a redistribution of 
matter power across all nonlinearly evolving perturbation modes.
This manifests itself in two observable effects in Fig.~
\ref{fig:sp_matterpower}(A)/(B) for all redshifts $z\leq10$: 
Firstly, an intermediate coupling regime emerges that is well described by
$P(k) \propto k^{-1}$. Secondly, the power suppression regime steepens even
further and continues to travel outward, leaving a distinct cutoff in $P(k)$ at
high $k$. Changing the mass parameter influences the transition scale between
both regimes.

\subsubsection{\label{subsub:small_scale}The Suppression Scale}

We recall from Sec.~\ref{subsub:linear_regime} that the Jeans length only
applies in the linear regime and can therefore not explain the observed
transition from matter coupling to suppression.
On the other hand, the uncertainty principle remains applicable even under nonlinear 
evolution. Thus, we adapt the heuristic argument of Sec.~
\ref{subsub:linear_regime} and again identify the hydrodynamic velocity dispersion 
as a measure for the velocity uncertainty $\sigma_v$, but this time infer the 
dispersion directly from the simulation. 
The suppression scale $k_\text{s}$ then follows from:
\begin{equation}
    \label{eq:kcut}
    k_\text{s}(a) = \frac{2\pi}{\sigma_x}
    \approx 
    2\pi a\frac{m}{\hbar}\sqrt{\langle v^2\rangle -\langle v \rangle^2}
\end{equation}
cf. \eqref{eq:v}. Figure \ref{fig:sp_matterpower} illustrates its ensemble average 
$\langle  k_\text{s}\rangle(a)$ as vertical dashed lines for $z<5$. 
The correspondence between $\langle k_\text{s}\rangle(a)$ and the true
suppression scale is convincing. 
We conclude that the small scale evolution remains well 
explained by the uncertainty principle even in the nonlinear evolution regime. 

\subsubsection{\label{subsub:mid_scale} The Coupling Regime}

For scales $k < k_s(a)$ one expects FDM to quickly recover the evolution of
cold dark matter (CDM). The observed power-law behavior of $P(k)$ at scales
$k_\text{box}\ll k < k_s$ should therefore be an intrinsic property of CDM
and a consequence of it being collisionless. In fact, by analyzing the CDM evolution 
in a Lagrangian frame, \cite{Chen2020}
found recently that the small scale limit of the nonlinear CDM power spectrum in 
$(1+1)$ dimensions admits the asymptotic expansion:
\begin{equation*}
    P(k) \sim \frac{\mathcal{C}_0}{k} 
    + \frac{\mathcal{C}_1}{k^3}
    + \frac{\mathcal{C}_2}{k^5}
    + \dots
    + \frac{\mathcal{C}_n}{k^{2n+1}}
    \qquad (k \to \infty) \;,
\end{equation*}
with numerical constants $\mathcal{C}_n$ set by the statistical averages of 
odd derivatives of the Lagrangian displacement field. 
We refer to \cite{Chen2020} for more details.
Since higher order terms die out quickly for large enough $k$, the leading term
is sufficient to describe the small scale behavior of $P(k)$ and
our low redshift power spectra in Fig.~\ref{fig:sp_matterpower} recover the expected 
$1/k$ scaling. 
Obviously, the asymptotic behavior is violated once we approach FDM-modified scales, 
i.e. past $k_s(a)$.

\subsection{\label{sub:artificial}Asymptotic \texorpdfstring{$(1+1)$}{1D}-FDM Dynamics}

We now shift our attention to the subject of asymptotic dynamics and its
equilibrated final $(1+1)$-FDM states. 
To restrict the complexity of the analysis two simplifications are introduced.

Firstly, artificial, i.e. non-cosmologically motivated, initial conditions 
are employed as these allow us to freely set the degree of spatial localization.
This is important since spatially delocalized initial conditions, 
such as the gaussian random fields of Sec.~\ref{sub:cosmo}, usually come with 
multiple overdense regions that collapse into multiple, high mass clusters
which then undergo a subsequent merger and collision phase. It is these
violent, late time events that complicate the dynamics unnecessarily since they drive 
already relaxed clusters again out of equilibrium, therefore increasing the required 
integration time to re-relax into the asymptotic state. 
Moreover, starting from a localized configuration reduces the time to first
collapse.

Secondly, to assure swift relaxation times the scale factor, which acts as a
coupling constant to the nonlinear interaction, is fixed to $a=1$. Extensions of
our static space time results to the expanding FLRW scenario are deferred to
section \ref{subsub:flrw_background}.

\subsubsection{\label{subsub:setup_artificial}Simulation Setup}
More precisely, all simulations reported in this section depart from a gaussian 
initial density with zero initial velocity:
\begin{equation*}
    |\psi(x_3,0)|^2 \propto e^{-\frac{x_3^2}{2\sigma^2}}\;, \quad
    \text{Arg}[\psi(x_3,0)] = 0 \;.
\end{equation*}
The standard deviation is chosen as $\sigma = 6k_J^{-1}(1)$ cf. \eqref{eq:jeans_FCDM}
assuring instability of the entire spectrum of
$\psi$ right from the beginning of the evolution. In order to comply with the periodic
boundary conditions up to floating point precision, the box size is chosen as
$L_3=30\sigma\approx127$ and the numerical study is conducted for both $(1+1)$-SP
and the harmonically confined reduction model. 
In both cases  the number of grid points was chosen such
that the entire wave function spectrum $|\psi_k|^2$ stays resolved throughout the
integration. The integration is stopped
some time after \eqref{eq:virial_general} and/or \eqref{eq:wehrl} indicate the 
completion of the virialization and/or thermalization process.

All data is reported in dimensionless quantities. To get some sense of scale,
choosing the canonical FDM mass $m = \SI{e-22}{\electronvolt}$ and adopting identical 
cosmological parameters as in Sec.~\ref{subsub:setup_cosmo} implies
a box size of $L_3 \approx \SI{2.6}{\mega\parsec}$. 

Both reduction models undergo two distinct evolutionary phases --- a
relaxation period followed by a equilibrated epoch. Their individual
phenomenology, however, depends strongly on the non-local interaction underlying
each $(1+1)$-FDM representation. Below we summarize key
aspects of both phases for each reduction model. 

\subsubsection{\label{subsub:relaxation_artificial}Relaxation}
\begin{figure*}
    \includegraphics{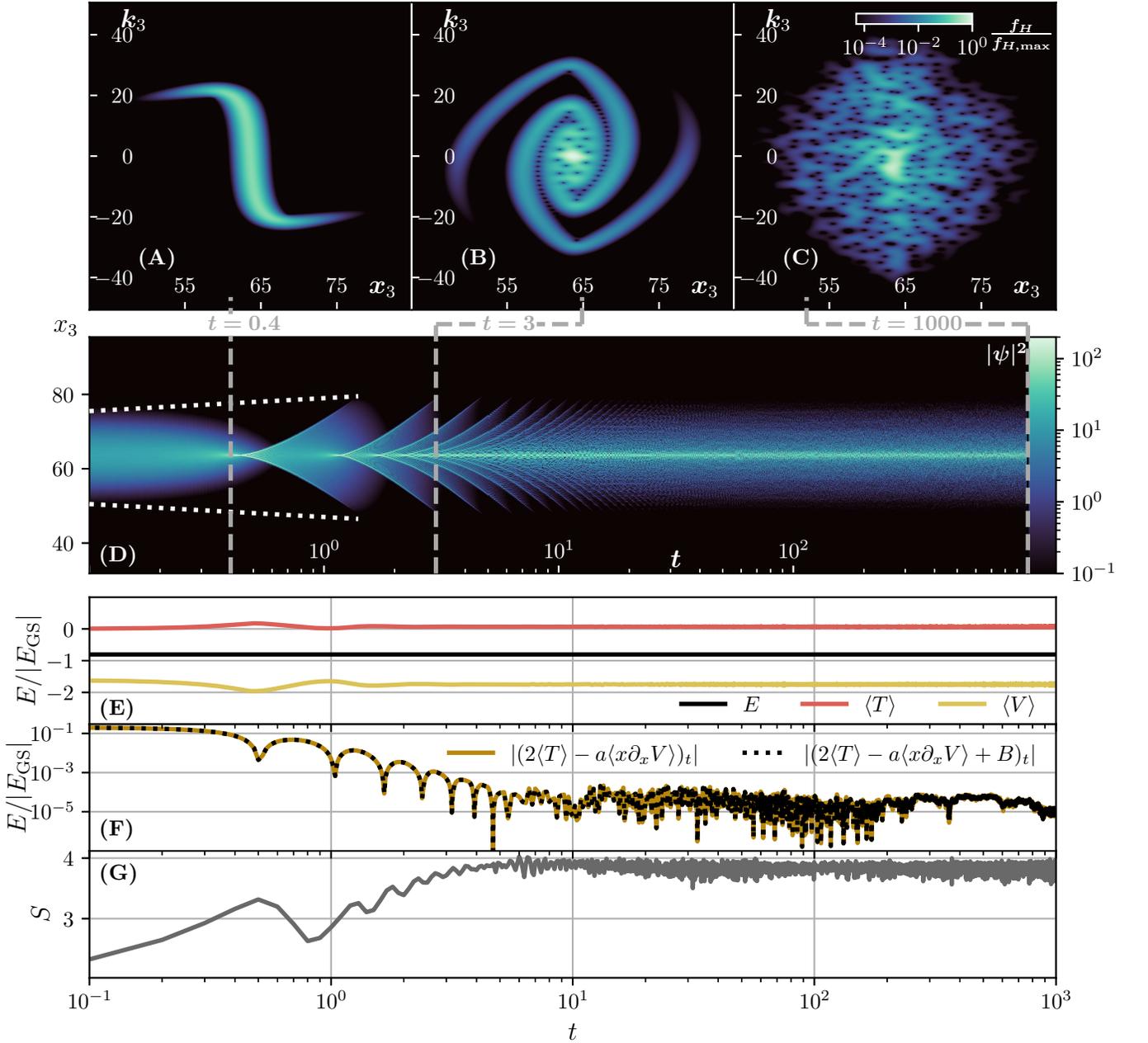}
    \caption{\label{fig:sp_evolution}%
    (Color online)
    Relaxation process of an unstable gaussian density profile under $(1+1)$-SP
    visualized in multiple observables. Panel (A)-(C): Husimi phase space
    distribution, \eqref{eq:smooth_f}, at characteristic stages of the
    evolution. For all panels a constant smoothing scale of $\sigma_x=1/\sqrt{2}$
    was used assuring equal resolution in the spatial and reciprocal domain.
    Panel~(A):
    Moment close to shell crossing, i.e. the instance in time when the initially flat 
    phase space sheet is perpendicular to the spatial axis and both inflowing matter 
    streams cross for the first time.  
    Panel~(B): 
    After first collapse, a recurring process of partial matter outflow from the 
    crossing site accompanied by a subsequent matter re-collapse takes place. 
    The result is a phase space spiral, characteristic for a one-dimensional, 
    collisionless N-body systems, see \cite{Binney2004, Pietroni2018}. 
    Panel~(C): 
    The coarse-grained Husimi distribution, however, attains a quasi-stationary form 
    which we associate with the asymptotic, equilibrated system state. 
    Panel~(D):
    Spatio-temporal evolution of the matter density $|\psi|^2$. No
    qualitatively new features appear in the density past $t\approx5$ and matter
    stays organized in a central, high-density core embedded in a halo of
    fluctuations. White dotted lines visualize the (slight) density expansion in
    $x$-space due to violent relaxation (see main text).
    Panel~(E) 
    Energy evolution. The conservation of \eqref{eq:energy_conservation} is apparent. 
    Panel~(F): 
    Assessment of the relaxation process in terms of the quantum virial theorem,
    \eqref{eq:virial_general}. The absolute deviation from
    \eqref{eq:virial_general} decays quickly until $t\approx 5$ when a lower limit is 
    reached. 
    Panel~(G): 
    Assessment of the relaxation process in terms of the proposed maximization of the 
    entropy functional defined in \eqref{eq:wehrl}. 
    The entropy evolves in a not strictly monotonically increasing fashion until it 
    saturates at $t\approx5$. 
    Comparing (F) and (G) reveals that virialization and
    thermalization occur on the same time scale.
    All energies were normalized to a $M=L_3$ $(1+1)$-SP ground state.
}
\end{figure*}

\paragraph{\label{par:relaxation_sp}$(1+1)$ Schr\"odinger-Poisson}
Figure \ref{fig:sp_evolution} illustrates the relaxation process under
$(1+1)$-SP in multiple observables. 

For $t\lessapprox 5$ relaxation takes place and is characterized by recurring
cycles of (i) infalling matter sheets crossing at the origin,
(ii) overshooting the crossing site at $x_3=L_3/2$, (iii) decelerating until a 
turn-around radius is hit and finally (iv) recollapsing
towards the origin. These cycles do not occur in a strictly sequential manner
but are increasingly superposed and thus induce a characteristic 
spiralization of the phase space distribution in Fig.~\ref{fig:sp_evolution}(B). 
Notice that space regions exist in which multiple inward and outward propagating matter
streams coexists simultaneously.

The structure of the early phase space distributions, Fig.~\ref{fig:sp_evolution} 
(A)/(B) is qualitatively in good accordance with the evolution of one dimensional
collisionless N-body systems, e.g. \cite{Binney2004}, and furthermore show the
natural signature of phase mixing and (less pronounced) violent relaxation. While
phase mixing manifests itself in the ever tighter spiralization of Husimi's
distribution, violent relaxation induces a small yet observable increase in the
occupied phase space volume. The expansion in $k_3$-direction is best seen by
comparing Fig.~\ref{fig:sp_evolution}(A)/(B) whereas the white dotted lines in
Fig.~\ref{fig:sp_evolution}(D) show the spatial expansion. 
As for collisionless $N$-body systems, reason for this expansion is the 
time-dependency of the gravitational potential which dies out quickly after only 
$1-2$ crossing cycles.

Each cycle stage is also observable in the energy components in Fig.
\ref{fig:sp_evolution}(E), where an increase in the kinetic energy follows from the 
steep gradient of $\psi$ when matter streams intersect and are therefore maximally 
localized in space. 
Once matter flows outward the system is less bound thus increasing the expectation 
value of the potential energy.

Only $3-4$ cycles are required to (i) minimize the absolute deviation from the quantum
virial theorem in Fig.~\ref{fig:sp_evolution}(F) and (ii) saturate the value of
the entropy functional in Fig.~\ref{fig:sp_evolution}(G) around $t\lessapprox 5$. 
Thus, the thermalization and virialization time scale are essentially
identical and both metrics capture the convergence into the asymptotic
state equally well in this example. Furthermore, the boundary term $B$,
\eqref{eq:boundary_term}, is negligible in this setting as the long interaction
range of $G_{\Laplace_1}^\text{p}$ does not allow ejected matter clumps to
propagate till the domain boundaries. Hence $\psi$ and its derivatives persist
to be small at $x_3 = 0$.

\begin{figure*}
    \includegraphics{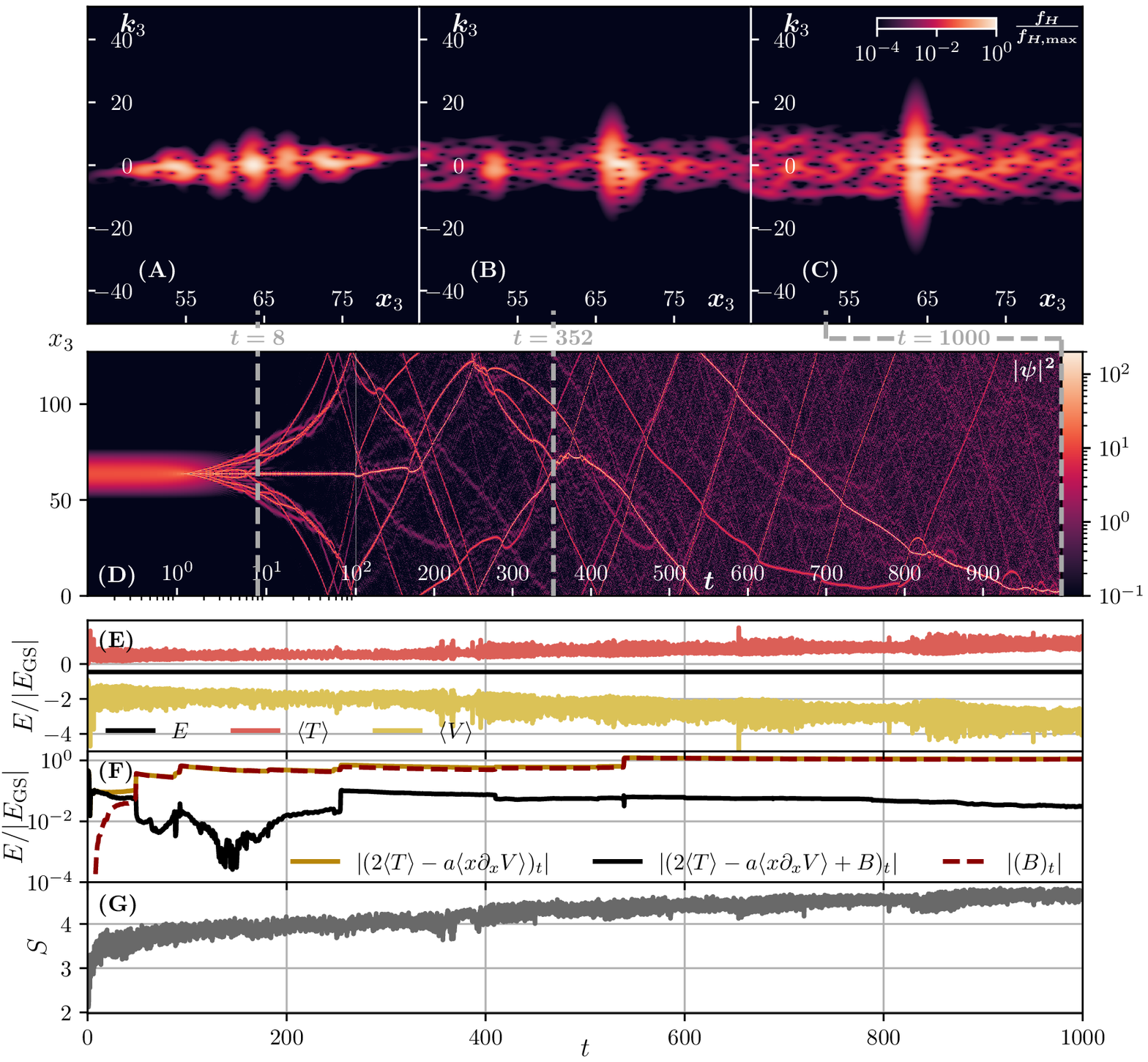}
    \caption{\label{fig:confined_evolution}%
    (Color online)
    Relaxation process of an unstable gaussian density profile under
    strong confinement, i.e. $\epsilon=0.01$, in multiple observables. 
    Panel~(D): 
    Spatio-temporal evolution of $|\psi|^2$. 
    One finds a stark contrast in the evolution of the strongly confined reduction 
    model compared to the unconfined scenario in Fig.~\ref{fig:sp_evolution}. 
    In particular, non-diffusive, stable
    excitations are ejected from the collapse sight. Their bulk velocity is
    sufficient to leave the small interaction range of $U^\text{p}_\text{conf}$ 
    and thus propagate freely through the entire domain without re-collapsing to
    the domain center. 
    The subsequent evolution may then be summarized as series of
    inelastic solitary wave interactions involving different mass ratios and during 
    which high mass excitations slowly consume small mass excitations until a final
    solitary wave persists at $t=1000$. As a 
    byproduct a completely delocalized fluctuation background
    emerges that increases in magnitude up to $\mathcal{O}(1)$. 
    Panel~(A)-(C): 
    Spatially re-centered Husimi distributions, see \eqref{eq:smooth_f}. 
    $\sigma_x$ chosen as in Fig.~\ref{fig:sp_evolution}.
    Panel~(A): Contrary to $(1+1)$-SP no
    phase space spiral develops. Instead circular, solitary excitations manage
    to leave the central gravitational potential. 
    Panel~(B): Example of a excitation merger taking
    place in a spatially delocalized, fluctuating background.
    Panel~(C):
    Quasi-stationary, final state after all solitary waves merged into a single high
    mass stationary configuration. 
    Panel~(E): 
    Energy evolution. Again total energy conservation is apparent. Variations in
    the potential and kinetic energy originate from solitary interactions
    and the associate expel of excess energy during their mergers.
    Panel~(F): 
    Deviation from the virial theorem in \eqref{eq:virial_general}. 
    Note without the boundary term the system would depart from 
    virial equilibrium. All energies in (E)/(F) are
    normalized to the ground state shown in Fig.~\ref{fig:confined_final}.
    Panel~(G): 
    Entropy evolution cf. \eqref{eq:wehrl}. Thermalization takes
    significantly longer compared to $(1+1)$-SP and is only completed at
    $t\approx900$ when all solitary excitations have been consumed. The
    thermalization time is again comparable to the virialization time.
}
\end{figure*}

\paragraph{\label{par:relaxation_confined}Strong Harmonic Confinement}
Figure \ref{fig:confined_evolution} depicts the situation under strong, harmonic
confinement with $\epsilon=0.01$. Again, a relaxation and quasi-stationary phase
may be identified. Their respective duration, however, is opposite to
$(1+1)$-SP. 

During relaxation, i.e. when $t\lessapprox950$, the system
exhibits a short phase of matter emission in response to the violent collapse
of the initial conditions, best seen in spatio-temporal evolution of
$|\psi|^2$ in Fig.~\ref{fig:confined_evolution}(D), around $t\approx2$. There, 
multiple, stable density excitations of various masses depart from the 
position of first collapse, propagate outward, overcome
the central gravitational potential at $x_3 = L_3/2$ and proceed to travel 
towards the domain boundaries as unbound excitations. 
Closer inspection reveals the non-diffusive, form
invariant nature of these excitations --- solitary waves.

The remaining part of the relaxation phase may then be summarized as a series of
inelastic solitary wave encounters, akin to Sec.~\ref{par:collision}.
Recall during these encounters
matter and energy is transferred from the low to high mass solitary wave. 

Once the kinetic energy of a low mass stationary excitation is insufficient to escape
the gravitational well of a high mass solitary wave, a merger takes place,
Fig.~\ref{fig:confined_evolution}(B). 
The matter of both waves then reorganizes into a single
gravitationally bound structure while expelling excess energy in form of small
scale background fluctuations --- the signature of gravitational cooling,
\cite{Seidel1994}. 
These become visible as completely delocalized background in which all solitary
waves are embedded. 

As the gas of solitary excitations continues to rarefy, 
the background grows in magnitude up to $\mathcal{O}(1)$ until all 
stationary states have been consumed by a single
high mass solitary wave, see Fig.~\ref{eq:boundary_term}(C). 
At this point the asymptotic, relaxed system configuration is reached. 

Inspection of the virialization theorem, Fig.~\ref{fig:confined_evolution}(F), 
and the entropy evolution, Fig.~\ref{fig:confined_evolution}(G), reveal that both 
observables capture the relaxation process equally well and report virialization or 
thermalization around $t\approx950$ respectively. 
Importantly, since the dynamics spans over the entire domain, the boundary term in 
\eqref{eq:boundary_term} cannot be neglected. 
In fact, a naive application of the quantum virial theorem omitting the boundary
term would suggest a departure from the equilibrium state.

Let us close this section by mentioning two imperfections of the reported
data. Firstly, $|\psi|^2$ experiences an unphysical symmetry breaking 
around $x_3=L_3/2$ in its evolution past $t=100$ being induced by small numerical 
errors. The same symmetry breaking is also apparent in Fig.
\ref{fig:sp_evolution} for $(1+1)$-SP.
Assessing the situation in more detail reveals an
absolute momentum drift of $\langle p \rangle = 10^{-3}$. 
Given the long integration time and
the high degree of mobility seen in $|\psi|^2$, we find this momentum
conservation violation still to be acceptable.

Secondly, the evolution of the virial theorem experiences unphysical jumps when
significant amounts of matter travels across the domain boundary, e.g. at
$t\approx250$. These upticks originate from sudden changes in the
boundary term, \eqref{eq:boundary_term}, being insufficiently resolved
in the continuous time averaging of \eqref{eq:virial_general}. Clearly,
increasing the sampling rate of $\psi$ around these events mediates this
problem. However, one cannot anticipate a priori when matter flows past the
periodic boundary, making the non-stationary virial theorem cumbersome 
to work with in practice.

\subsubsection{\label{subsub:final_artificial}Final States}

\paragraph{\label{par:final_sp}$(1+1)$ Schr\"odinger-Poisson}
Past $t\approx 5$, one finds the system in a quasi-stationary configuration. 
We coin the attained state quasi-stationary since the entire matter density still 
undergoes significant time-dependent variation yet does not produce 
qualitatively new features in the spatio-temporal evolution of
$|\psi|^2$ cf.  Fig.~\ref{fig:sp_evolution}(D). 
What remains is a circular 
phase space distribution comprised of a high density core and halo of
fluctuations surrounding it, see Fig.~\ref{fig:sp_evolution}(B).
One may think of this phase space distribution
as the result of smoothing out the fine-grained
filament structure of an ever tighter spiralized distribution on the scale of 
Heisenberg's uncertainty principle $\sigma_x\sigma_k=\frac{1}{2}$.

\begin{figure}
    \begin{overpic}[permille]{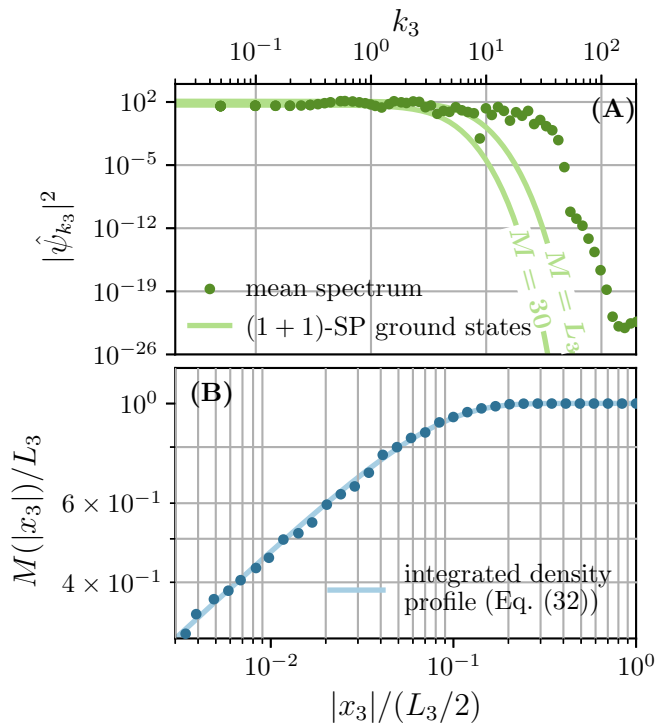}
        \put(647,165){\normalsize{(\eqref{eq:halo_model})}}
    \end{overpic}
    \caption{\label{fig:uniform_final}
        (Color online)
        Analysis of the quasi-stationary state of $(1+1)$-SP extracted from
        Fig.~\ref{fig:sp_evolution}. 
        Panel~(A): Comparison of the mean wave 
        function spectrum obtained by averaging 100 wave functions past $t=990$
        alongside two $(1+1)$-SP ground state spectra. The latter were generated by
        means of a gradient descent, Sec.~\ref{subsub:solitons}. The poor
        correspondence between the spectra suggests that the minimal energy
        solutions to $(1+1)$-SP does not act as dynamical attractors in the evolution.
        Panel~(B): 
        Integrated and normalized FDM density obtained from $|\psi|^2$ at
        $t=1000$ after radial averaging. 
        One finds good correspondence with the halo density model of
        \eqref{eq:halo_model} with $\gamma \approx 0.62$.
    }
\end{figure}

To answer whether the core coincides with a $(1+1)$-SP ground state, we consider the
mean spectral composition of $|\psi_{k_3}|^2$ obtained by averaging 100
quasi-stationary wave functions past $t>990$, i.e. well inside the relaxed
epoch. Fig.~\ref{fig:uniform_final}(A) compares the result with 
the spectrum of a $(1+1)$-SP ground state of maximum mass $M=L_3$. 
The ground states were generated independently, as discussed in Sec. 
\ref{subsub:solitons}.Evidently, no convincing agreement is achieved and the time 
asymptotic spectrum appears generally too broad for any viable ground state with 
$M<L_3$. Changing the width and location of the averaging time interval does not 
yield any improvements.

We conclude although $(1+1)$-SP realizes a cored density profile, this core is
not a ground state configuration of its Hamiltonian 
--- a qualitative difference to $(3+1)$-SP.
Nevertheless, there is still something to be learned about the obtained long term 
density distribution. 
The complementary, classical view point of Sec.~\ref{par:scm} suggests to
compare the results of $(1+1)$-SP with predictions for one-dimensional,
collisionless $N$-body systems, in particular density profiles for
dark matter haloes.

For the situation at hand
\cite{Binney2004} observed how phase mixing and violent relaxation drives such
system towards power-law densities 
$\rho(x_3) \propto |x_3|^{-\gamma}$ with $\gamma \simeq 0.5$. 
Inspired by Einasto's profile, \cite{Einasto1965}, 
the authors of \cite{Schulz2013} extended this halo model by an exponential 
suppression factor dominant past a cut-off radius $r_0$. 
Following this argumentation, one expects:
\begin{equation}
    \label{eq:halo_model}
    \rho(x_3) \propto 
    |x_3|^{-\gamma} 
    \exp\left(-\left(\frac{|x_3|}{r_0}\right)^{2-\gamma}\right)
\end{equation}
for a $d=1$ dimensional CDM halo located at $x_3=0$.

In accordance with \cite{Garny2018}, we consider the integrated and normalized
halo mass $M(|x_3|)/L = \frac{2}{L}\int_0^{|x_3|}\text{d}x_3\rho(x_3)$
instead of $\rho(x_3)$, thus sparing us to choose a particular value of the
smoothing scale $\sigma_x$ --- recall the necessity of smoothing to obtain a
satisfactory Vlasov-Schr\"odinger correspondence. Figure
\ref{fig:uniform_final}(B) depicts how this halo model compares to the simulated
$(1+1)$-SP density at $t=1000$. We observe a satisfactory correspondence with
the fit model of \eqref{eq:halo_model} at $\gamma \approx 0.62$. Better results
are achievable in the limit $\hbar \to 0$, see \cite{Garny2018}.

\paragraph{\label{par:final_confined}$(1+1)$ Strong Harmonic Confinement}
\begin{figure}
    \includegraphics{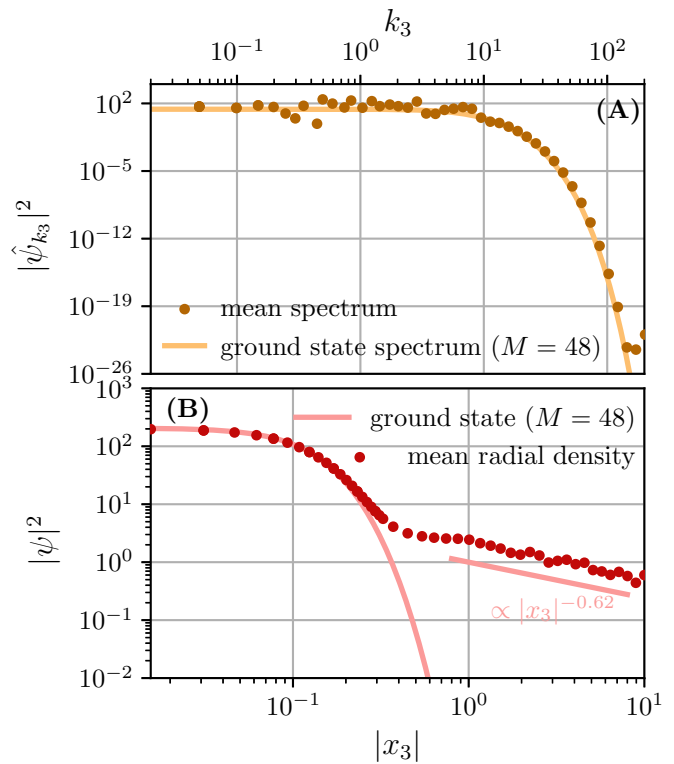}
    \caption{\label{fig:confined_final}
        (Color online)
        Analysis of the quasi-stationary state under strong harmonic
        confinement with $\epsilon=0.01$ extracted from
        Fig.~\ref{fig:sp_evolution}. 
        Panel~(A): Comparison of the mean wave 
        function spectrum obtained by averaging 100 wave functions past $t=990$
        alongside the best matching ground state of mass $M=48$ cf. Fig.
        \ref{fig:gradient_descent}. 
        Also note that the remaining deviations from the ground state spectrum are
        confined to $|k_3|<10$ --- exactly the wave number regime in which the
        delocalized background density is situated in Fig.
        \ref{fig:confined_evolution}.
        Panel~(B): 
        Radially and temporally averaged density. Time averaging as in
        (A). Again a clear distinction between
        ground state core and adjacent halo density can be made. Moreover, the halo
        surrounding the core indicates the same power law scaling found for
        the $(1+1)$-SP halo of Fig.~\ref{fig:uniform_final}.
    }
\end{figure}
Repeating the spectral analysis in Fig.~\ref{fig:confined_final}(A) yields convincing
accordance between the mean wave function spectrum and a strongly confined
ground state of mass $M=48$. The remaining spectral disturbances which were not 
completely suppressed by the time-averaging are confined to $|k_3|<10$. 
Comparison with the quasi-stationary phase space distribution in Fig.~
\ref{fig:confined_evolution}(C) shows they originate from the delocalized background 
oscillations. 

We conclude that the inelastic excitation dynamics experienced during relaxation does 
in fact drive the system towards a single high mass ground state. 
One may regard the minimal energy solution as the fixed point in the long term 
evolution under strong confinement 
--- a result in stark contrast the our observations for
$(1+1)$-SP but qualitatively close to the $(3+1)$-FDM phenomenology.

The halo matter undergoes spatial variation which diminish after averaging
multiple radial density profiles in the same time window used for Fig.
\ref{fig:confined_final}(A). The resulting mean density of Fig.
\ref{fig:confined_final}(B) then indicates a power law halo profile outside the
core consistent with the decay behavior of the numerically obtained $(1+1)$-SP
halo. This is to be expected, since there is no reason to assume a strongly confined, 
one dimensional halo organizes into a canonical $1/|x_3|^3$-NFW profile \cite{Navarro1996}. 
Intuitively, the behavior of a dark matter halo should
be influenced by (i) the interaction in the far field and (ii) dimension
dependent effects such as geometrical dilution. 

\begin{figure}
    \includegraphics{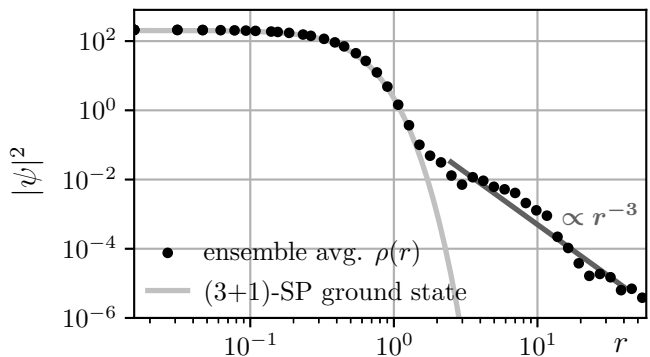}
    \caption{\label{fig:radial_final}
        (Color online)
        Ensemble averaged radial density deduced from $\mathcal{N}=20$
        realisations of different mass, initial gaussians.
        The integration was performed under the assumption of spherical symmetry.
        Notice that in addition to the
        ground state (\textcolor{red}{\q{solitonic}}) core, 
        \textcolor{red}{%
            the remaining matter organizes into a power law halo, 
            $\rho \propto r^\gamma$, at larger radii. A fit yields 
            $\gamma\approx-3.1$. Within the limits of accuracy of our data,
            we deem this to be consistent with the asymptotic behavior of the 
            NFW halo density, i.e., $\rho \sim r^{-3}$, shown as gray solid line.
        }
    }
\end{figure}

To substantiate this claim, we impose spherical symmetry to the full-fledged
problem in \eqref{eq:sp3d} and adopt the regularization approach of
\cite{Dong2011}. The choice of spherical symmetry presents yet another reduction of 
dimensionality to $(1+1)$.
Doing so allows us to implement the same $1/r$-far field
behavior as in our confinement model while making the density dilution in
radial direction manifest. Excess matter radiated away during the relaxation
process is suppressed with a complex absorbing potential situated at the domain
boundary, e.g. \cite{Guzman2004,Schwabe2016}. We refer to Fig.~\ref{fig:radial_final} for the relaxed, mean density profile
obtained from a ensemble of $\mathcal{N}=20$ gaussian initial conditions of
various masses. Application of the scaling symmetry,
\eqref{eq:scaling_symmetry}, allows us to rescale each realization to a common peak 
density.
Evidently, matter not included in the ground state core now organizes into a NFW
profile. 

\subsubsection{\label{subsub:organization}Control of Self-Organization Processes}

The foregoing discussion highlighted the superiority of the transversal confinement
model in mapping the $(3+1)$-FDM phenomenology to one dimensional analogues ---
while both models indicate reasonable accordance with classical predictions for
the outer halo density, it is only under confinement that the equilibrated state
evolves towards a ground state core.

In fact, the observed self-organization principle of $(1+1)$-FDM 
under strong confinement is not new. The author of \cite{zakharov1988}
showed how for a class of focusing, \emph{local} nonlinearities of
the NLSE perturbed uniform initial conditions have a single soliton as dynamical
attractor. 
More precisely, the perturbed initial conditions develop a number of small mass
solitons which subsequently merge into a single high-mass soliton at late times. 
This phenomenon was coined \emph{soliton turbulence} and it was argued it is 
"thermodynamically favorable" for the system to develop in this particular way. 
The authors of \cite{Jordan2000} later put these findings on more theoretical 
grounds by developing a statistical theory around a mean-field approximation of the
nonlinear Hamiltonian obeying a maximum entropy principle. 

The problem of \emph{non-local} interactions was considered in the context of 
nonlinear optics by \cite{Picozzi2011}. 
Numerical and analytical arguments showed that the dynamics
is mainly driven by the ratio between the interaction range $R$ and the soliton size
$\sigma$: If the interaction range is too large, matter far away
from a potential soliton, but still within interaction range, 
contributes significantly to the convolution integral. Consequently, the
delicate potential required to form a soliton gets averaged out by the
surrounding fluctuations. Hence, one expect soliton-turbulence-like behavior for
$R \ll \sigma$. In case of $R \geq \sigma$, the system organized
into a "spatially localized incoherent structure" coined \emph{incoherent soliton}. 
Their results resemble our findings for the quasi-stationary state of
$(1+1)$-SP.

A limit not yet discussed, is the weak confinement regime, i.e. $\epsilon
\to \infty$. It is intuitively clear that in this case the interaction kernel
$U^{\text{p}}_{\text{conf}}$ should approach $G^\text{p}_{\Laplace_1}$. A more
careful analysis shows 
\begin{equation*}
    U^{\text{p}}_{\text{conf}}(x_3, x_3') \sim
    \frac{1}{2\pi\epsilon^2}G^\text{p}_{\Laplace_1}(x_3, x_3')
    \qquad (\epsilon \to \infty) \;.
\end{equation*}
Increasing $\epsilon$ should therefore allow us to observe a transition from the
soliton to incoherent soliton turbulence regime. To keep relaxation times
comparable we also substitute $a \to 2\pi\epsilon^2 a$ so that the effective
nonlinear coupling stays unity. 

\begin{figure*}
    \includegraphics{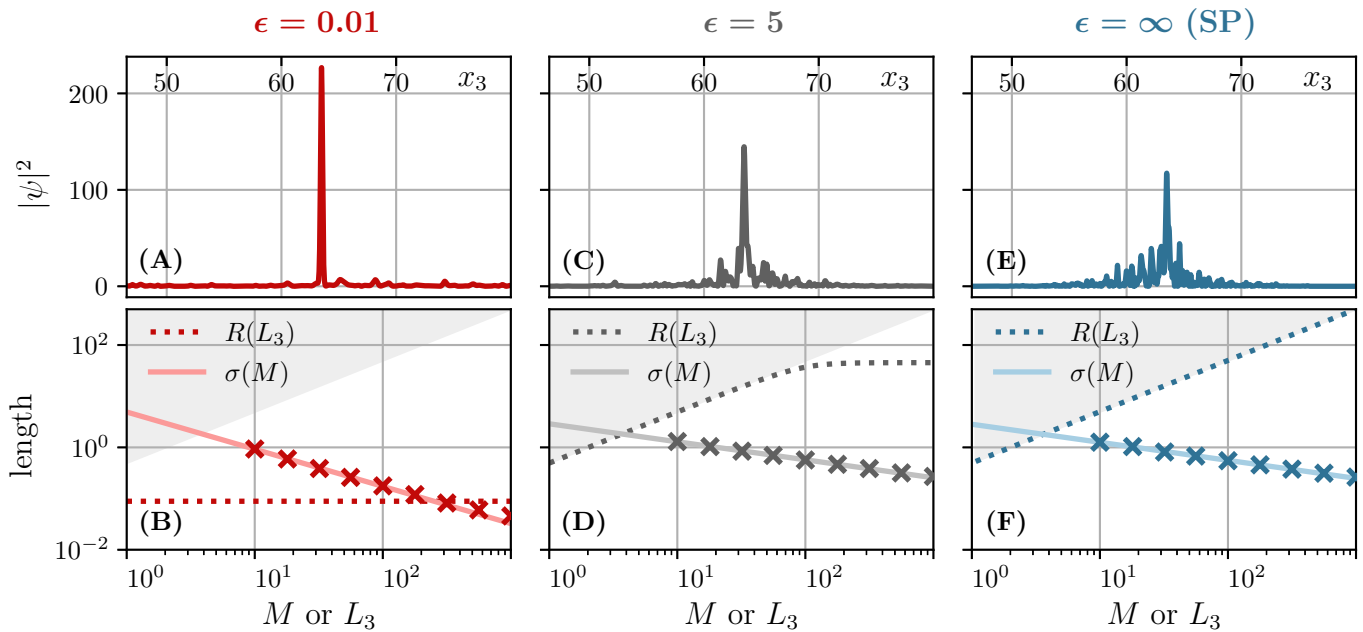}
    \caption{\label{fig:transition}
    (Color online)
    Overview of asymptotic states as a function of the confinement parameter
    $\epsilon$. 
    From left to right, we have (A)-(B) analyzing the strong
    confinement limit ($\epsilon = 10^{-2}$),
    (C)-(D) assessing the weak confinement regime ($\epsilon = 5$) and 
    (E)-(F) evaluating the uniform reduction scenario in the limit 
    $\epsilon\to\infty$, i.e. $(1+1)$-SP. Upper panels: Snapshot of the
    attained quasi-stationary states. Lower panels: Comparison of the
    interaction range $R(L_3)$, \eqref{eq:range}, and the soliton size $\sigma(M)$,
    \eqref{eq:soliton_size}. Crosses denote soliton sizes directly inferred from
    the gradient descent of various mass solitons cf. Fig.
    \ref{fig:spatial_extend}. Solitons with sizes inside the gray shaded area
    do not exist as they violate periodic boundary conditions.
    For $\epsilon>1$, we find a transition
    away from turbulent soliton dynamics toward a 
    "incoherent soliton" configuration,  i.e. a highly fluctuating state comprised 
    of many density maxima beating against each other in real space. 
    As argued in \cite{Picozzi2011}, this regime is
    entered once the interaction range $R(L_3)$ is significantly larger
    than the soliton size $\sigma(M)$. Even in the best case scenario for
    (D) and (F), i.e. when a soliton of maximal size could form, 
    one still finds $\sigma_\text{max}/R(L_3\approx127) < 0.1$ --- far outside the
    soliton regime. By contrast, the strong confinement scenario of Fig.
    \ref{fig:confined_final}, or equivalently (A), realizes 
    $\sigma(M=48)/R(L_3\approx127) > 1$ and is therefore well inside the soliton 
    regime.
}
\end{figure*}

Figure \ref{fig:transition} compares 
the asymptotic state obtained under strong, weak, and no confinement alongside
the respective interaction range $R(L)$, \eqref{eq:range}, and soliton
extent $\sigma(M)$, \eqref{eq:soliton_size}. 

We find confinement parameters
larger than unity to quickly approach quasi-stationary states comprised 
of many density maxima beating against each other around the origin. These
configurations are qualitatively identical to the $(1+1)$-SP case. Comparing the
maximal, boundary condition compatible soliton size $\sigma_\text{max}$ with the
interaction range $R(L_3)$ at the chosen domain size $L_3\approx127$ shows
$\sigma/R < 0.1$, which according to \cite{Picozzi2011} implies "incoherent
soliton" dynamics, and in particular no solitonic attractor. On the other hand
comparing both length scales for $\epsilon=0.01$, where a $M=48$ soliton is
formed, we have $\sigma/R > 1$, consistent with the soliton turbulence regime.

\subsubsection{\label{subsub:flrw_background}Space-time Expansion}

The foregoing results of Sec.~\ref{subsub:organization} allow us to extend
the discussion to non-static background cosmologies. We first note
that the interaction range, as defined in \eqref{eq:range}, is independent of the 
nonlinear coupling constant. The soliton size $\sigma(M)$, on the other hand,
is. This is intuitively clear: Decreasing the nonlinear coupling increases the
importance of the diffusive character of kinetic term in the Hamiltonian --- we
approach a free Schr\"odinger equation. Hence, the radius at which the focusing
nature of the non-linearity balances the kinetic term is expected to increase as
well. 

Under strong confinement cf. Fig.~\ref{fig:transition}(B) an expanding
background cosmology would therefore drive the system even further into the
soliton turbulence regime $\sigma \gg R(L)$. 

\begin{figure}
    \includegraphics{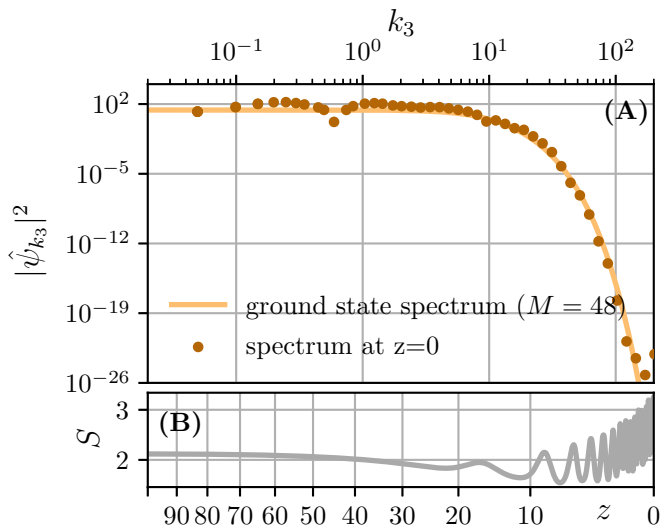}
    \caption{\label{fig:confined_final_expanding}
        (Color online)
        Strong confinement model with $\epsilon=0.01$ undergoing collapse in
        a background cosmology as in Sec.~\ref{subsub:setup_cosmo}.
        Panel~(A):
        Final wave function spectrum at $z=0$ together with the same
        $M=48$ ground state of Fig.~\ref{fig:confined_final}.
        Panel~(B):
        Entropy evolution. It is evident that relaxation is not completed at
        $z=0$.
    }
\end{figure}

Without confinement cf. Fig.~\ref{fig:transition}(F) the typical increase 
of $\sigma(M)$ experienced by starting from reasonable initial redshifts, say $z=100$, 
is insufficient to realize $\sigma_\text{max} \approx R(L)$, around which a
transition to the soliton turbulence regime should occur. 
Note that here $\sigma_\text{max} > R(L)$ 
is never achievable as it would violate the periodic boundary conditions.
That said, allowing for a time dependent coupling constant has, aside from
numerical implications, also influence on the relaxation time. Preliminary
analysis shows that although the strong confinement scenario including a FLRW
background does trend towards a ground state (\textcolor{red}{\q{solitonic}}) 
spectrum, relaxation is not completed at $z=0$. 
Figure \ref{fig:confined_final_expanding} illustrates this result. 
A full-fledged analysis of the expanding model is left to future work.

\section{\label{sec:concl} Conclusion}

Purpose of this work was to conduct an extensive numerical study on the
applicability of the Fuzzy Dark matter (FDM) model in one spatial dimension.
Particular emphasis was put on (i) properties of system's long term evolution,
(ii) the structure of the relaxed, asymptotic system state and how it
compares to the core-halo structure of $(3+1)$-FDM, as well as (iii) 
which model parameters may be used to control its phenomenology.
To this end, we derived two distinct one-dimensional FDM models by either
allowing for a complete delocalization of matter in the transversal plane, 
$(1+1)$-SP, or by confining all matter along one spatial
direction. While both models realize long range interactions free of
singularities, it is only under strong confinement
that the nonlocal interaction recovers the desired $-1/r$ interaction at large distances.

We proceeded to investigate the mean cosmic structure growth in
an ensemble of FDM-only, flat FLRW-universes obeying $(1+1)$-SP and starting
from cosmological initial conditions. By following the evolution of the matter
power spectrum until present time, two distinct spectral ranges were
identified: 
Firstly, a suppression range, in which the power spectrum is smoothed out by the
uncertainty principle. Secondly, a coupling regime, where the redistribution of 
matter across nonlinearly evolving modes leads to a scale-free spectrum
consistent with considerations for $(1+1)$-CDM.
The transition scale from coupling to suppression followed by a
self-consistent application of the uncertainty principle.

The analysis of the asymptotic system state was conducted under simplifying
assumptions, i.e., a static background cosmology and spatially localized
initial conditions. Our study suggests that $(1+1)$-SP relies on 
violent relaxation and phase mixing to approach its equilibrated, i.e. 
thermalized and virialized, system state.
Although the realized asymptotic configuration does form a core-halo structure, the
central core cannot be identified with the ground state solution of the
$(1+1)$-SP Hamiltonian --- a stark contrast to $(3+1)$-FDM for which the minimum
energy solution acts as dynamical attractor in the long term evolution. The halo
density, on the other hand, was found to be consistent with structure of one
dimensional CDM halos.

By contrast, the evolution under strong confinement favours the $(3+1)$-FDM typical
relaxation mechanism of gravitational cooling. Ultimately, the evolution then 
converges into a virialized and thermalized system state comprised of a single 
high-mass ground state solution embedded in a delocalized fluctuation background that
emerges from a series of inelastic ground state interactions.
The analysis of the halo density suggested identical CDM-like behavior as
for $(1+1)$-SP. We conclude that under the chosen simulation conditions the
strongly confined reduction model is superior in mapping the three-dimensional
phenomenology to one spatial dimension.

To understand the reason for the qualitative difference between the asymptotic 
behavior of $(1+1)$-SP and the confinement model, we investigated the weak 
confinement limit of our reduction. 
In accordance with arguments from nonlinear optics, see e.g. \cite{Picozzi2011}, we
found the system to converge towards a high-mass ground state if the effective
interaction range is (considerably) smaller than the spatial extent of the
ground state. For $(1+1)$-SP and arbitrary but fixed coupling constant, 
no ground state exists that satisfies this condition. 
It is for this reason, that we conjectured our static space-time analysis remains 
valid even for nontrivial background cosmologies.
Although our work focused on an interaction kernel that resembles 
Newtonian gravity, results from nonlinear optics suggest that  the dependence of the 
asymptotic state on the interaction range is a property also applicable for other 
long range potentials. For instance, the authors of \cite{Picozzi2011}
implemented a gaussian interaction kernel in \eqref{eq:nlse1d}, while \cite{Segev2015}
employed a screened Poisson equation as field equation. The latter approach
implies an exponentially decaying Green’s function.

Our work may be extended in multiple regards. From a physical perspective, a
full-fledged investigation of cosmological initial 
conditions in various cosmological expansion models is still pending and the
phenomenology under strong confinement is presumably not exhausted by our discussion. 
In this context, we mention the properties of the delocalized fluctuation
background as it may be possible to understand it as an ensemble of small scale plane
waves obeying a dispersion relation akin to Bogoliubov's excitation spectrum for
Bose-Einstein condensates \cite{Stringari2016,Maddaloni2000}.

Moreover, additional conceptional optimizations of our confinement approach are worth
exploring. For instance, our work only focused on a global, statically set
confinement parameter. However, incorporating the confinement ansatz directly into
SP's generating action is expected to yield additional information on the
spatio-temporal evolution of the confinement strength itself, thereby allowing
it to be set self-consistently and dependent on the wavefunction evolution
\cite{Salasnich2002}.

We also remark on our ongoing effort to optimize our numerical approach by
means of a more efficient basis-function method or splitting schemes with 
intrinsic error estimates. With this we hope to (i) achieve a fully adaptive
spatio-temporal grid that is sensitive to nonlinear evolution and (ii) pave the
way for a higher dimensional analysis. The latter should allow the investigation
of additional relaxation channels unique to FDM, especially the emission of
quantized vortices that may play an important role for the asymptotic evolution
on top of gravitational cooling.

Altogether, we hope that our investigation will lead to a better cross-fertilisation, 
see also \cite{Paredes2020}, between cosmology, statistical mechanics 
\cite{Kolmogorov91, Kobayashi2005}, nonlinear dynamics 
\cite{Tabor1989, Lakshmanan2003}, nonlinear wave optics 
\cite{Picozzi2011, Segev2015, Roger2016, Navarrete2017}, and the quantum evolution of,
e.g., Bose-Einstein condensates 
\cite{Jain2007, Girelli2008, Proment2009, Lahaye2009, Plestid2018, Combescot2017, 
Berezhiani2019}, with the scope to obtain further insight into the complex dynamics 
of the SP model and to search for possible laboratory experiments for the 
implementation and analogue simulation of FDM models.

\begin{acknowledgments}
The authors acknowledge support by the state of Baden-W\"urttemberg, Germany 
through bwHPC. Our special gratitude goes to Luca Amendola for his feedback to this
work, Javier Madro\~nero for various conversations on NLSE dynamics and to Jens
Niemeyer for the insightful discussion on the subject in general.
\end{acknowledgments}

\appendix
\section{\label{sec:convergence}Convergence and Stability}

Let us now give a numerical justification for the accuracy and stability claims
mentioned in Sec.~\ref{sub:summary_of_properties}. 
To this end, we conduct a convergence and stability study for the simulation 
scenarios of Sec.~\ref{subsub:setup_cosmo} (delocalized random field
including space-time expansion) and \ref{subsub:setup_artificial} (unstable
gaussian initial conditions in a static background cosmology).

The authors are not aware of a general analytical result in any of these cases. Thus,
we compute a reference solution $\psi_\text{ref}$ on a fine spatio-temporal grid 
$\{N_\text{ref},\Delta t_\text{ref}\}$ and measure the error of $\psi$ relative
to $\psi_\text{ref}$ via
$\Delta \epsilon = \norm{\psi_\text{ref} - \psi}_2 / \norm{\psi_\text{ref}}_2$.
Note the error $\Delta \epsilon$ is a function of $N$, $\Delta t$ and integration time
$t$. Since the results are qualitatively identical for $(1+1)$-SP and the
confinement model we only report data for the former.

\begin{figure*}
    \includegraphics{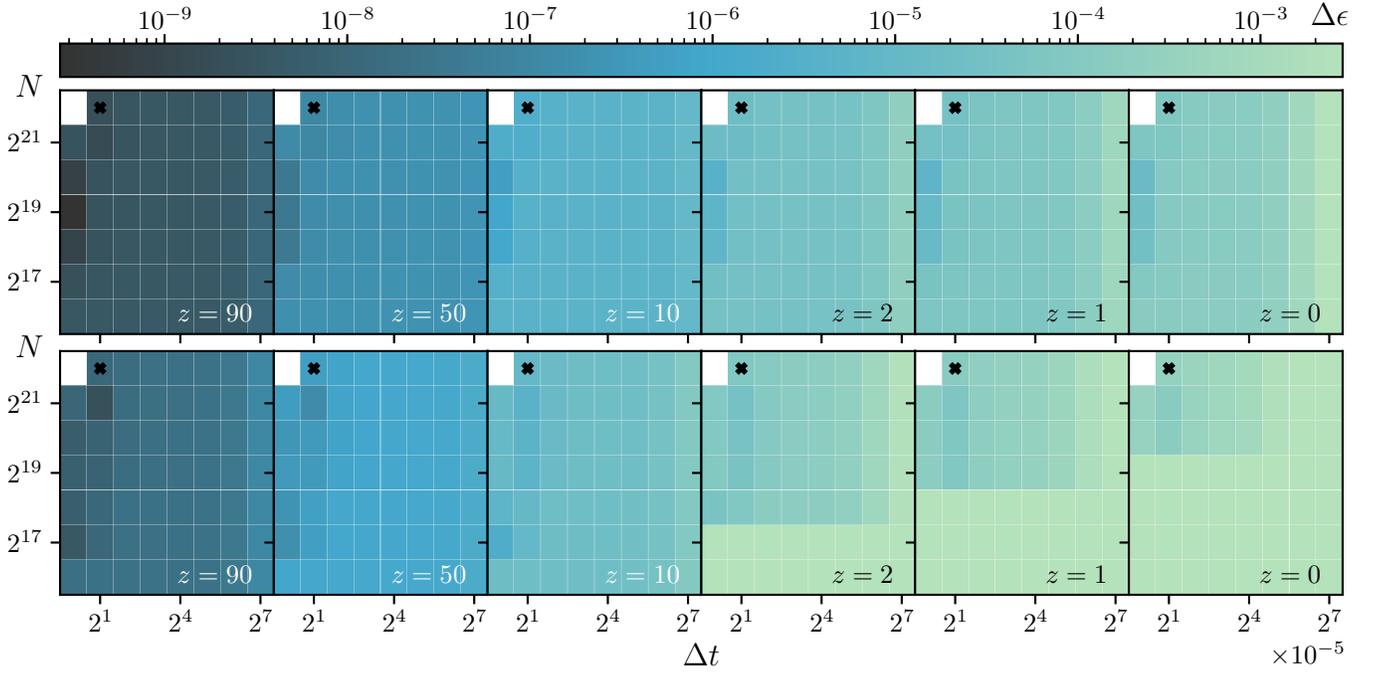}
    \caption{\label{fig:convergence_matrix}
        (Color online)
        Evolution of the relative error $\Delta \epsilon$ as a function of the 
        spatio-temporal grid.
        The convergence study adopts the initial conditions and parameters of Sec.~
        \ref{subsub:setup_cosmo} using
        ${m_1 = \SI{e-23}{\electronvolt}}$ in the first row and 
        ${m_2 = \SI{e-22}{\electronvolt}}$ in the second row.
        Space-time expansion is turned on.
        The white cells corresponds to the numerical reference solution 
        $\psi_\text{ref}$ relative to which $\Delta \epsilon$ is computed. 
        The black cross represents the spatio-temporal grid used for Fig.~
        \ref{fig:sp_matterpower}.
    }
\end{figure*}

We begin with the cosmological simulation scenario of Sec.~\ref{sub:cosmo},
i.e. a random field evolving in a dynamic FLRW background.
Figure \ref{fig:convergence_matrix} depicts the dependence of the numerical
error as a function of the spatio-temporal grid parameters $\{N, \Delta t\}$
relative to the reference grid $\Delta t_\text{ref} = 10^{-5}$ and 
$N_\text{ref}=2^{22}$ (the white cells). 
The grid parameters of Sec.~\ref{sub:cosmo} are marked with a black cross.
To assure comparability, all $56$ $\{N,\Delta t\}$ combinations are
initialized with every $N/N_\text{ref}$ point of the same reference gaussian 
random field $\delta(x)$. 

In the $m=m_1$ case, we find a high degree of uniformity in $N$ at fixed 
$\Delta t$ throughout the integration. This is the result of the spectral
accuracy of the employed spatial discretization. 
Variations of the relative error $\Delta\epsilon$
in $N$ at $\Delta t=1-2 \times 10^{-5}$ are common once we reach the convergence 
plateau but may also be induced by a lack of fidelity close to the reference
solution. More pronounced is the loss of accuracy in $\Delta t$ direction 
(at any considered $N$) and we conclude the overall inaccuracy is dominated 
by the temporal error. Solutions $\Delta t \leq 8\times 10^{-5}$ can be 
considered as converged.

The situation for the $m=m_2$ scenario is qualitatively identical with
an additional caveat at low red shifts. Here, spatial grids with $N <
2^{20}$ proof to be insufficient to resolve the entire spectrum of
$|\hat\psi_{k_3}|$. In this case spatio-temporal grids with 
$N>2^{20}$ and $\Delta t\leq8\times10^{-5}$ are deemed sufficient to achieve 
convergence.

\begin{figure}
    \includegraphics{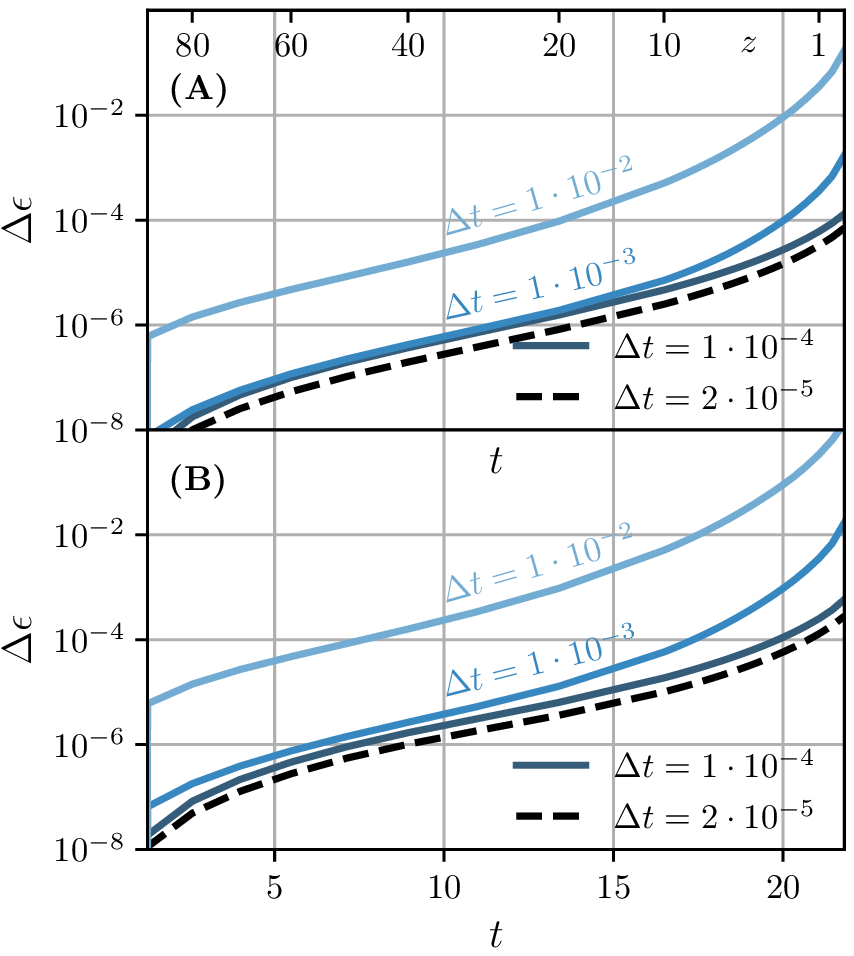}
    \caption{\label{fig:stability}
        (Color online)
        Evolution of the numerical error $\Delta \epsilon$ as a function of the 
        integration time.
        The convergence study adopts the initial conditions and parameters of Sec.~
        \ref{subsub:setup_cosmo} 
        with ${m_1 = \SI{e-23}{\electronvolt}}$ in panel (A) and 
        ${m_2 = \SI{e-22}{\electronvolt}}$ in panel (B). 
        Space-time expansion $a(t)$ is turned on.
        The space-time grid of Sec.~\ref{sub:artificial} corresponds to the
        black, dashed line.
    }
\end{figure}

Concerning temporal accuracy and overall stability, we refer to Fig.
\ref{fig:stability} which evaluates the time dependence of the numerical error 
for various time increments $\Delta t$ at fixed $N=N_\text{ref}=2^{22}$. The reference
solution is identical to the one used in Fig.~\ref{fig:convergence_matrix} and
the spatio-temporal grid of Sec.~\ref{sub:cosmo} is depicted as red, dashed
line.

Evidently, allowing for an dynamic cosmological background yields a
numerical error that evolves roughly exponentially.
Nevertheless, errors are still of acceptable size at present time which is why
we deem our numerical treatment of the non-autonomous Hamiltonian
as acceptable for the purposes of this work.
Decreasing the time step beyond $\Delta t < 10^{-4}$ results in no significant gain in
accuracy. Note that the non-converged time steps, i.e. $\Delta t = 10^{-2},
10^{-3}$ approach the convergence plateau with roughly quadratic speed.
This can be inferred from their relative offset which evaluates to about
$2$ orders of magnitude and is expected given that Strang
splitting cf. Sec.~\ref{sub:evolution} is second order accurate in time.

\begin{figure}
    \includegraphics{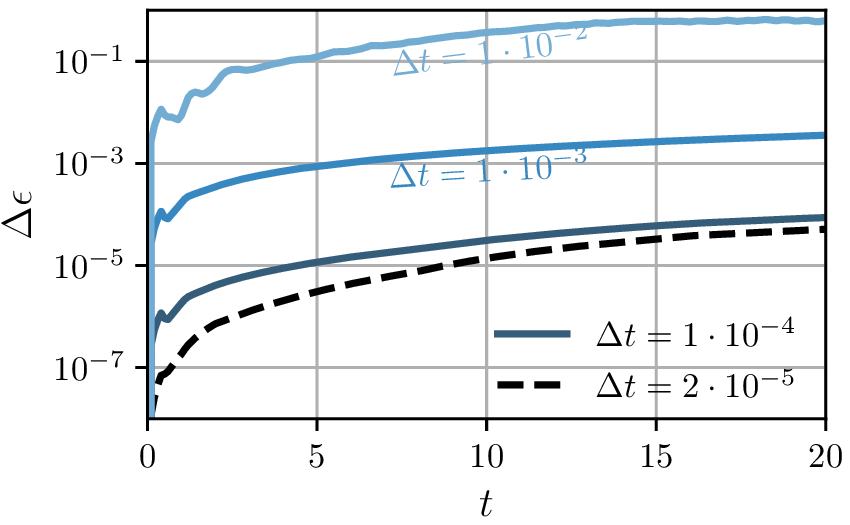}
    \caption{\label{fig:stability_gauss}
        (Color online)
        Evolution of the numerical error $\Delta \epsilon$ as a function of the 
        integration time. 
        The convergence study adopts the initial conditions and parameters of 
        Sec.~ \ref{subsub:setup_artificial}.
        Space-time expansion is turned off, i.e., $a=1$.
        The space-time grid of Sec.~\ref{sub:artificial} corresponds to the
        black, dashed line.
    }
\end{figure}

Let us contrast the error evolution of the fully cosmological case with the
static simulation conditions of Sec.~\ref{sub:artificial} cf. Fig.
\ref{fig:stability_gauss}. Here, the smaller dimensionless box size allows us to 
reduce the number of spatial grid points required to fully resolve the spectrum of
$\psi$. This in turn makes a smaller reference time increment possible. Figure
\ref{fig:stability_gauss} therefore uses a reference grid with $\Delta
t_\text{ref} = 10^{-6}$ and $N_\text{ref} = 2^{13}$. Again, the grid parameters
used in Sec.~\ref{sub:artificial} correspond to the red, dashed line.

As for the cosmological evolution, quadratic accuracy is achieved for
non-converged time steps approaching the plateau. A notable difference, however,
is the growth behavior of $\Delta \epsilon(t)$ which only
evolves linearly in time if the scale factor remains static. 
It is for this reason that we can extend the
integration time up to $t=1000$ cf. Sec.~\ref{sub:relaxation} without losing 
reliability of our data.


\begin{thebibliography}{85}%
\makeatletter
\providecommand \@ifxundefined [1]{%
 \@ifx{#1\undefined}
}%
\providecommand \@ifnum [1]{%
 \ifnum #1\expandafter \@firstoftwo
 \else \expandafter \@secondoftwo
 \fi
}%
\providecommand \@ifx [1]{%
 \ifx #1\expandafter \@firstoftwo
 \else \expandafter \@secondoftwo
 \fi
}%
\providecommand \natexlab [1]{#1}%
\providecommand \enquote  [1]{``#1''}%
\providecommand \bibnamefont  [1]{#1}%
\providecommand \bibfnamefont [1]{#1}%
\providecommand \citenamefont [1]{#1}%
\providecommand \href@noop [0]{\@secondoftwo}%
\providecommand \href [0]{\begingroup \@sanitize@url \@href}%
\providecommand \@href[1]{\@@startlink{#1}\@@href}%
\providecommand \@@href[1]{\endgroup#1\@@endlink}%
\providecommand \@sanitize@url [0]{\catcode `\\12\catcode `\$12\catcode
  `\&12\catcode `\#12\catcode `\^12\catcode `\_12\catcode `\%12\relax}%
\providecommand \@@startlink[1]{}%
\providecommand \@@endlink[0]{}%
\providecommand \url  [0]{\begingroup\@sanitize@url \@url }%
\providecommand \@url [1]{\endgroup\@href {#1}{\urlprefix }}%
\providecommand \urlprefix  [0]{URL }%
\providecommand \Eprint [0]{\href }%
\providecommand \doibase [0]{https://doi.org/}%
\providecommand \selectlanguage [0]{\@gobble}%
\providecommand \bibinfo  [0]{\@secondoftwo}%
\providecommand \bibfield  [0]{\@secondoftwo}%
\providecommand \translation [1]{[#1]}%
\providecommand \BibitemOpen [0]{}%
\providecommand \bibitemStop [0]{}%
\providecommand \bibitemNoStop [0]{.\EOS\space}%
\providecommand \EOS [0]{\spacefactor3000\relax}%
\providecommand \BibitemShut  [1]{\csname bibitem#1\endcsname}%
\let\auto@bib@innerbib\@empty
%</preamble>
\bibitem [{\citenamefont {Fetter}\ and\ \citenamefont
  {Walecka}(2003)}]{Fetter2003}%
  \BibitemOpen
  \bibfield  {author} {\bibinfo {author} {\bibfnamefont {A.~L.}\ \bibnamefont
  {Fetter}}\ and\ \bibinfo {author} {\bibfnamefont {J.~D.}\ \bibnamefont
  {Walecka}},\ }\href {http://www.worldcat.org/isbn/0486428273} {\emph
  {\bibinfo {title} {{Quantum Theory of Many-Particle Systems}}}}\ (\bibinfo
  {publisher} {Dover Publications},\ \bibinfo {year} {2003})\BibitemShut
  {NoStop}%
\bibitem [{\citenamefont {Mahan}(2000)}]{Mahan2000}%
  \BibitemOpen
  \bibfield  {author} {\bibinfo {author} {\bibfnamefont {G.~D.}\ \bibnamefont
  {Mahan}},\ }\href@noop {} {\emph {\bibinfo {title} {{Many Particle Physics
  (Physics of Solids and Liquids)}}}}\ (\bibinfo  {publisher} {Kluwer
  Academic/Plenum Publishers, New York},\ \bibinfo {year} {2000})\BibitemShut
  {NoStop}%
\bibitem [{\citenamefont {Dalfovo}\ \emph {et~al.}(1999)\citenamefont
  {Dalfovo}, \citenamefont {Giorgini}, \citenamefont {Pitaevskii},\ and\
  \citenamefont {Stringari}}]{Dalfovo1999}%
  \BibitemOpen
  \bibfield  {author} {\bibinfo {author} {\bibfnamefont {F.}~\bibnamefont
  {Dalfovo}}, \bibinfo {author} {\bibfnamefont {S.}~\bibnamefont {Giorgini}},
  \bibinfo {author} {\bibfnamefont {L.~P.}\ \bibnamefont {Pitaevskii}},\ and\
  \bibinfo {author} {\bibfnamefont {S.}~\bibnamefont {Stringari}},\ }\bibfield
  {title} {\bibinfo {title} {{Theory of Bose-Einstein condensation in trapped
  gases}},\ }\href {https://doi.org/10.1103/revmodphys.71.463} {\bibfield
  {journal} {\bibinfo  {journal} {Reviews of Modern Physics}\ }\textbf
  {\bibinfo {volume} {71}},\ \bibinfo {pages} {463} (\bibinfo {year}
  {1999})}\BibitemShut {NoStop}%
\bibitem [{\citenamefont {Pitaevskii}\ and\ \citenamefont
  {Stringari}(2016)}]{Stringari2016}%
  \BibitemOpen
  \bibfield  {author} {\bibinfo {author} {\bibfnamefont {L.~P.}\ \bibnamefont
  {Pitaevskii}}\ and\ \bibinfo {author} {\bibfnamefont {S.}~\bibnamefont
  {Stringari}},\ }\href@noop {} {\emph {\bibinfo {title} {{Bose-Einstein
  Condensation and Superfluidity}}}}\ (\bibinfo  {publisher} {Oxford University
  Press},\ \bibinfo {year} {2016})\BibitemShut {NoStop}%
\bibitem [{\citenamefont {Picozzi}\ and\ \citenamefont
  {Garnier}(2011)}]{Picozzi2011}%
  \BibitemOpen
  \bibfield  {author} {\bibinfo {author} {\bibfnamefont {A.}~\bibnamefont
  {Picozzi}}\ and\ \bibinfo {author} {\bibfnamefont {J.}~\bibnamefont
  {Garnier}},\ }\bibfield  {title} {\bibinfo {title} {{Incoherent Soliton
  Turbulence in Nonlocal Nonlinear Media}},\ }\bibfield  {journal} {\bibinfo
  {journal} {Physical Review Letters}\ }\textbf {\bibinfo {volume} {107}},\
  \href {https://doi.org/10.1103/physrevlett.107.233901}
  {10.1103/physrevlett.107.233901} (\bibinfo {year} {2011})\BibitemShut
  {NoStop}%
\bibitem [{\citenamefont {Bekenstein}\ \emph {et~al.}(2015)\citenamefont
  {Bekenstein}, \citenamefont {Schley}, \citenamefont {Mutzafi}, \citenamefont
  {Rotschild},\ and\ \citenamefont {Segev}}]{Segev2015}%
  \BibitemOpen
  \bibfield  {author} {\bibinfo {author} {\bibfnamefont {R.}~\bibnamefont
  {Bekenstein}}, \bibinfo {author} {\bibfnamefont {R.}~\bibnamefont {Schley}},
  \bibinfo {author} {\bibfnamefont {M.}~\bibnamefont {Mutzafi}}, \bibinfo
  {author} {\bibfnamefont {C.}~\bibnamefont {Rotschild}},\ and\ \bibinfo
  {author} {\bibfnamefont {M.}~\bibnamefont {Segev}},\ }\bibfield  {title}
  {\bibinfo {title} {{Optical simulations of gravitational effects in the
  Newton--Schr{\"o}dinger system}},\ }\href {https://doi.org/10.1038/nphys3451}
  {\bibfield  {journal} {\bibinfo  {journal} {Nature Physics}\ }\textbf
  {\bibinfo {volume} {11}},\ \bibinfo {pages} {872} (\bibinfo {year}
  {2015})}\BibitemShut {NoStop}%
\bibitem [{\citenamefont {Roger}\ \emph {et~al.}(2016)\citenamefont {Roger},
  \citenamefont {Maitland}, \citenamefont {Wilson}, \citenamefont {Westerberg},
  \citenamefont {Vocke}, \citenamefont {Wright},\ and\ \citenamefont
  {Faccio}}]{Roger2016}%
  \BibitemOpen
  \bibfield  {author} {\bibinfo {author} {\bibfnamefont {T.}~\bibnamefont
  {Roger}}, \bibinfo {author} {\bibfnamefont {C.}~\bibnamefont {Maitland}},
  \bibinfo {author} {\bibfnamefont {K.}~\bibnamefont {Wilson}}, \bibinfo
  {author} {\bibfnamefont {N.}~\bibnamefont {Westerberg}}, \bibinfo {author}
  {\bibfnamefont {D.}~\bibnamefont {Vocke}}, \bibinfo {author} {\bibfnamefont
  {E.~M.}\ \bibnamefont {Wright}},\ and\ \bibinfo {author} {\bibfnamefont
  {D.}~\bibnamefont {Faccio}},\ }\bibfield  {title} {\bibinfo {title} {{Optical
  analogues of the Newton--Schr{\"o}dinger equation and boson star
  evolution}},\ }\href {https://doi.org/10.1038/ncomms13492} {\bibfield
  {journal} {\bibinfo  {journal} {Nature Communications}\ }\textbf {\bibinfo
  {volume} {7}},\ \bibinfo {pages} {13492} (\bibinfo {year}
  {2016})}\BibitemShut {NoStop}%
\bibitem [{\citenamefont {Navarrete}\ \emph {et~al.}(2017)\citenamefont
  {Navarrete}, \citenamefont {Paredes}, \citenamefont {Salgueiro},\ and\
  \citenamefont {Michinel}}]{Navarrete2017}%
  \BibitemOpen
  \bibfield  {author} {\bibinfo {author} {\bibfnamefont {A.}~\bibnamefont
  {Navarrete}}, \bibinfo {author} {\bibfnamefont {A.}~\bibnamefont {Paredes}},
  \bibinfo {author} {\bibfnamefont {J.~R.}\ \bibnamefont {Salgueiro}},\ and\
  \bibinfo {author} {\bibfnamefont {H.}~\bibnamefont {Michinel}},\ }\bibfield
  {title} {\bibinfo {title} {{Spatial solitons in thermo-optical media from the
  nonlinear Schr{\"o}dinger-Poisson equation and dark-matter analogs}},\
  }\bibfield  {journal} {\bibinfo  {journal} {Physical Review A}\ }\textbf
  {\bibinfo {volume} {95}},\ \href {https://doi.org/10.1103/physreva.95.013844}
  {10.1103/physreva.95.013844} (\bibinfo {year} {2017})\BibitemShut {NoStop}%
\bibitem [{\citenamefont {Di{\'{o}}si}(1984)}]{Diosi1984}%
  \BibitemOpen
  \bibfield  {author} {\bibinfo {author} {\bibfnamefont {L.}~\bibnamefont
  {Di{\'{o}}si}},\ }\bibfield  {title} {\bibinfo {title} {{Gravitation and
  quantum-mechanical localization of macro-objects}},\ }\href
  {https://doi.org/10.1016/0375-9601(84)90397-9} {\bibfield  {journal}
  {\bibinfo  {journal} {Physics Letters A}\ }\textbf {\bibinfo {volume}
  {105}},\ \bibinfo {pages} {199} (\bibinfo {year} {1984})}\BibitemShut
  {NoStop}%
\bibitem [{\citenamefont {Ruffini}\ and\ \citenamefont
  {Bonazzola}(1969)}]{Ruffini1969}%
  \BibitemOpen
  \bibfield  {author} {\bibinfo {author} {\bibfnamefont {R.}~\bibnamefont
  {Ruffini}}\ and\ \bibinfo {author} {\bibfnamefont {S.}~\bibnamefont
  {Bonazzola}},\ }\bibfield  {title} {\bibinfo {title} {{Systems of
  Self-Gravitating Particles in General Relativity and the Concept of an
  Equation of State}},\ }\href {https://doi.org/10.1103/PhysRev.187.1767}
  {\bibfield  {journal} {\bibinfo  {journal} {Phys. Rev.}\ }\textbf {\bibinfo
  {volume} {187}},\ \bibinfo {pages} {1767} (\bibinfo {year}
  {1969})}\BibitemShut {NoStop}%
\bibitem [{\citenamefont {Paredes~Galan}\ \emph {et~al.}(2019)\citenamefont
  {Paredes~Galan}, \citenamefont {Olivieri},\ and\ \citenamefont
  {Michinel}}]{Paredes2020}%
  \BibitemOpen
  \bibfield  {author} {\bibinfo {author} {\bibfnamefont {A.}~\bibnamefont
  {Paredes~Galan}}, \bibinfo {author} {\bibfnamefont {D.}~\bibnamefont
  {Olivieri}},\ and\ \bibinfo {author} {\bibfnamefont {H.}~\bibnamefont
  {Michinel}},\ }\bibfield  {title} {\bibinfo {title} {{From optics to dark
  matter: A review on nonlinear Schrödinger–Poisson systems}},\ }\href
  {https://doi.org/10.1016/j.physd.2019.132301} {\bibfield  {journal} {\bibinfo
   {journal} {Physica D: Nonlinear Phenomena}\ }\textbf {\bibinfo {volume}
  {403}},\ \bibinfo {pages} {132301} (\bibinfo {year} {2019})}\BibitemShut
  {NoStop}%
\bibitem [{\citenamefont {Hu}\ \emph {et~al.}(2000)\citenamefont {Hu},
  \citenamefont {Barkana},\ and\ \citenamefont {Gruzinov}}]{Hu2000}%
  \BibitemOpen
  \bibfield  {author} {\bibinfo {author} {\bibfnamefont {W.}~\bibnamefont
  {Hu}}, \bibinfo {author} {\bibfnamefont {R.}~\bibnamefont {Barkana}},\ and\
  \bibinfo {author} {\bibfnamefont {A.}~\bibnamefont {Gruzinov}},\ }\bibfield
  {title} {\bibinfo {title} {{Fuzzy Cold Dark Matter: The Wave Properties of
  Ultralight Particles}},\ }\href {https://doi.org/10.1103/physrevlett.85.1158}
  {\bibfield  {journal} {\bibinfo  {journal} {Physical Review Letters}\
  }\textbf {\bibinfo {volume} {85}},\ \bibinfo {pages} {1158} (\bibinfo {year}
  {2000})}\BibitemShut {NoStop}%
\bibitem [{\citenamefont {Schive}\ \emph {et~al.}(2014)\citenamefont {Schive},
  \citenamefont {Chiueh},\ and\ \citenamefont {Broadhurst}}]{Schive2014}%
  \BibitemOpen
  \bibfield  {author} {\bibinfo {author} {\bibfnamefont {H.-Y.}\ \bibnamefont
  {Schive}}, \bibinfo {author} {\bibfnamefont {T.}~\bibnamefont {Chiueh}},\
  and\ \bibinfo {author} {\bibfnamefont {T.}~\bibnamefont {Broadhurst}},\
  }\bibfield  {title} {\bibinfo {title} {{Cosmic structure as the quantum
  interference of a coherent dark wave}},\ }\href
  {https://doi.org/10.1038/nphys2996} {\bibfield  {journal} {\bibinfo
  {journal} {Nature Physics}\ }\textbf {\bibinfo {volume} {10}},\ \bibinfo
  {pages} {496} (\bibinfo {year} {2014})}\BibitemShut {NoStop}%
\bibitem [{\citenamefont {Bullock}\ and\ \citenamefont
  {Boylan-Kolchin}(2017)}]{Bullock2017}%
  \BibitemOpen
  \bibfield  {author} {\bibinfo {author} {\bibfnamefont {J.~S.}\ \bibnamefont
  {Bullock}}\ and\ \bibinfo {author} {\bibfnamefont {M.}~\bibnamefont
  {Boylan-Kolchin}},\ }\bibfield  {title} {\bibinfo {title} {{Small-Scale
  Challenges to the {$\Lambda$}CDM Paradigm}},\ }\bibfield  {journal} {\bibinfo
   {journal} {Annual Review of Astronomy and Astrophysics}\ }\href
  {https://doi.org/10.1146/annurev-astro-091916-055313}
  {10.1146/annurev-astro-091916-055313} (\bibinfo {year} {2017})\BibitemShut
  {NoStop}%
\bibitem [{\citenamefont {Moore}(1994)}]{Moore1994}%
  \BibitemOpen
  \bibfield  {author} {\bibinfo {author} {\bibfnamefont {B.}~\bibnamefont
  {Moore}},\ }\bibfield  {title} {\bibinfo {title} {{Evidence against
  dissipation-less dark matter from observations of galaxy haloes}},\ }\href
  {https://doi.org/10.1038/370629a0} {\bibfield  {journal} {\bibinfo  {journal}
  {Nature}\ }\textbf {\bibinfo {volume} {370}},\ \bibinfo {pages} {629}
  (\bibinfo {year} {1994})}\BibitemShut {NoStop}%
\bibitem [{\citenamefont {de~Blok}(2010)}]{deBlok2010}%
  \BibitemOpen
  \bibfield  {author} {\bibinfo {author} {\bibfnamefont {W.~J.~G.}\
  \bibnamefont {de~Blok}},\ }\bibfield  {title} {\bibinfo {title} {{The
  Core-Cusp Problem}},\ }\href@noop {} {\bibfield  {journal} {\bibinfo
  {journal} {Advances in Astronomy}\ }\textbf {\bibinfo {volume} {2010}},\
  \bibinfo {pages} {789293} (\bibinfo {year} {2010})}\BibitemShut {NoStop}%
\bibitem [{\citenamefont {Mocz}\ \emph {et~al.}(2017)\citenamefont {Mocz},
  \citenamefont {Vogelsberger}, \citenamefont {Robles}, \citenamefont {Zavala},
  \citenamefont {Boylan-Kolchin}, \citenamefont {Fialkov},\ and\ \citenamefont
  {Hernquist}}]{Mocz2017}%
  \BibitemOpen
  \bibfield  {author} {\bibinfo {author} {\bibfnamefont {P.}~\bibnamefont
  {Mocz}}, \bibinfo {author} {\bibfnamefont {M.}~\bibnamefont {Vogelsberger}},
  \bibinfo {author} {\bibfnamefont {V.~H.}\ \bibnamefont {Robles}}, \bibinfo
  {author} {\bibfnamefont {J.}~\bibnamefont {Zavala}}, \bibinfo {author}
  {\bibfnamefont {M.}~\bibnamefont {Boylan-Kolchin}}, \bibinfo {author}
  {\bibfnamefont {A.}~\bibnamefont {Fialkov}},\ and\ \bibinfo {author}
  {\bibfnamefont {L.}~\bibnamefont {Hernquist}},\ }\bibfield  {title} {\bibinfo
  {title} {{Galaxy formation with BECDM {\textendash} I. Turbulence and
  relaxation of idealized haloes}},\ }\href
  {https://doi.org/10.1093/mnras/stx1887} {\bibfield  {journal} {\bibinfo
  {journal} {Monthly Notices of the Royal Astronomical Society}\ }\textbf
  {\bibinfo {volume} {471}},\ \bibinfo {pages} {4559} (\bibinfo {year}
  {2017})}\BibitemShut {NoStop}%
\bibitem [{\citenamefont {Porayko}\ \emph {et~al.}(2018)\citenamefont
  {Porayko}, \citenamefont {Zhu}, \citenamefont {Levin}, \citenamefont {Hui},
  \citenamefont {Hobbs}, \citenamefont {Grudskaya}, \citenamefont {Postnov},
  \citenamefont {Bailes}, \citenamefont {Bhat}, \citenamefont {Coles},
  \citenamefont {Dai}, \citenamefont {Dempsey}, \citenamefont {Keith},
  \citenamefont {Kerr}, \citenamefont {Kramer}, \citenamefont {Lasky},
  \citenamefont {Manchester}, \citenamefont {Os\l{}owski}, \citenamefont
  {Parthasarathy}, \citenamefont {Ravi}, \citenamefont {Reardon}, \citenamefont
  {Rosado}, \citenamefont {Russell}, \citenamefont {Shannon}, \citenamefont
  {Spiewak}, \citenamefont {van Straten}, \citenamefont {Toomey}, \citenamefont
  {Wang}, \citenamefont {Wen},\ and\ \citenamefont {You}}]{Porayko2018}%
  \BibitemOpen
  \bibfield  {author} {\bibinfo {author} {\bibfnamefont {N.~K.}\ \bibnamefont
  {Porayko}}, \bibinfo {author} {\bibfnamefont {X.}~\bibnamefont {Zhu}},
  \bibinfo {author} {\bibfnamefont {Y.}~\bibnamefont {Levin}}, \bibinfo
  {author} {\bibfnamefont {L.}~\bibnamefont {Hui}}, \bibinfo {author}
  {\bibfnamefont {G.}~\bibnamefont {Hobbs}}, \bibinfo {author} {\bibfnamefont
  {A.}~\bibnamefont {Grudskaya}}, \bibinfo {author} {\bibfnamefont
  {K.}~\bibnamefont {Postnov}}, \bibinfo {author} {\bibfnamefont
  {M.}~\bibnamefont {Bailes}}, \bibinfo {author} {\bibfnamefont {N.~D.~R.}\
  \bibnamefont {Bhat}}, \bibinfo {author} {\bibfnamefont {W.}~\bibnamefont
  {Coles}}, \bibinfo {author} {\bibfnamefont {S.}~\bibnamefont {Dai}}, \bibinfo
  {author} {\bibfnamefont {J.}~\bibnamefont {Dempsey}}, \bibinfo {author}
  {\bibfnamefont {M.~J.}\ \bibnamefont {Keith}}, \bibinfo {author}
  {\bibfnamefont {M.}~\bibnamefont {Kerr}}, \bibinfo {author} {\bibfnamefont
  {M.}~\bibnamefont {Kramer}}, \bibinfo {author} {\bibfnamefont {P.~D.}\
  \bibnamefont {Lasky}}, \bibinfo {author} {\bibfnamefont {R.~N.}\ \bibnamefont
  {Manchester}}, \bibinfo {author} {\bibfnamefont {S.}~\bibnamefont
  {Os\l{}owski}}, \bibinfo {author} {\bibfnamefont {A.}~\bibnamefont
  {Parthasarathy}}, \bibinfo {author} {\bibfnamefont {V.}~\bibnamefont {Ravi}},
  \bibinfo {author} {\bibfnamefont {D.~J.}\ \bibnamefont {Reardon}}, \bibinfo
  {author} {\bibfnamefont {P.~A.}\ \bibnamefont {Rosado}}, \bibinfo {author}
  {\bibfnamefont {C.~J.}\ \bibnamefont {Russell}}, \bibinfo {author}
  {\bibfnamefont {R.~M.}\ \bibnamefont {Shannon}}, \bibinfo {author}
  {\bibfnamefont {R.}~\bibnamefont {Spiewak}}, \bibinfo {author} {\bibfnamefont
  {W.}~\bibnamefont {van Straten}}, \bibinfo {author} {\bibfnamefont
  {L.}~\bibnamefont {Toomey}}, \bibinfo {author} {\bibfnamefont
  {J.}~\bibnamefont {Wang}}, \bibinfo {author} {\bibfnamefont {L.}~\bibnamefont
  {Wen}},\ and\ \bibinfo {author} {\bibfnamefont {X.}~\bibnamefont {You}}
  (\bibinfo {collaboration} {PPTA Collaboration}),\ }\bibfield  {title}
  {\bibinfo {title} {{Parkes Pulsar Timing Array constraints on ultralight
  scalar-field dark matter}},\ }\href
  {https://doi.org/10.1103/PhysRevD.98.102002} {\bibfield  {journal} {\bibinfo
  {journal} {Phys. Rev. D}\ }\textbf {\bibinfo {volume} {98}},\ \bibinfo
  {pages} {102002} (\bibinfo {year} {2018})}\BibitemShut {NoStop}%
\bibitem [{\citenamefont {Amorisco}\ and\ \citenamefont
  {Loeb}(2018)}]{Amorisco2018}%
  \BibitemOpen
  \bibfield  {author} {\bibinfo {author} {\bibfnamefont {N.~C.}\ \bibnamefont
  {Amorisco}}\ and\ \bibinfo {author} {\bibfnamefont {A.}~\bibnamefont
  {Loeb}},\ }\href@noop {} {\bibinfo {title} {{First constraints on Fuzzy Dark
  Matter from the dynamics of stellar streams in the Milky Way}}} (\bibinfo
  {year} {2018}),\ \Eprint {https://arxiv.org/abs/1808.00464} {arXiv:1808.00464
  [astro-ph.GA]} \BibitemShut {NoStop}%
\bibitem [{\citenamefont {Lidz}\ and\ \citenamefont {Hui}(2018)}]{Lidz2018}%
  \BibitemOpen
  \bibfield  {author} {\bibinfo {author} {\bibfnamefont {A.}~\bibnamefont
  {Lidz}}\ and\ \bibinfo {author} {\bibfnamefont {L.}~\bibnamefont {Hui}},\
  }\bibfield  {title} {\bibinfo {title} {{The Implications of a
  Pre-reionization 21 cm Absorption Signal for Fuzzy Dark Matter}},\ }\href
  {https://doi.org/10.1103/PhysRevD.98.023011} {\bibfield  {journal} {\bibinfo
  {journal} {Phys. Rev. D}\ }\textbf {\bibinfo {volume} {98}},\ \bibinfo
  {pages} {023011} (\bibinfo {year} {2018})}\BibitemShut {NoStop}%
\bibitem [{\citenamefont {Niemeyer}(2020)}]{Niemeyer2020}%
  \BibitemOpen
  \bibfield  {author} {\bibinfo {author} {\bibfnamefont {J.~C.}\ \bibnamefont
  {Niemeyer}},\ }\bibfield  {title} {\bibinfo {title} {{Small-scale structure
  of fuzzy and axion-like dark matter}},\ }\href
  {https://doi.org/10.1016/j.ppnp.2020.103787} {\bibfield  {journal} {\bibinfo
  {journal} {Progress in Particle and Nuclear Physics}\ }\textbf {\bibinfo
  {volume} {113}},\ \bibinfo {pages} {103787} (\bibinfo {year}
  {2020})}\BibitemShut {NoStop}%
\bibitem [{\citenamefont {Widrow}\ and\ \citenamefont
  {Kaiser}(1993)}]{Widrow1993}%
  \BibitemOpen
  \bibfield  {author} {\bibinfo {author} {\bibfnamefont {L.~M.}\ \bibnamefont
  {Widrow}}\ and\ \bibinfo {author} {\bibfnamefont {N.}~\bibnamefont
  {Kaiser}},\ }\bibfield  {title} {\bibinfo {title} {{Using the Schroedinger
  Equation to Simulate Collisionless Matter}},\ }\href
  {https://doi.org/10.1086/187073} {\bibfield  {journal} {\bibinfo  {journal}
  {The Astrophysical Journal}\ }\textbf {\bibinfo {volume} {416}},\ \bibinfo
  {pages} {L71} (\bibinfo {year} {1993})}\BibitemShut {NoStop}%
\bibitem [{\citenamefont {Uhlemann}\ \emph {et~al.}(2014)\citenamefont
  {Uhlemann}, \citenamefont {Kopp},\ and\ \citenamefont
  {Haugg}}]{Uhlemann2014}%
  \BibitemOpen
  \bibfield  {author} {\bibinfo {author} {\bibfnamefont {C.}~\bibnamefont
  {Uhlemann}}, \bibinfo {author} {\bibfnamefont {M.}~\bibnamefont {Kopp}},\
  and\ \bibinfo {author} {\bibfnamefont {T.}~\bibnamefont {Haugg}},\ }\bibfield
   {title} {\bibinfo {title} {{Schr{\"o}dinger method as {N}-body double and UV
  completion of dust}},\ }\bibfield  {journal} {\bibinfo  {journal} {Physical
  Review D}\ }\textbf {\bibinfo {volume} {90}},\ \href
  {https://doi.org/10.1103/physrevd.90.023517} {10.1103/physrevd.90.023517}
  (\bibinfo {year} {2014})\BibitemShut {NoStop}%
\bibitem [{\citenamefont {Kopp}\ \emph {et~al.}(2017)\citenamefont {Kopp},
  \citenamefont {Vattis},\ and\ \citenamefont {Skordis}}]{Kopp2017}%
  \BibitemOpen
  \bibfield  {author} {\bibinfo {author} {\bibfnamefont {M.}~\bibnamefont
  {Kopp}}, \bibinfo {author} {\bibfnamefont {K.}~\bibnamefont {Vattis}},\ and\
  \bibinfo {author} {\bibfnamefont {C.}~\bibnamefont {Skordis}},\ }\bibfield
  {title} {\bibinfo {title} {{Solving the Vlasov equation in two spatial
  dimensions with the Schr{\"o}dinger method}},\ }\bibfield  {journal}
  {\bibinfo  {journal} {Physical Review D}\ }\textbf {\bibinfo {volume} {96}},\
  \href {https://doi.org/10.1103/physrevd.96.123532}
  {10.1103/physrevd.96.123532} (\bibinfo {year} {2017})\BibitemShut {NoStop}%
\bibitem [{\citenamefont {Mocz}\ \emph {et~al.}(2018)\citenamefont {Mocz},
  \citenamefont {Lancaster}, \citenamefont {Fialkov}, \citenamefont {Becerra},\
  and\ \citenamefont {Chavanis}}]{Mocz2018}%
  \BibitemOpen
  \bibfield  {author} {\bibinfo {author} {\bibfnamefont {P.}~\bibnamefont
  {Mocz}}, \bibinfo {author} {\bibfnamefont {L.}~\bibnamefont {Lancaster}},
  \bibinfo {author} {\bibfnamefont {A.}~\bibnamefont {Fialkov}}, \bibinfo
  {author} {\bibfnamefont {F.}~\bibnamefont {Becerra}},\ and\ \bibinfo {author}
  {\bibfnamefont {P.-H.}\ \bibnamefont {Chavanis}},\ }\bibfield  {title}
  {\bibinfo {title} {{Schrödinger-Poisson{\textendash}Vlasov-Poisson
  correspondence}},\ }\bibfield  {journal} {\bibinfo  {journal} {Physical
  Review D}\ }\textbf {\bibinfo {volume} {97}},\ \href
  {https://doi.org/10.1103/physrevd.97.083519} {10.1103/physrevd.97.083519}
  (\bibinfo {year} {2018})\BibitemShut {NoStop}%
\bibitem [{\citenamefont {Eberhardt}\ \emph {et~al.}(2020)\citenamefont
  {Eberhardt}, \citenamefont {Banerjee}, \citenamefont {Kopp},\ and\
  \citenamefont {Abel}}]{Eberhardt2020}%
  \BibitemOpen
  \bibfield  {author} {\bibinfo {author} {\bibfnamefont {A.}~\bibnamefont
  {Eberhardt}}, \bibinfo {author} {\bibfnamefont {A.}~\bibnamefont {Banerjee}},
  \bibinfo {author} {\bibfnamefont {M.}~\bibnamefont {Kopp}},\ and\ \bibinfo
  {author} {\bibfnamefont {T.}~\bibnamefont {Abel}},\ }\bibfield  {title}
  {\bibinfo {title} {{Investigating the use of field solvers for simulating
  classical systems}},\ }\bibfield  {journal} {\bibinfo  {journal} {Physical
  Review D}\ }\textbf {\bibinfo {volume} {101}},\ \href
  {https://doi.org/10.1103/physrevd.101.043011} {10.1103/physrevd.101.043011}
  (\bibinfo {year} {2020})\BibitemShut {NoStop}%
\bibitem [{\citenamefont {Guzm{\'{a}}n}\ and\ \citenamefont
  {Ure{\~{n}}a-L{\'{o}}pez}(2004)}]{Guzman2004}%
  \BibitemOpen
  \bibfield  {author} {\bibinfo {author} {\bibfnamefont {F.~S.}\ \bibnamefont
  {Guzm{\'{a}}n}}\ and\ \bibinfo {author} {\bibfnamefont {L.~A.}\ \bibnamefont
  {Ure{\~{n}}a-L{\'{o}}pez}},\ }\bibfield  {title} {\bibinfo {title}
  {{Evolution of the Schr{\"o}dinger-Newton system for a self-gravitating
  scalar field}},\ }\bibfield  {journal} {\bibinfo  {journal} {Physical Review
  D}\ }\textbf {\bibinfo {volume} {69}},\ \href
  {https://doi.org/10.1103/physrevd.69.124033} {10.1103/physrevd.69.124033}
  (\bibinfo {year} {2004})\BibitemShut {NoStop}%
\bibitem [{\citenamefont {Guzman}\ and\ \citenamefont
  {Urena-Lopez}(2006)}]{Guzman2006}%
  \BibitemOpen
  \bibfield  {author} {\bibinfo {author} {\bibfnamefont {F.~S.}\ \bibnamefont
  {Guzman}}\ and\ \bibinfo {author} {\bibfnamefont {L.~A.}\ \bibnamefont
  {Urena-Lopez}},\ }\bibfield  {title} {\bibinfo {title} {{Gravitational
  Cooling of Self-gravitating Bose Condensates}},\ }\href
  {https://doi.org/10.1086/504508} {\bibfield  {journal} {\bibinfo  {journal}
  {The Astrophysical Journal}\ }\textbf {\bibinfo {volume} {645}},\ \bibinfo
  {pages} {814} (\bibinfo {year} {2006})}\BibitemShut {NoStop}%
\bibitem [{\citenamefont {Schwabe}\ \emph {et~al.}(2016)\citenamefont
  {Schwabe}, \citenamefont {Niemeyer},\ and\ \citenamefont
  {Engels}}]{Schwabe2016}%
  \BibitemOpen
  \bibfield  {author} {\bibinfo {author} {\bibfnamefont {B.}~\bibnamefont
  {Schwabe}}, \bibinfo {author} {\bibfnamefont {J.~C.}\ \bibnamefont
  {Niemeyer}},\ and\ \bibinfo {author} {\bibfnamefont {J.~F.}\ \bibnamefont
  {Engels}},\ }\bibfield  {title} {\bibinfo {title} {{Simulations of solitonic
  core mergers in ultralight axion dark matter cosmologies}},\ }\bibfield
  {journal} {\bibinfo  {journal} {Physical Review D}\ }\textbf {\bibinfo
  {volume} {94}},\ \href {https://doi.org/10.1103/physrevd.94.043513}
  {10.1103/physrevd.94.043513} (\bibinfo {year} {2016})\BibitemShut {NoStop}%
\bibitem [{\citenamefont {Seidel}\ and\ \citenamefont
  {Suen}(1994)}]{Seidel1994}%
  \BibitemOpen
  \bibfield  {author} {\bibinfo {author} {\bibfnamefont {E.}~\bibnamefont
  {Seidel}}\ and\ \bibinfo {author} {\bibfnamefont {W.-M.}\ \bibnamefont
  {Suen}},\ }\bibfield  {title} {\bibinfo {title} {{Formation of solitonic
  stars through gravitational cooling}},\ }\href
  {https://doi.org/10.1103/physrevlett.72.2516} {\bibfield  {journal} {\bibinfo
   {journal} {Physical Review Letters}\ }\textbf {\bibinfo {volume} {72}},\
  \bibinfo {pages} {2516} (\bibinfo {year} {1994})}\BibitemShut {NoStop}%
\bibitem [{\citenamefont {Navarro}\ \emph {et~al.}(1996)\citenamefont
  {Navarro}, \citenamefont {Frenk},\ and\ \citenamefont {White}}]{Navarro1996}%
  \BibitemOpen
  \bibfield  {author} {\bibinfo {author} {\bibfnamefont {J.~F.}\ \bibnamefont
  {Navarro}}, \bibinfo {author} {\bibfnamefont {C.~S.}\ \bibnamefont {Frenk}},\
  and\ \bibinfo {author} {\bibfnamefont {S.~D.~M.}\ \bibnamefont {White}},\
  }\bibfield  {title} {\bibinfo {title} {{The Structure of Cold Dark Matter
  Halos}},\ }\href {https://doi.org/10.1086/177173} {\bibfield  {journal}
  {\bibinfo  {journal} {The Astrophysical Journal}\ }\textbf {\bibinfo {volume}
  {462}},\ \bibinfo {pages} {563} (\bibinfo {year} {1996})}\BibitemShut
  {NoStop}%
\bibitem [{\citenamefont {Modugno}\ \emph {et~al.}(2004)\citenamefont
  {Modugno}, \citenamefont {Tozzo},\ and\ \citenamefont
  {Dalfovo}}]{Modugno2004}%
  \BibitemOpen
  \bibfield  {author} {\bibinfo {author} {\bibfnamefont {M.}~\bibnamefont
  {Modugno}}, \bibinfo {author} {\bibfnamefont {C.}~\bibnamefont {Tozzo}},\
  and\ \bibinfo {author} {\bibfnamefont {F.}~\bibnamefont {Dalfovo}},\
  }\bibfield  {title} {\bibinfo {title} {{Role of transverse excitations in the
  instability of Bose-Einstein condensates moving in optical lattices}},\
  }\bibfield  {journal} {\bibinfo  {journal} {Physical Review A}\ }\textbf
  {\bibinfo {volume} {70}},\ \href {https://doi.org/10.1103/physreva.70.043625}
  {10.1103/physreva.70.043625} (\bibinfo {year} {2004})\BibitemShut {NoStop}%
\bibitem [{\citenamefont {Chavanis}(2012)}]{Chavanis2012}%
  \BibitemOpen
  \bibfield  {author} {\bibinfo {author} {\bibfnamefont {P.~H.}\ \bibnamefont
  {Chavanis}},\ }\bibfield  {title} {\bibinfo {title} {{Growth of perturbations
  in an expanding universe with Bose-Einstein condensate dark matter}},\ }\href
  {https://doi.org/10.1051/0004-6361/201116905} {\bibfield  {journal} {\bibinfo
   {journal} {Astronomy {\&} Astrophysics}\ }\textbf {\bibinfo {volume}
  {537}},\ \bibinfo {pages} {A127} (\bibinfo {year} {2012})}\BibitemShut
  {NoStop}%
\bibitem [{\citenamefont {Woo}\ and\ \citenamefont {Chiueh}(2009)}]{Woo2009}%
  \BibitemOpen
  \bibfield  {author} {\bibinfo {author} {\bibfnamefont {T.-P.}\ \bibnamefont
  {Woo}}\ and\ \bibinfo {author} {\bibfnamefont {T.}~\bibnamefont {Chiueh}},\
  }\bibfield  {title} {\bibinfo {title} {{High-Resolution Simulation on
  Structure Formation with Extremely Light Bosonic Dark Matter}},\ }\href
  {https://doi.org/10.1088/0004-637x/697/1/850} {\bibfield  {journal} {\bibinfo
   {journal} {The Astrophysical Journal}\ }\textbf {\bibinfo {volume} {697}},\
  \bibinfo {pages} {850} (\bibinfo {year} {2009})}\BibitemShut {NoStop}%
\bibitem [{\citenamefont {Li}\ \emph {et~al.}(2019)\citenamefont {Li},
  \citenamefont {Hui},\ and\ \citenamefont {Bryan}}]{Li2019}%
  \BibitemOpen
  \bibfield  {author} {\bibinfo {author} {\bibfnamefont {X.}~\bibnamefont
  {Li}}, \bibinfo {author} {\bibfnamefont {L.}~\bibnamefont {Hui}},\ and\
  \bibinfo {author} {\bibfnamefont {G.~L.}\ \bibnamefont {Bryan}},\ }\bibfield
  {title} {\bibinfo {title} {{Numerical and perturbative computations of the
  fuzzy dark matter model}},\ }\bibfield  {journal} {\bibinfo  {journal}
  {Physical Review D}\ }\textbf {\bibinfo {volume} {99}},\ \href
  {https://doi.org/10.1103/physrevd.99.063509} {10.1103/physrevd.99.063509}
  (\bibinfo {year} {2019})\BibitemShut {NoStop}%
\bibitem [{\citenamefont {Madelung}(1927)}]{Madelung1927}%
  \BibitemOpen
  \bibfield  {author} {\bibinfo {author} {\bibfnamefont {E.}~\bibnamefont
  {Madelung}},\ }\bibfield  {title} {\bibinfo {title} {{Quantentheorie in
  hydrodynamischer Form}},\ }\href {https://doi.org/10.1007/bf01400372}
  {\bibfield  {journal} {\bibinfo  {journal} {Zeitschrift f\"ur Physik}\
  }\textbf {\bibinfo {volume} {40}},\ \bibinfo {pages} {322} (\bibinfo {year}
  {1927})}\BibitemShut {NoStop}%
\bibitem [{\citenamefont {Kolmogorov}(1991)}]{Kolmogorov91}%
  \BibitemOpen
  \bibfield  {author} {\bibinfo {author} {\bibfnamefont {A.~N.}\ \bibnamefont
  {Kolmogorov}},\ }\bibfield  {title} {\bibinfo {title} {{The local structure
  of turbulence in incompressible viscous fluid for very large Reynolds
  numbers}},\ }\href {https://doi.org/10.1098/rspa.1991.0075} {\bibfield
  {journal} {\bibinfo  {journal} {Proceedings of the Royal Society of London.
  Series A: Mathematical and Physical Sciences}\ }\textbf {\bibinfo {volume}
  {434}},\ \bibinfo {pages} {9} (\bibinfo {year} {1991})}\BibitemShut {NoStop}%
\bibitem [{\citenamefont {Kobayashi}\ and\ \citenamefont
  {Tsubota}(2005)}]{Kobayashi2005}%
  \BibitemOpen
  \bibfield  {author} {\bibinfo {author} {\bibfnamefont {M.}~\bibnamefont
  {Kobayashi}}\ and\ \bibinfo {author} {\bibfnamefont {M.}~\bibnamefont
  {Tsubota}},\ }\bibfield  {title} {\bibinfo {title} {{Kolmogorov Spectrum of
  Superfluid Turbulence: Numerical Analysis of the Gross-Pitaevskii Equation
  with a Small-Scale Dissipation}},\ }\bibfield  {journal} {\bibinfo  {journal}
  {Physical Review Letters}\ }\textbf {\bibinfo {volume} {94}},\ \href
  {https://doi.org/10.1103/physrevlett.94.065302}
  {10.1103/physrevlett.94.065302} (\bibinfo {year} {2005})\BibitemShut
  {NoStop}%
\bibitem [{\citenamefont {Baggaley}\ \emph {et~al.}(2012)\citenamefont
  {Baggaley}, \citenamefont {Laurie},\ and\ \citenamefont
  {Barenghi}}]{Baggaley2012}%
  \BibitemOpen
  \bibfield  {author} {\bibinfo {author} {\bibfnamefont {A.~W.}\ \bibnamefont
  {Baggaley}}, \bibinfo {author} {\bibfnamefont {J.}~\bibnamefont {Laurie}},\
  and\ \bibinfo {author} {\bibfnamefont {C.~F.}\ \bibnamefont {Barenghi}},\
  }\bibfield  {title} {\bibinfo {title} {Vortex-density fluctuations, energy
  spectra, and vortical regions in superfluid turbulence},\ }\bibfield
  {journal} {\bibinfo  {journal} {Physical Review Letters}\ }\textbf {\bibinfo
  {volume} {109}},\ \href {https://doi.org/10.1103/physrevlett.109.205304}
  {10.1103/physrevlett.109.205304} (\bibinfo {year} {2012})\BibitemShut
  {NoStop}%
\bibitem [{\citenamefont {Wallstrom}(1994)}]{Wallstrom1994}%
  \BibitemOpen
  \bibfield  {author} {\bibinfo {author} {\bibfnamefont {T.~C.}\ \bibnamefont
  {Wallstrom}},\ }\bibfield  {title} {\bibinfo {title} {{Inequivalence between
  the Schrödinger equation and the Madelung hydrodynamic equations}},\ }\href
  {https://doi.org/10.1103/physreva.49.1613} {\bibfield  {journal} {\bibinfo
  {journal} {Physical Review A}\ }\textbf {\bibinfo {volume} {49}},\ \bibinfo
  {pages} {1613} (\bibinfo {year} {1994})}\BibitemShut {NoStop}%
\bibitem [{\citenamefont {Kellogg}(1967)}]{Kellogg1967}%
  \BibitemOpen
  \bibfield  {author} {\bibinfo {author} {\bibfnamefont {O.~D.}\ \bibnamefont
  {Kellogg}},\ }\href {https://doi.org/10.1007/978-3-642-86748-4} {\emph
  {\bibinfo {title} {{Foundations of Potential Theory}}}}\ (\bibinfo
  {publisher} {Springer Berlin Heidelberg},\ \bibinfo {year}
  {1967})\BibitemShut {NoStop}%
\bibitem [{\citenamefont {Marshall}(2000)}]{Marshall2000}%
  \BibitemOpen
  \bibfield  {author} {\bibinfo {author} {\bibfnamefont {S.~L.}\ \bibnamefont
  {Marshall}},\ }\bibfield  {title} {\bibinfo {title} {{A periodic Green
  function for calculation of coloumbic lattice potentials}},\ }\href
  {https://doi.org/10.1088/0953-8984/12/21/304} {\bibfield  {journal} {\bibinfo
   {journal} {Journal of Physics: Condensed Matter}\ }\textbf {\bibinfo
  {volume} {12}},\ \bibinfo {pages} {4575} (\bibinfo {year}
  {2000})}\BibitemShut {NoStop}%
\bibitem [{\citenamefont {{Gradshteyn, I. S. and Ryzhik, I.
  M.}}(2014)}]{Gradshteyn2014}%
  \BibitemOpen
  \bibfield  {author} {\bibinfo {author} {\bibnamefont {{Gradshteyn, I. S. and
  Ryzhik, I. M.}}},\ }\href@noop {} {\emph {\bibinfo {title} {{Table of
  Integrals, Series, and Products}}}}\ (\bibinfo  {publisher} {Elsevier LTD,
  Oxford},\ \bibinfo {year} {2014})\BibitemShut {NoStop}%
\bibitem [{\citenamefont {Garny}\ and\ \citenamefont
  {Konstandin}(2018)}]{Garny2018}%
  \BibitemOpen
  \bibfield  {author} {\bibinfo {author} {\bibfnamefont {M.}~\bibnamefont
  {Garny}}\ and\ \bibinfo {author} {\bibfnamefont {T.}~\bibnamefont
  {Konstandin}},\ }\bibfield  {title} {\bibinfo {title} {{Gravitational
  collapse in the Schr{\"o}dinger-Poisson system}},\ }\href
  {https://doi.org/10.1088/1475-7516/2018/01/009} {\bibfield  {journal}
  {\bibinfo  {journal} {Journal of Cosmology and Astroparticle Physics}\
  }\textbf {\bibinfo {volume} {2018}}\bibinfo  {number} { (01)},\ \bibinfo
  {pages} {009}}\BibitemShut {NoStop}%
\bibitem [{\citenamefont {Zimmermann}\ \emph {et~al.}(2019)\citenamefont
  {Zimmermann}, \citenamefont {Pietroni}, \citenamefont {Madro{\~{n}}ero},
  \citenamefont {Amendola},\ and\ \citenamefont {Wimberger}}]{Zimmermann2019}%
  \BibitemOpen
\bibfield  {number} {  }\bibfield  {author} {\bibinfo {author} {\bibfnamefont
  {T.}~\bibnamefont {Zimmermann}}, \bibinfo {author} {\bibfnamefont
  {M.}~\bibnamefont {Pietroni}}, \bibinfo {author} {\bibfnamefont
  {J.}~\bibnamefont {Madro{\~{n}}ero}}, \bibinfo {author} {\bibfnamefont
  {L.}~\bibnamefont {Amendola}},\ and\ \bibinfo {author} {\bibfnamefont
  {S.}~\bibnamefont {Wimberger}},\ }\bibfield  {title} {\bibinfo {title} {{A
  Quantum Model for the Dynamics of Cold Dark Matter}},\ }\href
  {https://doi.org/10.3390/condmat4040089} {\bibfield  {journal} {\bibinfo
  {journal} {Condensed Matter}\ }\textbf {\bibinfo {volume} {4}},\ \bibinfo
  {pages} {89} (\bibinfo {year} {2019})}\BibitemShut {NoStop}%
\bibitem [{\citenamefont {Garny}\ \emph {et~al.}(2020)\citenamefont {Garny},
  \citenamefont {Konstandin},\ and\ \citenamefont {Rubira}}]{Garny2020}%
  \BibitemOpen
  \bibfield  {author} {\bibinfo {author} {\bibfnamefont {M.}~\bibnamefont
  {Garny}}, \bibinfo {author} {\bibfnamefont {T.}~\bibnamefont {Konstandin}},\
  and\ \bibinfo {author} {\bibfnamefont {H.}~\bibnamefont {Rubira}},\
  }\bibfield  {title} {\bibinfo {title} {{The Schrödinger-Poisson method for
  Large-Scale Structure}},\ }\href
  {https://doi.org/10.1088/1475-7516/2020/04/003} {\bibfield  {journal}
  {\bibinfo  {journal} {Journal of Cosmology and Astroparticle Physics}\
  }\textbf {\bibinfo {volume} {2020}}\bibinfo  {number} { (04)},\ \bibinfo
  {pages} {003}}\BibitemShut {NoStop}%
\bibitem [{\citenamefont {Bao}\ \emph {et~al.}(2013)\citenamefont {Bao},
  \citenamefont {Jian}, \citenamefont {Mauser},\ and\ \citenamefont
  {Zhang}}]{Bao2013}%
  \BibitemOpen
\bibfield  {number} {  }\bibfield  {author} {\bibinfo {author} {\bibfnamefont
  {W.}~\bibnamefont {Bao}}, \bibinfo {author} {\bibfnamefont {H.}~\bibnamefont
  {Jian}}, \bibinfo {author} {\bibfnamefont {N.~J.}\ \bibnamefont {Mauser}},\
  and\ \bibinfo {author} {\bibfnamefont {Y.}~\bibnamefont {Zhang}},\ }\bibfield
   {title} {\bibinfo {title} {{Dimension Reduction of the Schrödinger Equation
  with Coulomb and Anisotropic Confining Potentials}},\ }\href
  {https://doi.org/10.1137/13091436x} {\bibfield  {journal} {\bibinfo
  {journal} {{SIAM} Journal on Applied Mathematics}\ }\textbf {\bibinfo
  {volume} {73}},\ \bibinfo {pages} {2100} (\bibinfo {year}
  {2013})}\BibitemShut {NoStop}%
\bibitem [{\citenamefont {Bloch}\ \emph {et~al.}(2008)\citenamefont {Bloch},
  \citenamefont {Dalibard},\ and\ \citenamefont {Zwerger}}]{Bloch2008}%
  \BibitemOpen
  \bibfield  {author} {\bibinfo {author} {\bibfnamefont {I.}~\bibnamefont
  {Bloch}}, \bibinfo {author} {\bibfnamefont {J.}~\bibnamefont {Dalibard}},\
  and\ \bibinfo {author} {\bibfnamefont {W.}~\bibnamefont {Zwerger}},\
  }\bibfield  {title} {\bibinfo {title} {{Many-Body Physics with Ultracold
  Gases}},\ }\href {https://doi.org/10.1103/RevModPhys.80.885} {\bibfield
  {journal} {\bibinfo  {journal} {Rev. Mod. Phys.}\ }\textbf {\bibinfo {volume}
  {80}},\ \bibinfo {pages} {885} (\bibinfo {year} {2008})}\BibitemShut
  {NoStop}%
\bibitem [{\citenamefont {Olshanii}(1998)}]{Olshanii1998}%
  \BibitemOpen
  \bibfield  {author} {\bibinfo {author} {\bibfnamefont {M.}~\bibnamefont
  {Olshanii}},\ }\bibfield  {title} {\bibinfo {title} {{Atomic Scattering in
  the Presence of an External Confinement and a Gas of Impenetrable Bosons}},\
  }\href {https://doi.org/10.1103/PhysRevLett.81.938} {\bibfield  {journal}
  {\bibinfo  {journal} {Phys. Rev. Lett.}\ }\textbf {\bibinfo {volume} {81}},\
  \bibinfo {pages} {938} (\bibinfo {year} {1998})}\BibitemShut {NoStop}%
\bibitem [{\citenamefont {Salasnich}\ \emph {et~al.}(2002)\citenamefont
  {Salasnich}, \citenamefont {Parola},\ and\ \citenamefont
  {Reatto}}]{Salasnich2002}%
  \BibitemOpen
  \bibfield  {author} {\bibinfo {author} {\bibfnamefont {L.}~\bibnamefont
  {Salasnich}}, \bibinfo {author} {\bibfnamefont {A.}~\bibnamefont {Parola}},\
  and\ \bibinfo {author} {\bibfnamefont {L.}~\bibnamefont {Reatto}},\
  }\bibfield  {title} {\bibinfo {title} {{Effective wave equations for the
  dynamics of cigar-shaped and disk-shaped Bose condensates}},\ }\href
  {https://doi.org/10.1103/PhysRevA.65.043614} {\bibfield  {journal} {\bibinfo
  {journal} {Phys. Rev. A}\ }\textbf {\bibinfo {volume} {65}},\ \bibinfo
  {pages} {043614} (\bibinfo {year} {2002})}\BibitemShut {NoStop}%
\bibitem [{\citenamefont {Wimberger}\ \emph
  {et~al.}(2005{\natexlab{a}})\citenamefont {Wimberger}, \citenamefont
  {Mannella}, \citenamefont {Morsch},\ and\ \citenamefont
  {Arimondo}}]{Wimberger2005}%
  \BibitemOpen
  \bibfield  {author} {\bibinfo {author} {\bibfnamefont {S.}~\bibnamefont
  {Wimberger}}, \bibinfo {author} {\bibfnamefont {R.}~\bibnamefont {Mannella}},
  \bibinfo {author} {\bibfnamefont {O.}~\bibnamefont {Morsch}},\ and\ \bibinfo
  {author} {\bibfnamefont {E.}~\bibnamefont {Arimondo}},\ }\bibfield  {title}
  {\bibinfo {title} {{Resonant Nonlinear Quantum Transport for a Periodically
  Kicked Bose Condensate}},\ }\href
  {https://doi.org/10.1103/PhysRevLett.94.130404} {\bibfield  {journal}
  {\bibinfo  {journal} {Phys. Rev. Lett.}\ }\textbf {\bibinfo {volume} {94}},\
  \bibinfo {pages} {130404} (\bibinfo {year} {2005}{\natexlab{a}})}\BibitemShut
  {NoStop}%
\bibitem [{\citenamefont {Wimberger}\ \emph
  {et~al.}(2005{\natexlab{b}})\citenamefont {Wimberger}, \citenamefont
  {Mannella}, \citenamefont {Morsch}, \citenamefont {Arimondo}, \citenamefont
  {Kolovsky},\ and\ \citenamefont {Buchleitner}}]{Wimberger2005b}%
  \BibitemOpen
  \bibfield  {author} {\bibinfo {author} {\bibfnamefont {S.}~\bibnamefont
  {Wimberger}}, \bibinfo {author} {\bibfnamefont {R.}~\bibnamefont {Mannella}},
  \bibinfo {author} {\bibfnamefont {O.}~\bibnamefont {Morsch}}, \bibinfo
  {author} {\bibfnamefont {E.}~\bibnamefont {Arimondo}}, \bibinfo {author}
  {\bibfnamefont {A.~R.}\ \bibnamefont {Kolovsky}},\ and\ \bibinfo {author}
  {\bibfnamefont {A.}~\bibnamefont {Buchleitner}},\ }\bibfield  {title}
  {\bibinfo {title} {{Nonlinearity-induced destruction of resonant tunneling in
  the Wannier-Stark problem}},\ }\href
  {https://doi.org/10.1103/PhysRevA.72.063610} {\bibfield  {journal} {\bibinfo
  {journal} {Phys. Rev. A}\ }\textbf {\bibinfo {volume} {72}},\ \bibinfo
  {pages} {063610} (\bibinfo {year} {2005}{\natexlab{b}})}\BibitemShut
  {NoStop}%
\bibitem [{\citenamefont {Owen}\ \emph {et~al.}(1965)\citenamefont {Owen},
  \citenamefont {Abramowitz},\ and\ \citenamefont {Stegun}}]{Owen1965}%
  \BibitemOpen
  \bibfield  {author} {\bibinfo {author} {\bibfnamefont {D.~B.}\ \bibnamefont
  {Owen}}, \bibinfo {author} {\bibfnamefont {M.}~\bibnamefont {Abramowitz}},\
  and\ \bibinfo {author} {\bibfnamefont {I.~A.}\ \bibnamefont {Stegun}},\
  }\bibfield  {title} {\bibinfo {title} {{Handbook of Mathematical Functions
  with Formulas, Graphs, and Mathematical Tables}},\ }\href
  {https://doi.org/10.2307/1266136} {\bibfield  {journal} {\bibinfo  {journal}
  {Technometrics}\ }\textbf {\bibinfo {volume} {7}},\ \bibinfo {pages} {78}
  (\bibinfo {year} {1965})}\BibitemShut {NoStop}%
\bibitem [{\citenamefont {Bao}\ and\ \citenamefont {Du}(2004)}]{Bao2004}%
  \BibitemOpen
  \bibfield  {author} {\bibinfo {author} {\bibfnamefont {W.}~\bibnamefont
  {Bao}}\ and\ \bibinfo {author} {\bibfnamefont {Q.}~\bibnamefont {Du}},\
  }\bibfield  {title} {\bibinfo {title} {{Computing the Ground State Solution
  of Bose--Einstein Condensates by a Normalized Gradient Flow}},\ }\href
  {https://doi.org/10.1137/s1064827503422956} {\bibfield  {journal} {\bibinfo
  {journal} {{SIAM} Journal on Scientific Computing}\ }\textbf {\bibinfo
  {volume} {25}},\ \bibinfo {pages} {1674} (\bibinfo {year}
  {2004})}\BibitemShut {NoStop}%
\bibitem [{\citenamefont {Bao}\ \emph {et~al.}(2006)\citenamefont {Bao},
  \citenamefont {Chern},\ and\ \citenamefont {Lim}}]{Bao2006}%
  \BibitemOpen
  \bibfield  {author} {\bibinfo {author} {\bibfnamefont {W.}~\bibnamefont
  {Bao}}, \bibinfo {author} {\bibfnamefont {I.-L.}\ \bibnamefont {Chern}},\
  and\ \bibinfo {author} {\bibfnamefont {F.~Y.}\ \bibnamefont {Lim}},\
  }\bibfield  {title} {\bibinfo {title} {{Efficient and spectrally accurate
  numerical methods for computing ground and first excited states in
  Bose{\textendash}Einstein condensates}},\ }\href
  {https://doi.org/10.1016/j.jcp.2006.04.019} {\bibfield  {journal} {\bibinfo
  {journal} {Journal of Computational Physics}\ }\textbf {\bibinfo {volume}
  {219}},\ \bibinfo {pages} {836} (\bibinfo {year} {2006})}\BibitemShut
  {NoStop}%
\bibitem [{\citenamefont {Choquard}\ \emph {et~al.}(2008)\citenamefont
  {Choquard}, \citenamefont {Stubbe},\ and\ \citenamefont
  {Vuffray}}]{Choquard2008}%
  \BibitemOpen
  \bibfield  {author} {\bibinfo {author} {\bibfnamefont {P.}~\bibnamefont
  {Choquard}}, \bibinfo {author} {\bibfnamefont {J.}~\bibnamefont {Stubbe}},\
  and\ \bibinfo {author} {\bibfnamefont {M.}~\bibnamefont {Vuffray}},\
  }\bibfield  {title} {\bibinfo {title} {{Stationary solutions of the
  Schr{\"o}dinger-Newton model---an ODE approach}},\ }\href@noop {} {\bibfield
  {journal} {\bibinfo  {journal} {Differential and integral equations}\
  }\textbf {\bibinfo {volume} {21}},\ \bibinfo {pages} {665} (\bibinfo {year}
  {2008})}\BibitemShut {NoStop}%
\bibitem [{\citenamefont {Pang}(2006)}]{Pang2006}%
  \BibitemOpen
  \bibfield  {author} {\bibinfo {author} {\bibfnamefont {T.}~\bibnamefont
  {Pang}},\ }\href {https://doi.org/10.1017/cbo9780511800870} {\emph {\bibinfo
  {title} {{An Introduction to Computational Physics}}}}\ (\bibinfo
  {publisher} {Cambridge University Press},\ \bibinfo {year}
  {2006})\BibitemShut {NoStop}%
\bibitem [{\citenamefont {Drazin}(1989)}]{Drazin1989}%
  \BibitemOpen
  \bibfield  {author} {\bibinfo {author} {\bibfnamefont {P.~G.}\ \bibnamefont
  {Drazin}},\ }\href@noop {} {\emph {\bibinfo {title} {{Solitons: an
  introduction}}}}\ (\bibinfo  {publisher} {Cambridge University Press},\
  \bibinfo {address} {Cambridge England New York},\ \bibinfo {year}
  {1989})\BibitemShut {NoStop}%
\bibitem [{\citenamefont {Lynden-Bell}(1967)}]{Lynden-Bell1967}%
  \BibitemOpen
  \bibfield  {author} {\bibinfo {author} {\bibfnamefont {D.}~\bibnamefont
  {Lynden-Bell}},\ }\bibfield  {title} {\bibinfo {title} {{Statistical
  Mechanics of Violent Relaxation in Stellar Systems}},\ }\href
  {https://doi.org/10.1093/mnras/136.1.101} {\bibfield  {journal} {\bibinfo
  {journal} {Monthly Notices of the Royal Astronomical Society}\ }\textbf
  {\bibinfo {volume} {136}},\ \bibinfo {pages} {101} (\bibinfo {year}
  {1967})}\BibitemShut {NoStop}%
\bibitem [{\citenamefont {Binney}(2004)}]{Binney2004}%
  \BibitemOpen
  \bibfield  {author} {\bibinfo {author} {\bibfnamefont {J.}~\bibnamefont
  {Binney}},\ }\bibfield  {title} {\bibinfo {title} {{Discreteness effects in
  cosmological N-body simulations}},\ }\href
  {https://doi.org/10.1111/j.1365-2966.2004.07699.x} {\bibfield  {journal}
  {\bibinfo  {journal} {Monthly Notices of the Royal Astronomical Society}\
  }\textbf {\bibinfo {volume} {350}},\ \bibinfo {pages} {939} (\bibinfo {year}
  {2004})}\BibitemShut {NoStop}%
\bibitem [{\citenamefont {Weislinger}\ and\ \citenamefont
  {Olivier}(2009)}]{Weislinger2009}%
  \BibitemOpen
  \bibfield  {author} {\bibinfo {author} {\bibfnamefont {E.}~\bibnamefont
  {Weislinger}}\ and\ \bibinfo {author} {\bibfnamefont {G.}~\bibnamefont
  {Olivier}},\ }\bibfield  {title} {\bibinfo {title} {{The classical and
  quantum mechanical virial theorem}},\ }\href
  {https://doi.org/10.1002/qua.560080842} {\bibfield  {journal} {\bibinfo
  {journal} {International Journal of Quantum Chemistry}\ }\textbf {\bibinfo
  {volume} {8}},\ \bibinfo {pages} {389} (\bibinfo {year} {2009})}\BibitemShut
  {NoStop}%
\bibitem [{\citenamefont {Esteve}\ \emph {et~al.}(2012)\citenamefont {Esteve},
  \citenamefont {Falceto},\ and\ \citenamefont {Giri}}]{Esteve2012}%
  \BibitemOpen
  \bibfield  {author} {\bibinfo {author} {\bibfnamefont {J.~G.}\ \bibnamefont
  {Esteve}}, \bibinfo {author} {\bibfnamefont {F.}~\bibnamefont {Falceto}},\
  and\ \bibinfo {author} {\bibfnamefont {P.~R.}\ \bibnamefont {Giri}},\
  }\bibfield  {title} {\bibinfo {title} {{Boundary contributions to the
  hypervirial theorem}},\ }\bibfield  {journal} {\bibinfo  {journal} {Physical
  Review A}\ }\textbf {\bibinfo {volume} {85}},\ \href
  {https://doi.org/10.1103/physreva.85.022104} {10.1103/physreva.85.022104}
  (\bibinfo {year} {2012})\BibitemShut {NoStop}%
\bibitem [{\citenamefont {Wehrl}(1979)}]{Wehrl1979}%
  \BibitemOpen
  \bibfield  {author} {\bibinfo {author} {\bibfnamefont {A.}~\bibnamefont
  {Wehrl}},\ }\bibfield  {title} {\bibinfo {title} {{On the relation between
  classical and quantum-mechanical entropy}},\ }\href
  {https://doi.org/10.1016/0034-4877(79)90070-3} {\bibfield  {journal}
  {\bibinfo  {journal} {Reports on Mathematical Physics}\ }\textbf {\bibinfo
  {volume} {16}},\ \bibinfo {pages} {353} (\bibinfo {year} {1979})}\BibitemShut
  {NoStop}%
\bibitem [{\citenamefont {Antoine}\ \emph {et~al.}(2013)\citenamefont
  {Antoine}, \citenamefont {Bao},\ and\ \citenamefont {Besse}}]{Antoine2013}%
  \BibitemOpen
  \bibfield  {author} {\bibinfo {author} {\bibfnamefont {X.}~\bibnamefont
  {Antoine}}, \bibinfo {author} {\bibfnamefont {W.}~\bibnamefont {Bao}},\ and\
  \bibinfo {author} {\bibfnamefont {C.}~\bibnamefont {Besse}},\ }\bibfield
  {title} {\bibinfo {title} {{Computational methods for the dynamics of the
  nonlinear Schr{\"o}dinger/Gross-Pitaevskii equations}},\ }\href
  {https://doi.org/10.1016/j.cpc.2013.07.012} {\bibfield  {journal} {\bibinfo
  {journal} {Computer Physics Communications}\ }\textbf {\bibinfo {volume}
  {184}},\ \bibinfo {pages} {2621} (\bibinfo {year} {2013})}\BibitemShut
  {NoStop}%
\bibitem [{\citenamefont {Zhang}\ \emph {et~al.}(2019)\citenamefont {Zhang},
  \citenamefont {Liu},\ and\ \citenamefont {Chu}}]{Zhang2019}%
  \BibitemOpen
  \bibfield  {author} {\bibinfo {author} {\bibfnamefont {J.}~\bibnamefont
  {Zhang}}, \bibinfo {author} {\bibfnamefont {H.}~\bibnamefont {Liu}},\ and\
  \bibinfo {author} {\bibfnamefont {M.-C.}\ \bibnamefont {Chu}},\ }\bibfield
  {title} {\bibinfo {title} {{Cosmological Simulation for Fuzzy Dark Matter
  Model}},\ }\bibfield  {journal} {\bibinfo  {journal} {Frontiers in Astronomy
  and Space Sciences}\ }\textbf {\bibinfo {volume} {5}},\ \href
  {https://doi.org/10.3389/fspas.2018.00048} {10.3389/fspas.2018.00048}
  (\bibinfo {year} {2019})\BibitemShut {NoStop}%
\bibitem [{\citenamefont {Sergio~Blanes}(2016)}]{Blanes2016}%
  \BibitemOpen
  \bibfield  {author} {\bibinfo {author} {\bibfnamefont {F.~C.}\ \bibnamefont
  {Sergio~Blanes}},\ }\href@noop {} {\emph {\bibinfo {title} {{A Concise
  Introduction to Geometric Numerical Integration}}}}\ (\bibinfo  {publisher}
  {Apple Academic Press Inc.},\ \bibinfo {year} {2016})\BibitemShut {NoStop}%
\bibitem [{\citenamefont {Dodelson}(2003)}]{Dodelson2003}%
  \BibitemOpen
  \bibfield  {author} {\bibinfo {author} {\bibfnamefont {S.}~\bibnamefont
  {Dodelson}},\ }\href@noop {} {\emph {\bibinfo {title} {{Modern Cosmology}}}}\
  (\bibinfo  {publisher} {Elsevier LTD, Oxford},\ \bibinfo {year}
  {2003})\BibitemShut {NoStop}%
\bibitem [{\citenamefont {Lewis}\ \emph {et~al.}(2000)\citenamefont {Lewis},
  \citenamefont {Challinor},\ and\ \citenamefont {Lasenby}}]{Lewis1999}%
  \BibitemOpen
  \bibfield  {author} {\bibinfo {author} {\bibfnamefont {A.}~\bibnamefont
  {Lewis}}, \bibinfo {author} {\bibfnamefont {A.}~\bibnamefont {Challinor}},\
  and\ \bibinfo {author} {\bibfnamefont {A.}~\bibnamefont {Lasenby}},\
  }\bibfield  {title} {\bibinfo {title} {{Efficient computation of CMB
  anisotropies in closed FRW models}},\ }\href {https://doi.org/10.1086/309179}
  {\bibfield  {journal} {\bibinfo  {journal} {ApJ}\ }\textbf {\bibinfo {volume}
  {538}},\ \bibinfo {pages} {473} (\bibinfo {year} {2000})},\ \Eprint
  {https://arxiv.org/abs/astro-ph/9911177} {arXiv:astro-ph/9911177 [astro-ph]}
  \BibitemShut {NoStop}%
%%CITATION = ASTRO-PH/9911177;%%
\bibitem [{\citenamefont {Chen}\ and\ \citenamefont
  {Pietroni}(2020)}]{Chen2020}%
  \BibitemOpen
  \bibfield  {author} {\bibinfo {author} {\bibfnamefont {S.-F.}\ \bibnamefont
  {Chen}}\ and\ \bibinfo {author} {\bibfnamefont {M.}~\bibnamefont
  {Pietroni}},\ }\bibfield  {title} {\bibinfo {title} {{Asymptotic expansions
  for Large Scale Structure}},\ }\href
  {https://doi.org/10.1088/1475-7516/2020/06/033} {\bibfield  {journal}
  {\bibinfo  {journal} {Journal of Cosmology and Astroparticle Physics}\
  }\textbf {\bibinfo {volume} {2020}}\bibinfo  {number} { (06)},\ \bibinfo
  {pages} {033}}\BibitemShut {NoStop}%
\bibitem [{\citenamefont {Pietroni}(2018)}]{Pietroni2018}%
  \BibitemOpen
\bibfield  {number} {  }\bibfield  {author} {\bibinfo {author} {\bibfnamefont
  {M.}~\bibnamefont {Pietroni}},\ }\bibfield  {title} {\bibinfo {title}
  {Structure formation beyond shell-crossing: nonperturbative expansions and
  late-time attractors},\ }\href
  {https://doi.org/10.1088/1475-7516/2018/06/028} {\bibfield  {journal}
  {\bibinfo  {journal} {Journal of Cosmology and Astroparticle Physics}\
  }\textbf {\bibinfo {volume} {2018}}\bibinfo  {number} { (06)},\ \bibinfo
  {pages} {028}}\BibitemShut {NoStop}%
\bibitem [{\citenamefont {Einasto}(1965)}]{Einasto1965}%
  \BibitemOpen
\bibfield  {number} {  }\bibfield  {author} {\bibinfo {author} {\bibfnamefont
  {J.}~\bibnamefont {Einasto}},\ }\bibfield  {title} {\bibinfo {title} {{On the
  Construction of a Composite Model for the Galaxy and on the Determination of
  the System of Galactic Parameters}},\ }\href@noop {} {\bibfield  {journal}
  {\bibinfo  {journal} {Trudy Astrofizicheskogo Instituta Alma-Ata}\ }\textbf
  {\bibinfo {volume} {5}},\ \bibinfo {pages} {87} (\bibinfo {year}
  {1965})}\BibitemShut {NoStop}%
\bibitem [{\citenamefont {Schulz}\ \emph {et~al.}(2013)\citenamefont {Schulz},
  \citenamefont {Dehnen}, \citenamefont {Jungman},\ and\ \citenamefont
  {Tremaine}}]{Schulz2013}%
  \BibitemOpen
  \bibfield  {author} {\bibinfo {author} {\bibfnamefont {A.~E.}\ \bibnamefont
  {Schulz}}, \bibinfo {author} {\bibfnamefont {W.}~\bibnamefont {Dehnen}},
  \bibinfo {author} {\bibfnamefont {G.}~\bibnamefont {Jungman}},\ and\ \bibinfo
  {author} {\bibfnamefont {S.}~\bibnamefont {Tremaine}},\ }\bibfield  {title}
  {\bibinfo {title} {{Gravitational Collapse in One Dimension}},\ }\href
  {https://doi.org/10.1093/mnras/stt073} {\bibfield  {journal} {\bibinfo
  {journal} {Monthly Notices of the Royal Astronomical Society}\ }\textbf
  {\bibinfo {volume} {431}},\ \bibinfo {pages} {49} (\bibinfo {year}
  {2013})}\BibitemShut {NoStop}%
\bibitem [{\citenamefont {Dong}(2011)}]{Dong2011}%
  \BibitemOpen
  \bibfield  {author} {\bibinfo {author} {\bibfnamefont {X.}~\bibnamefont
  {Dong}},\ }\bibfield  {title} {\bibinfo {title} {{A short note on simplified
  pseudospectral methods for computing ground state and dynamics of spherically
  symmetric Schrödinger-Poisson-Slater system}},\ }\href
  {https://doi.org/10.1016/j.jcp.2011.07.026} {\bibfield  {journal} {\bibinfo
  {journal} {Journal of Computational Physics}\ }\textbf {\bibinfo {volume}
  {230}},\ \bibinfo {pages} {7917} (\bibinfo {year} {2011})}\BibitemShut
  {NoStop}%
\bibitem [{\citenamefont {Zakharov}\ \emph {et~al.}(1988)\citenamefont
  {Zakharov}, \citenamefont {Pushkarev}, \citenamefont {Shvets},\ and\
  \citenamefont {Yan’kov}}]{zakharov1988}%
  \BibitemOpen
  \bibfield  {author} {\bibinfo {author} {\bibfnamefont {V.}~\bibnamefont
  {Zakharov}}, \bibinfo {author} {\bibfnamefont {A.}~\bibnamefont {Pushkarev}},
  \bibinfo {author} {\bibfnamefont {V.}~\bibnamefont {Shvets}},\ and\ \bibinfo
  {author} {\bibfnamefont {V.}~\bibnamefont {Yan’kov}},\ }\bibfield  {title}
  {\bibinfo {title} {{Soliton turbulence}},\ }\href@noop {} {\bibfield
  {journal} {\bibinfo  {journal} {JETP Lett}\ }\textbf {\bibinfo {volume}
  {48}},\ \bibinfo {pages} {79} (\bibinfo {year} {1988})}\BibitemShut {NoStop}%
\bibitem [{\citenamefont {Jordan}\ and\ \citenamefont
  {Josserand}(2000)}]{Jordan2000}%
  \BibitemOpen
  \bibfield  {author} {\bibinfo {author} {\bibfnamefont {R.}~\bibnamefont
  {Jordan}}\ and\ \bibinfo {author} {\bibfnamefont {C.}~\bibnamefont
  {Josserand}},\ }\bibfield  {title} {\bibinfo {title} {{Self-organization in
  nonlinear wave turbulence}},\ }\href
  {https://doi.org/10.1103/PhysRevE.61.1527} {\bibfield  {journal} {\bibinfo
  {journal} {Phys. Rev. E}\ }\textbf {\bibinfo {volume} {61}},\ \bibinfo
  {pages} {1527} (\bibinfo {year} {2000})}\BibitemShut {NoStop}%
\bibitem [{\citenamefont {Maddaloni}\ \emph {et~al.}(2000)\citenamefont
  {Maddaloni}, \citenamefont {Modugno}, \citenamefont {Fort}, \citenamefont
  {Minardi},\ and\ \citenamefont {Inguscio}}]{Maddaloni2000}%
  \BibitemOpen
  \bibfield  {author} {\bibinfo {author} {\bibfnamefont {P.}~\bibnamefont
  {Maddaloni}}, \bibinfo {author} {\bibfnamefont {M.}~\bibnamefont {Modugno}},
  \bibinfo {author} {\bibfnamefont {C.}~\bibnamefont {Fort}}, \bibinfo {author}
  {\bibfnamefont {F.}~\bibnamefont {Minardi}},\ and\ \bibinfo {author}
  {\bibfnamefont {M.}~\bibnamefont {Inguscio}},\ }\bibfield  {title} {\bibinfo
  {title} {{Collective Oscillations of Two Colliding Bose-Einstein
  Condensates}},\ }\href {https://doi.org/10.1103/physrevlett.85.2413}
  {\bibfield  {journal} {\bibinfo  {journal} {Physical Review Letters}\
  }\textbf {\bibinfo {volume} {85}},\ \bibinfo {pages} {2413} (\bibinfo {year}
  {2000})}\BibitemShut {NoStop}%
\bibitem [{\citenamefont {Tabor}(1989)}]{Tabor1989}%
  \BibitemOpen
  \bibfield  {author} {\bibinfo {author} {\bibfnamefont {M.}~\bibnamefont
  {Tabor}},\ }\href@noop {} {\emph {\bibinfo {title} {{Chaos and Integrability
  in Nonlinear Dynamics}}}}\ (\bibinfo  {publisher} {John Wiley \& Sons},\
  \bibinfo {address} {New York},\ \bibinfo {year} {1989})\BibitemShut {NoStop}%
\bibitem [{\citenamefont {Lakshmanan}\ and\ \citenamefont
  {Rajaseekar}(2003)}]{Lakshmanan2003}%
  \BibitemOpen
  \bibfield  {author} {\bibinfo {author} {\bibfnamefont {M.}~\bibnamefont
  {Lakshmanan}}\ and\ \bibinfo {author} {\bibfnamefont {S.}~\bibnamefont
  {Rajaseekar}},\ }\href@noop {} {\emph {\bibinfo {title} {{Nonlinear Dynamics:
  Integrability, Chaos and Patterns}}}}\ (\bibinfo  {publisher} {Springer
  Verlag},\ \bibinfo {address} {Heidelberg},\ \bibinfo {year}
  {2003})\BibitemShut {NoStop}%
\bibitem [{\citenamefont {Jain}\ \emph {et~al.}(2007)\citenamefont {Jain},
  \citenamefont {Weinfurtner}, \citenamefont {Visser},\ and\ \citenamefont
  {Gardiner}}]{Jain2007}%
  \BibitemOpen
  \bibfield  {author} {\bibinfo {author} {\bibfnamefont {P.}~\bibnamefont
  {Jain}}, \bibinfo {author} {\bibfnamefont {S.}~\bibnamefont {Weinfurtner}},
  \bibinfo {author} {\bibfnamefont {M.}~\bibnamefont {Visser}},\ and\ \bibinfo
  {author} {\bibfnamefont {C.~W.}\ \bibnamefont {Gardiner}},\ }\bibfield
  {title} {\bibinfo {title} {{Analogue model of a FRW universe in Bose-Einstein
  condensates: Application of the classical field method}},\ }\href
  {https://doi.org/10.1103/PhysRevA.76.033616} {\bibfield  {journal} {\bibinfo
  {journal} {Phys. Rev. A}\ }\textbf {\bibinfo {volume} {76}},\ \bibinfo
  {pages} {033616} (\bibinfo {year} {2007})}\BibitemShut {NoStop}%
\bibitem [{\citenamefont {Girelli}\ \emph {et~al.}(2008)\citenamefont
  {Girelli}, \citenamefont {Liberati},\ and\ \citenamefont
  {Sindoni}}]{Girelli2008}%
  \BibitemOpen
  \bibfield  {author} {\bibinfo {author} {\bibfnamefont {F.}~\bibnamefont
  {Girelli}}, \bibinfo {author} {\bibfnamefont {S.}~\bibnamefont {Liberati}},\
  and\ \bibinfo {author} {\bibfnamefont {L.}~\bibnamefont {Sindoni}},\
  }\bibfield  {title} {\bibinfo {title} {{Gravitational dynamics in Bose
  Einstein condensates}},\ }\href {https://doi.org/10.1103/PhysRevD.78.084013}
  {\bibfield  {journal} {\bibinfo  {journal} {Phys. Rev. D}\ }\textbf {\bibinfo
  {volume} {78}},\ \bibinfo {pages} {084013} (\bibinfo {year}
  {2008})}\BibitemShut {NoStop}%
\bibitem [{\citenamefont {Proment}\ \emph {et~al.}(2009)\citenamefont
  {Proment}, \citenamefont {Nazarenko},\ and\ \citenamefont
  {Onorato}}]{Proment2009}%
  \BibitemOpen
  \bibfield  {author} {\bibinfo {author} {\bibfnamefont {D.}~\bibnamefont
  {Proment}}, \bibinfo {author} {\bibfnamefont {S.}~\bibnamefont {Nazarenko}},\
  and\ \bibinfo {author} {\bibfnamefont {M.}~\bibnamefont {Onorato}},\
  }\bibfield  {title} {\bibinfo {title} {{Quantum turbulence cascades in the
  Gross-Pitaevskii model}},\ }\href
  {https://doi.org/10.1103/PhysRevA.80.051603} {\bibfield  {journal} {\bibinfo
  {journal} {Phys. Rev. A}\ }\textbf {\bibinfo {volume} {80}},\ \bibinfo
  {pages} {051603} (\bibinfo {year} {2009})}\BibitemShut {NoStop}%
\bibitem [{\citenamefont {Lahaye}\ \emph {et~al.}(2009)\citenamefont {Lahaye},
  \citenamefont {Menotti}, \citenamefont {Santos}, \citenamefont {Lewenstein},\
  and\ \citenamefont {Pfau}}]{Lahaye2009}%
  \BibitemOpen
  \bibfield  {author} {\bibinfo {author} {\bibfnamefont {T.}~\bibnamefont
  {Lahaye}}, \bibinfo {author} {\bibfnamefont {C.}~\bibnamefont {Menotti}},
  \bibinfo {author} {\bibfnamefont {L.}~\bibnamefont {Santos}}, \bibinfo
  {author} {\bibfnamefont {M.}~\bibnamefont {Lewenstein}},\ and\ \bibinfo
  {author} {\bibfnamefont {T.}~\bibnamefont {Pfau}},\ }\bibfield  {title}
  {\bibinfo {title} {{The physics of dipolar bosonic quantum gases}},\ }\href
  {https://doi.org/10.1088/0034-4885/72/12/126401} {\bibfield  {journal}
  {\bibinfo  {journal} {Reports on Progress in Physics}\ }\textbf {\bibinfo
  {volume} {72}},\ \bibinfo {pages} {126401} (\bibinfo {year}
  {2009})}\BibitemShut {NoStop}%
\bibitem [{\citenamefont {Plestid}\ \emph {et~al.}(2018)\citenamefont
  {Plestid}, \citenamefont {Mahon},\ and\ \citenamefont
  {O'Dell}}]{Plestid2018}%
  \BibitemOpen
  \bibfield  {author} {\bibinfo {author} {\bibfnamefont {R.}~\bibnamefont
  {Plestid}}, \bibinfo {author} {\bibfnamefont {P.}~\bibnamefont {Mahon}},\
  and\ \bibinfo {author} {\bibfnamefont {D.~H.~J.}\ \bibnamefont {O'Dell}},\
  }\bibfield  {title} {\bibinfo {title} {Violent relaxation in quantum fluids
  with long-range interactions},\ }\href
  {https://doi.org/10.1103/PhysRevE.98.012112} {\bibfield  {journal} {\bibinfo
  {journal} {Phys. Rev. E}\ }\textbf {\bibinfo {volume} {98}},\ \bibinfo
  {pages} {012112} (\bibinfo {year} {2018})}\BibitemShut {NoStop}%
\bibitem [{\citenamefont {Combescot}\ \emph {et~al.}(2017)\citenamefont
  {Combescot}, \citenamefont {Combescot},\ and\ \citenamefont
  {Dubin}}]{Combescot2017}%
  \BibitemOpen
  \bibfield  {author} {\bibinfo {author} {\bibfnamefont {M.}~\bibnamefont
  {Combescot}}, \bibinfo {author} {\bibfnamefont {R.}~\bibnamefont
  {Combescot}},\ and\ \bibinfo {author} {\bibfnamefont {F.}~\bibnamefont
  {Dubin}},\ }\bibfield  {title} {\bibinfo {title} {{Bose{\textendash}Einstein
  condensation and indirect excitons: a review}},\ }\href
  {https://doi.org/10.1088/1361-6633/aa50e3} {\bibfield  {journal} {\bibinfo
  {journal} {Reports on Progress in Physics}\ }\textbf {\bibinfo {volume}
  {80}},\ \bibinfo {pages} {066501} (\bibinfo {year} {2017})}\BibitemShut
  {NoStop}%
\bibitem [{\citenamefont {Berezhiani}\ and\ \citenamefont
  {Khoury}(2019)}]{Berezhiani2019}%
  \BibitemOpen
  \bibfield  {author} {\bibinfo {author} {\bibfnamefont {L.}~\bibnamefont
  {Berezhiani}}\ and\ \bibinfo {author} {\bibfnamefont {J.}~\bibnamefont
  {Khoury}},\ }\bibfield  {title} {\bibinfo {title} {{Emergent long-range
  interactions in Bose-Einstein condensates}},\ }\href
  {https://doi.org/10.1103/PhysRevD.99.076003} {\bibfield  {journal} {\bibinfo
  {journal} {Phys. Rev. D}\ }\textbf {\bibinfo {volume} {99}},\ \bibinfo
  {pages} {076003} (\bibinfo {year} {2019})}\BibitemShut {NoStop}%
\end{thebibliography}
\end{document}